\documentclass[11pt,a4paper]{article}

\usepackage{float}
\usepackage{jheppub} 
                     
\usepackage{braket,slashed,bm}
\usepackage{array,multirow}
\usepackage[normalem]{ulem}
\usepackage{xcolor,cancel,youngtab}

\usepackage[T1]{fontenc} 

\usepackage{mathrsfs}

\usepackage{booktabs}
\usepackage{adjustbox}
\usepackage{mathtools}

\usepackage{soul}

\usepackage{tikz}
\usetikzlibrary{arrows,decorations.pathmorphing,backgrounds,positioning,fit,petri,automata,shadows,calendar,mindmap,
decorations.markings,calc}

\definecolor{labelkey}{rgb}{0,0.5,0.0}

\def\nn{\nonumber\\ }
\def\rd{{\rm d}}
\def\abs#1{\left| #1 \right| }

\def\hyp{\mathsf{y}}
\def\tr{{\rm Tr}\,}

\def\gcb{{\overline g_{1}}}
\def\gcw{{\overline g_{2}}}
\def\gcg{{\overline g_{3}}}
\def\gcZ{{\overline g_{Z}}}

\def\tc{{\overline \theta}}
\def\ec{{\overline e}}
\def\sc{{\overline s}}

\def\ckin{c_{H,\text{kin}}}

\def\q{\mathsf{q}}

\renewcommand{\O}{\mathcal{O}}
\newcommand{\op}[3]{\O^{#2,#3}_{#1}}

\newcommand{\wc}[3]{L^{#2,#3}_{#1}}
\newcommand{\lcc}[3]{L_{#1}^{#2,#3}}
\newcommand{\dlcc}[3]{{\dot L}_{#1}^{#2,#3}}
\newcommand{\wcc}[3]{c_{#1}^{#2,#3}}
\newcommand{\dwcc}[3]{{\dot c}_{#1}^{#2,#3}}

\newcommand{\hc}{\mathrm{h.c.}}

{\count255=\time\divide\count255 by 60 \xdef\hourmin{\number\count255}
  \multiply\count255 by-60\advance\count255 by\time
  \xdef\hourmin{\hourmin:\ifnum\count255<10 0\fi\the\count255}}

\begin{document}

\title{Low-Energy Effective Field Theory below the Electroweak Scale: Operators and Matching}

\author[a]{Elizabeth E.~Jenkins,}

\author[a]{Aneesh V.~Manohar,}

\author[a,1]{Peter Stoffer}\note{Corresponding author.}

\affiliation[a]{Department of Physics, University of California at San Diego, 9500 Gilman Drive,\\ La Jolla, CA 92093-0319, USA}

\abstract{The gauge-invariant operators up to dimension six in the low-energy effective field theory below the electroweak scale are classified.  There are 
70 Hermitian dimension-five and 3631 Hermitian dimension-six operators that conserve baryon and lepton number, as well as 
$\Delta B= \pm \Delta L = \pm 1$, $\Delta L=\pm 2$, and $\Delta L=\pm 4$ operators.  The matching onto these operators from the Standard Model Effective Field Theory (SMEFT) up to order $1/\Lambda^2$ is computed at tree level.  SMEFT imposes constraints on the coefficients of the low-energy effective theory, which can be checked experimentally to determine whether the electroweak gauge symmetry is broken by a single fundamental scalar doublet as in SMEFT.  Our results, when combined with the one-loop anomalous dimensions of the low-energy theory 
and the one-loop anomalous dimensions of SMEFT, allow one to compute the low-energy implications of new physics to leading-log accuracy, and combine them consistently with high-energy LHC constraints.

}
\maketitle



\section{Introduction}
\label{sec:Intro}

Experimental results to date are overwhelmingly consistent with the predictions of the Standard Model (SM) with electroweak gauge symmetry spontaneously broken by a fundamental scalar doublet, and a Higgs boson with a mass 
$\sim 125$\,GeV. The absence of new particles at energies up to $\sim 1$\,TeV allows one to parametrize the effects of new physics at LHC energies by higher-dimension gauge-invariant local operators built out of SM fields.
The resulting effective field theory (EFT) is known as the Standard Model Effective Field Theory (SMEFT).  The SMEFT Lagrangian contains the usual SM Lagrangian at dimension four, plus a complete set of independent higher-dimension operators.

At dimension five, SMEFT contains a single lepton-number-violating $\Delta L = 2$ operator and its Hermitian conjugate $\Delta L=-2$ operator, each in a single irreducible flavor representation.  For three generations of fermions, the irreducible flavor representation has 6 components.  The dimension-five operators give rise to dimension-three Majorana mass terms for the left-handed neutrinos in the spontaneously broken theory.  Neutrino oscillation experiments require these neutrino masses to be very small, so the suppression scale of the dimension-five operators is necessarily very large.  Because the dimension-five operators violate lepton number, the lepton-number violation scale $\Lambda_{\slashed{L}}$ that suppresses the dimension-five operators can be naturally much larger than the scale $\Lambda$ that suppresses $\Delta B = \Delta L =0$ operators.

The dimension-six operators of SMEFT are classified in Refs.~\cite{Buchmuller:1985jz,Grzadkowski:2010es}.  For three generations of fermions, there are 2499 independent dimension-six operators (151 irreducible flavor representations) that do not violate baryon number and lepton number~\cite{Alonso:2013hga}.  These dimension-six operators, which are suppressed by a factor $1/\Lambda^2$, give the dominant effects of new physics in SMEFT if $\Lambda \ll \Lambda_{\slashed{L}}$.  
Current LHC experiments are sensitive to $\Lambda$ in the $1-1000$\,TeV range, depending on the operator considered.

In addition to the 2499 dimension-six operators (for three generations of fermions), there are $273$ dimension-six $\Delta B = \Delta L = 1$ operators (7 irreducible flavor representations), and their Hermitian conjugates~\cite{Buchmuller:1985jz,Grzadkowski:2010es,Weinberg:1979sa,Wilczek:1979hc,Abbott:1980zj,Alonso:2014zka}.  These operators are important because they are the leading operators that permit proton decay in SMEFT.  Again, it is natural for both the scales of baryon-number violation and lepton-number violation, $\Lambda_{\slashed{B}}$ and $\Lambda_{\slashed{L}}$, to be much larger than 
$\Lambda$, so these operators can be very suppressed in comparison to the dominant 2499 dimension-six operators that do not violate baryon and lepton number.
SMEFT operators at dimension seven and eight also have been studied recently, and the number of operators in SMEFT at each mass dimension has been determined~\cite{Lehman:2014jma,Lehman:2015coa,Henning:2015alf,Kobach:2016ami,Liao:2016hru}. A comprehensive review on SMEFT can be found in Ref.~\cite{Brivio:2017vri}.

Most of the flavor constraints on the renormalizable SM arise from measurements of low-energy flavor-changing processes.  These low-energy decays can be computed using an EFT derived from the SM obtained by integrating out the massive electroweak gauge bosons $(W^\pm, Z)$, the Higgs boson $h$, and the chiral top quark fermion fields $(t_L, t_R)$.  The resulting low-energy effective field theory of the SM, which is essentially the Fermi theory of weak interactions, contains four-fermion operators at dimension six that give the leading contributions to flavor-changing charged-current weak decays such as $\mu \to e \nu_\mu \bar \nu_e$ or $b \to c e \bar \nu_e$. 
This low-energy EFT (LEFT) has been extensively applied to flavor physics such as $B$ and $K$ decays and mixing, and it provides some of the most accurate tests of the SM and constraints on new physics beyond the SM (for reviews, see \cite{PDG,Buchalla:1995vs}). The effects of new physics can be studied by introducing local-operator coefficients with values that differ from those obtained by matching to the SM.

The gauge group of LEFT is QCD $\times$ QED, and the fermions are the usual quarks and leptons, except that there is no top quark in the theory.  In this paper, we construct all the gauge-invariant operators in LEFT up to dimension six. To our knowledge, a complete classification of these LEFT operators has never been given in the literature. There are $\Delta L=\pm2$ dimension-three  Majorana-neutrino mass terms and $\Delta L=\pm2$ dimension-five neutrino dipole operators. Furthermore, there are 70 $\Delta B=\Delta L=0$ dimension-five quark and lepton dipole operators (10 irreducible flavor representations). At dimension six, we find 3631 $\Delta B = \Delta L =0$ operators (191 irreducible flavor representations), of which 1933 are $CP$-even and 1698 are $CP$-odd.   In addition, there are many dimension-six operators that violate lepton number and baryon number. We give the matching onto these operators at tree level from the SMEFT up to terms of order $v^2/\Lambda^2$. Such a matching has been presented in~\cite{Aebischer:2015fzz} for the subset of operators relevant for $B$-physics. We present here the entire matching equations including flavor-conserving operators.

The EFT framework allows one to search for beyond-the-standard-model (BSM) physics in a model-independent way: instead of testing the predictions of specific new physics models at the LHC or in low-energy experiments and ruling out one model after the other, the EFT approach allows one to obtain experimental constraints on the coefficients of the higher-dimensional operators, or, in the presence of a signal deviating from the SM prediction, to determine non-zero values of (linear combinations of) the operator coefficients. If one is interested in a specific model, one can match the model on the EFT and decide directly if the model is compatible with all experimental constraints.

As is well known, the operator coefficients in the EFT depend on the renormalization scale. In order to avoid the presence of large logarithms, one has to take into account the running and mixing of the operator coefficients from the high scale of BSM physics down to the scale of high-energy collider experiments, and further down to the scale of low-energy precision experiments. The leading effect is obtained from the divergent part of a one-loop calculation. 
The complete one-loop anomalous-dimension matrix of the dimension-six operators in the SMEFT is computed in Refs.~\cite{Grojean:2013kd,Jenkins:2013zja,Jenkins:2013wua,Alonso:2013hga,Alonso:2014zka}. Some results for parts of the anomalous-dimension matrix, with flavor neglected, also can be found in Refs.~\cite{Elias-Miro:2013gya,Elias-Miro:2013mua}. The structure of the renormalization-group mixing is non-trivial and has important implications for the flavor structure of SMEFT. The SMEFT renormalization-group equations (RGE) allow one to compute the running and mixing between the BSM scale down to the electroweak scale.

When going to energies below the electroweak scale, the running and mixing should be calculated in LEFT. In a subsequent publication~\cite{Jenkins:2017dyc}, we give the complete one-loop anomalous dimension matrix for LEFT up to terms of dimension six. It is well known that especially the QCD contribution to running and mixing below the electroweak scale is an important effect, see e.g.\ the review~\cite{Buchalla:1995vs}. Hence, parts of the RGE relevant for particular processes have been well studied in the literature and are known to higher order~\cite{Buchalla:1995vs,Cirigliano:2012ab,Dekens:2013zca,Bhattacharya:2015rsa,Aebischer:2015fzz,Davidson:2016edt,Crivellin:2017rmk,Cirigliano:2017azj,Celis:2017hod,Aebischer:2017gaw,Gonzalez-Alonso:2017iyc,Falkowski:2017pss}. In the case of $b \to s \gamma$, the three-loop matching and four-loop anomalous dimensions are known~\cite{Misiak:2004ew,Czakon:2006ss,Misiak:2017woa}, which is the highest order to which computations have been done. However, the systematic study of the entire RGE that we present in~\cite{Jenkins:2017dyc} is new. In particular, the RGE include non-linear terms quadratic in the dipole coefficients, as well as modifications to the RGE for the QCD and QED gauge couplings and fermion mass matrices due to higher-dimension operators in LEFT.

Combined with previous results on the SMEFT~\cite{Grojean:2013kd,Jenkins:2013zja,Jenkins:2013wua,Alonso:2013hga,Alonso:2014zka}, the calculation given here and 
in~\cite{Jenkins:2017dyc} allows one to compute low-energy consequences of BSM physics in a model-independ\-ent way at leading-log order, i.e.\ tree-level matching plus one-loop running. It also allows one to combine high-energy constraints from the LHC with low-energy constraints e.g.\ from hadronic decays in a unified framework. The results can be used in two different ways. If one assumes that BSM physics respects the electroweak symmetry breaking mechanism of the SM, then one can start with SMEFT operators at a high scale, run down to $M_Z$, match onto LEFT, and then run the LEFT operators down to the low-energy scale of the experimental observables, such as $\mu=m_b$ for $B$ decays. If one instead relaxes the assumption about electroweak symmetry breaking, then one can introduce LEFT operators with arbitrary coefficients at $M_Z$, and run down to low energies. Clearly, starting from the SMEFT imposes constraints on the LEFT coefficients that need not be satisfied in other BSM scenarios, such as Higgs Effective Field Theory (HEFT)~\cite{Feruglio:1992wf,Grinstein:2007iv}.  Experimental checks of these constraints test whether electroweak symmetry in the SM is broken by the Higgs mechanism with a fundamental scalar doublet. The dimension-five dipole operators are particularly interesting, because in the SM their coefficients are of order $\alpha_W m_q/(4\pi v^2) \sim G_F m_q \alpha_W/(4\pi)$, where $m_q$ is a light quark mass, and hence effectively the same size as one-loop dimension-six coefficients, whereas in SMEFT, they can be of order $v/\Lambda^2$ due to matching from the dimension-six dipole operators  $\psi^2 X H$, such as $Q_{dW}$. In contrast, dimension-five dipole operators in HEFT can be of order $1/\Lambda$~\cite{Gavela:2016bzc}. Hence, in this scenario effects that are quadratic in dimension-five LEFT coefficients are parametrically of the same order as effects linear in dimension-six LEFT coefficients. Our results presented here and in~\cite{Jenkins:2017dyc} include these contributions.

The organization of this paper is as follows.
In Section~\ref{sec:SMEFT}, we briefly review  SMEFT, focusing on the operators up to dimension six.  The complete operator basis of SMEFT up to dimension six is listed in Tables~\ref{tab:smeft5ops}, \ref{tab:smeft6ops}, and \ref{tab:smeft6baryonops} of Appendix~\ref{sec:SMEFTBasis}.  
We then consider spontaneously broken SMEFT at the electroweak symmetry-breaking scale $v$ and briefly review salient results of prior work relating the parameters of  SMEFT to the usual parameters of the spontaneously broken SM.  In addition, we discuss modifications of the charged and neutral fermion currents in the spontaneously broken SMEFT.  
The construction of the matching conditions of LEFT in the spontaneously broken SMEFT depends crucially on these modified weak charged and neutral currents. 

Section~\ref{sec:PowerCounting} discusses the power counting of LEFT.  
The expansion of the spontaneously broken SMEFT is in powers of $v/\Lambda$, where $v \sim 246$\,GeV is the vacuum expectation value of the Higgs scalar doublet, which spontaneously breaks the electroweak gauge symmetry down to $SU(3) \times U(1)_Q$, and $\Lambda$ is the scale of new physics. 
LEFT has a double expansion---in addition to the usual $p/v$ expansion of the low-energy weak interactions, it inherits the $v/\Lambda$ expansion of spontaneously broken SMEFT.  Here $p \ll M_{W,Z}$ is a typical low-energy scale such as $m_b$ or $m_\mu$. We explain how to power-count terms in LEFT in the presence of two expansion parameters. 
Section~\ref{sec:Matching} derives the power-counting rule for matching SMEFT onto LEFT at tree level.

In Section~\ref{sec:LEFT}, we classify all the $SU(3) \times U(1)_Q$ invariant operators of LEFT up to dimension six.  A complete operator basis of LEFT up to dimension six is constructed and presented in Tables~\ref{tab:oplist1} and \ref{tab:oplist2} of Appendix~\ref{sec:LEFTBasis}.  
We determine the tree-level matching conditions in SMEFT for all of the LEFT operators, tabulated in Appendix~\ref{sec:MatchingConditions}.  The usual computation of these matching conditions in the renormalizable SM is generalized to include all possible new-physics effects in SMEFT up to dimension-six operators.  Since the SMEFT has far fewer dimension-six operators than the LEFT, there are many relations among LEFT coefficients.  Predictions of this type have been given recently for $B$ decays in Ref.~\cite{Alonso:2014csa}.  The full set of predictions is obtained in this work.

Section~\ref{sec:FlavorPhysics} presents a number of applications of LEFT to well-known flavor-nonconserving processes, illustrating the advantages of using the LEFT operator basis for the analysis of low-energy flavor observables.
Conclusions are given in Sec.~\ref{sec:Conclusions}.


\section{SMEFT}
\label{sec:SMEFT}

Basic results on the SMEFT in the broken phase, which are needed  to compute the matching to the low-energy theory below the electroweak scale, are summarized in this section.  Throughout this paper, we use the notation of Ref.~\cite{Alonso:2013hga}.  

The SM Lagrangian is
\begin{align}
	\label{eq:SM}
	\mathcal{L} _{\rm SM} &= -\frac14 G_{\mu \nu}^A G^{A\mu \nu}-\frac14 W_{\mu \nu}^I W^{I \mu \nu} -\frac14 B_{\mu \nu} B^{\mu \nu}
		+ (D_\mu H^\dagger)(D^\mu H)
		+\sum_{\mathclap{\psi=q,u,d,l,e}} \overline \psi\, i \slashed{D} \, \psi\nn
		&-\lambda \left(H^\dagger H -\frac12 v^2\right)^2- \biggl[ H^{\dagger i} \overline d\, Y_d\, q_{i} + \widetilde H^{\dagger i} \overline u\, Y_u\, q_{i} + H^{\dagger i} \overline e\, Y_e\,  l_{i} + \hc \biggr]\nn
		&+ \frac{\theta_3 g_3^2}{32 \pi^2} \, G^A_{\mu \nu} \widetilde G^{A\, \mu \nu} + \frac{\theta_2 g_2^2}{32 \pi^2} \, W^I_{\mu \nu} \widetilde W^{I\, \mu \nu} + \frac{\theta_1 g_1^2}{32 \pi^2} \, B_{\mu \nu} \widetilde B^{\mu \nu} \, .
\end{align}
The gauge covariant derivative is $D_\mu = \partial_\mu + i g_3 T^A G^A_\mu + i g_2  t^I W^I_\mu + i g_1 Y B_\mu$, where $T^A$ are the $SU(3)$ generators,  $t^I=\tau^I/2$ are the $SU(2)$ generators, and $Y=\hyp$ is the $U(1)$ hypercharge generator.  $SU(2)$ indices $i,j,k$ and $I,J,K$ are in the fundamental and adjoint representations, respectively, and $SU(3)$ indices $A,B,C$ are in the adjoint representation.  $SU(3)$ indices in the fundamental representation are denoted by the Greek letters $\alpha, \beta, \gamma$.
$\widetilde H$ is defined by
\begin{align}
	\widetilde H_i &= \epsilon_{ij} H^{\dagger\, j} \, ,
\end{align}
where the $SU(2)$-invariant tensor $\epsilon_{ij}$ is defined by $\epsilon_{12}=1$ and $\epsilon_{ij}=-\epsilon_{ji}$, $i,j=1,2$.  Fermion fields $q$ and $l$ are left-handed fields, and $u$, $d$, and $e$ are right-handed fields.  Note that theta terms have been added to the SM Lagrangian in Eq.~\eqref{eq:SM} for completeness: they are needed for a splitting of the gauge terms into holomorphic and anti-holomorphic pieces~\cite{Alonso:2014rga}.
The fermion fields have weak-eigenstate indices $r=1,\ldots,n_g$, where $n_g=3$, and the Yukawa couplings are $n_g \times n_g$ matrices.

The SMEFT Lagrangian is the SM Lagrangian~\eqref{eq:SM} plus higher-dimension operators.
In this paper, we consider operators in SMEFT up to dimension six.  The number and quantum numbers of SMEFT operators at dimension five and six are given in Table~\ref{tab:nsmeft}.  An explicit list of the operators in the notation of Ref.~\cite{Grzadkowski:2010es} is presented in 
Tables~\ref{tab:smeft5ops}, \ref{tab:smeft6ops}, and \ref{tab:smeft6baryonops} of Appendix~\ref{sec:SMEFTBasis}.  The coefficients of the SMEFT Lagrangian will be denoted by $C_{\substack{ ll \\ prst}}$, etc.\ in the notation of Refs.~\cite{Jenkins:2013zja,Jenkins:2013wua,Alonso:2013hga}, where $p,r,s,t$ are weak-eigenstate indices,\footnote{Regrettably, there are not enough letters in the alphabet.  Thus, $t$ is a weak-eigenstate index, which can sometimes take the value $t=3$ or $t$, i.e. the top quark.  A similar problem occurs for $s$. Sorry. } and powers of $1/\Lambda$ are included in the coefficients $C$.

The dimension-five Lagrangian of SMEFT is given by the $\Delta L=\pm 2$ operators of Table~\ref{tab:smeft5ops}
\begin{align}
	\label{eq:smeft5}
	\mathcal{L}^{(5)} &= C_{\substack{5 \\ rs}}\epsilon^{ij} \epsilon^{kl}  (l_{ i r}^T C l_{k s}) H_j H_l  + \hc \, ,
\end{align}
where  $i,j,k,l$ are $SU(2)$ indices, $r,s$ are weak-eigenstate indices, and $C = i \gamma^2 \gamma^0$ is the charge-conjugation matrix. The coefficients $C_{\substack{5 \\pr}}$ are symmetric in the weak-eigenstate indices and of order $1/\Lambda$. For $n_g=3$ generations, $C_5$ 
has $n_g(n_g+1)/2=6$ complex entries.
The anomalous dimension for the dimension-five operator was computed in Refs.~\cite{Babu:1993qv,Antusch:2001ck}. On converting to the notation and normalization of Ref.~\cite{Jenkins:2013zja,Jenkins:2013wua,Alonso:2013hga}, it is given by
\begin{align}
	\begin{split}
		\dot C_{5} &= -\frac32 \left[ C_5 (Y_e^\dagger Y_e)+(Y_e^\dagger Y_e)^T C_5 \right] +4 \lambda\, C_5 - 3 g_2^2 \, C_5 + 2 \tr \!\left( 3 Y_u^\dagger Y_u + 3 Y_d^\dagger Y_d + Y_e^\dagger Y_e \right) C_5 \,,
	\end{split}
\end{align}
where $C_5$ and $Y_\psi$, $\psi=e,u,d,$ are matrices in flavor space, and we use the notation
\begin{align}
	\dot C \equiv 16\pi^2 \mu \frac{\rd}{\rd \mu} C \,.
\end{align}

The dimension-six Lagrangian divides into operators that conserve baryon number and lepton number, listed in Table~\ref{tab:smeft6ops}, and the operators with $\Delta B = \Delta L = \pm 1$ listed in Table~\ref{tab:smeft6baryonops}.
It is worth repeating that the scale $\Lambda$ of new physics does not have to be the same for the lepton- and baryon-number preserving and violating sectors in the SMEFT.  

\begin{table}
	\begin{align*}
		\setlength\arraycolsep{0.3cm}
		\begin{array}{c|c|c|c}
			\toprule
			d & \text{quantum numbers} & n_g=1  & n_g=3 \\
			\midrule
			5 &( \Delta L=  2) + {\rm h.c.}  & 1 + 1 & 6 + 6 \\
			6 & \Delta B = \Delta L=0 & 76 =53_+ + 23_- & 2499=1350_++1149_- \\
			6 & (\Delta B = \Delta L=  1)+{\rm h.c.} & 4+4 &   273+273\\
			\bottomrule
	\end{array}
\end{align*}
\caption{Number and quantum numbers of operators in SMEFT at dimensions five and six.  The first column gives the operator dimension $d$, and the second column gives the $\Delta B$ and $\Delta L$ quantum numbers. The third and fourth columns list the number of Hermitian operators in SMEFT for $n_g=1$ and $n_g=3$ generations of fermions, split according to their sign under $CP$. 
\label{tab:nsmeft}
}
\end{table}

\subsection{SMEFT in the Broken Phase} \label{sec:broken}

Electroweak symmetry breaking in SMEFT is modified by the presence of dimension-six operators.
The scalar field can be written in unitary gauge as
\begin{align}
	\label{eq:Hvev}
	H &= \frac{1}{\sqrt 2}
		\left(\begin{array}{c}
			0 \\
			\left[ 1+ \ckin \right]  h + v_T
		\end{array}\right) \,,
\end{align}
where
\begin{align}
	\label{eq:chkindef}
	\ckin &\equiv \left(C_{H\Box}-\frac14 C_{HD}\right)v^2 \,, \qquad
	v_T \equiv \left( 1+ \frac{3 C_H v^2}{8 \lambda} \right) v \,,
\end{align}
in the notation of Ref.~\cite{Alonso:2013hga}.
The rescaling of $h$ in Eq.~\eqref{eq:Hvev} is necessary so that $h$ has a conventionally normalized kinetic energy term (see Ref.~\cite{Alonso:2013hga}), and the vacuum expectation value (VEV) $v_T$ in SMEFT is not the same as $v$ in the Lagrangian Eq.~\eqref{eq:SM} due to the dimension-six contributions to the Higgs interactions,
\begin{align}
	\label{eq:HiggsDim6}
	\mathcal{L}^{(6)}_{H} &=  C_H \left(H^\dagger H \right)^3 + C_{H \Box} \left( H^\dagger H \right) \Box \left( H^\dagger H \right) + C_{HD} \left( H^\dagger D^\mu H \right)^* \left( H^\dagger D_\mu H \right)\,,
\end{align}
which contribute to the scalar potential and kinetic energy terms.

The fermion mass matrices in the SMEFT are modified by dimension-six operators~\cite{Alonso:2013hga}.  The $u$-quark, $d$-quark, and $e$-charged lepton mass matrices are
\begin{align}
	\label{eq:MassMatrices}
	\mathcal{L} &=  - \left[ M_\psi \right]_{rs} \overline \psi_{Rr} \psi_{Ls} + \hc \,,  &
	\left[ M_\psi \right]_{rs} &= \frac{v_T}{\sqrt 2} \left( \left[Y_\psi \right]_{rs}   - \frac12 v^2 C^*_{\substack{\psi H \\ sr}}   \right), & \psi&=u,d,e \,.
\end{align}
The Yukawa coupling matrices of the $h$ boson to the fermions $\mathcal{L}=- h\ \overline u \, \mathcal{Y}\, q + \ldots$
also are modified from those of the SM due to the same dimension-six operators~\cite{Alonso:2013hga}.  The Yukawa couplings in the spontaneously broken SMEFT are
\begin{align}
	\label{eq:Yukawas}
	\left[ {\cal Y}_\psi \right]_{rs} &= \frac{1}{\sqrt 2}  \left[ Y_\psi \right]_{rs}\left[ 1+ \ckin \right] - \frac3{2 \sqrt 2} v^2 C^*_{\substack{\psi H \\ sr}} 
		= \frac{1}{v_T}\left[ M_\psi \right]_{rs} \left[ 1+ \ckin  \right]   - \frac{v^2}{\sqrt 2} C^*_{\substack{\psi H \\ sr}} \,,
\end{align}
$\psi =u,d,e$, where $Y_\psi$ are the Yukawa couplings in the dimension-four SM Lagrangian, and
$\ckin$ and $v_T$ are defined in Eq.~\eqref{eq:chkindef}.  An important feature of SMEFT is that the dimension-six operators $Q_{\psi H}$ generically lead to $h$ boson Yukawa couplings that are no longer simply proportional to the fermion Dirac masses.

The left-handed neutrinos also acquire a Majorana mass matrix upon spontaneous symmetry breakdown from the dimension-five Lagrangian $\mathcal{L}^{(5)}$,
\begin{align}
\label{eq:nuMass}
\mathcal{L} &= - \frac12 \left[ M_\nu \right]_{rs} \left( \nu^T_{Lr} C \nu_{Ls} \right) + \hc \,, &
\left[ M_\nu \right]_{rs} = - C_{\substack{5 \\ rs}} v_T^2.
\end{align}
The Higgs boson couples to the neutrinos via $\mathcal{L} = h\ \left[ \mathcal{Y}_5 \right]_{rs} (\nu^T_{L r} C \nu_{L s} )+ \hc$, where
\begin{align}
\label{eq:nuYuk}
\left[ \mathcal{Y}_5 \right]_{rs} & \equiv v_T \left[ C_5 \right]_{rs} \left[ 1+ \ckin \right] 
\end{align}
is proportional to the Majorana-neutrino mass matrix when keeping only operators up to dimension six in SMEFT.
It is important to note, however, that dimension-seven operators contribute a correction to the above equation at relative order $v^2$, which generically results in a Majorana-neutrino Yukawa coupling $\mathcal{Y}_5$ that is not proportional to the Majorana-neutrino mass matrix.   

\subsection{Flavor Indices}\label{sec:indices}

The spontaneously broken SMEFT Lagrangian is written in terms of fields $q_r$, $l_r$, $u_r$, $d_r$ and $e_r$, where $r=1,\ldots,n_g=3$ is a generation (weak-eigenstate) index.  In this weak-eigenstate basis, the fermion mass matrices are not diagonal.  Transformation from the weak-eigenstate basis to the mass-eigenstate basis for the fermions in the SM results in quark and lepton mixing matrices appearing in the weak charged currents.  Similar effects occur in SMEFT, so care must be taken in switching from the fermion weak eigenstates to the mass eigenstates.  
 
One can make flavor transformations on the SMEFT fields that put the charged fermion  mass matrices in the form
\begin{align}
M_e &\to \text{diag}(m_e,m_\mu,m_\tau), &
M_d &\to  \text{diag}(m_d,m_s,m_b) V^\dagger ,  &
M_u &\to  \text{diag}(m_u,m_c,m_t),
\label{YukExpl}
\end{align}
where $V$ is a unitary mixing matrix, which is the CKM matrix in the SM (but not in SMEFT, see Sec.~\ref{sec:Gmass}).  We will assume that these flavor transformations have been performed.
Then, the weak-eigenstate index is the same as the mass-eigenstate index for the charged leptons, left- and right-handed $u$-type quarks, and right-handed $d$-type quarks.\footnote{Note that the Majorana-neutrino mass matrix $M_\nu$ has not been diagonalized. Throughout this work, we consider only neutrino weak eigenstates, so neutrinos $\nu_{Lr}$ are always weak eigenstates with indices $r=1,2,3$ corresponding to $r=e, \mu, \tau$, and the Majorana-neutrino mass matrix is not diagonal.}  For left-handed $d$-type quarks, one gets the usual relation between weak and mass eigenstates,
\begin{align}
d_{Lr} &= V_{rd}\, d_L + V_{rs}\, s_L + V_{rb}\, b_L \equiv V_{rx}\, d_{Lx}\,,
\label{convert}
\end{align}
where the left-hand side is a weak eigenstate, and the right-hand side is a linear combination of mass eigenstates. It should be clear from the context whether we are referring to weak or mass eigenstates, so we will not distinguish between them by using different symbols $d^\prime$ and $d$.

When we go from SMEFT to the low-energy theory, the $t$ quark is integrated out.  This procedure can be done by letting the weak-eigenstate index (which is the same as the mass-eigenstate index for $u$-type quarks) run over $r=1,\ldots,n_u=2$.  For $d$-type quarks, $r=1,\ldots,n_d=3$, and one converts between weak- and mass-eigenstate indices for left-handed $d$-type quarks using Eq.~(\ref{convert}).
For charged leptons, $r=1,\ldots,n_e=3$, and the weak-eigenstate index is the same as the mass-eigenstate index, whereas for neutrinos, $r=1,\ldots,n_\nu=3$ is always a weak-eigenstate index in this paper.  Treating $n_e$ and $n_\nu$ as independent variables allows our results to be used even if there are light sterile neutrinos or right-handed neutrinos, which have the same quantum numbers as SM neutrinos under $SU(3) \times U(1)_Q$. Furthermore, by using this notation, we do not have to worry about the mixing matrix $V$ when we compute the low-energy EFT Lagrangian. The diagonalization of the $u$- and $d$-quark mass matrices can be done in the low-energy theory, at which point the  matrix $V$ enters via Eq.~\eqref{convert}.  However, a word of warning is needed.  The CKM matrix $V$ that diagonalizes the mass matrices in Eq.~(\ref{YukExpl}) is \emph{not the same} as the mixing matrix $K$, defined in Sec.~\ref{sec:Gmass}, that enters the $W$ boson coupling to the weak charged quark current in SMEFT because of dimension-six operator contributions.  This point is explained in detail in the next subsection.

In  LEFT, we can use either weak-eigenstate indices or mass-eigenstate indices for the quarks and leptons. With our convention, the two agree for all fermions except left-handed $d$-type quarks.  A simple example is useful to illustrate how to convert from weak-eigenstate to mass-eigenstate indices in LEFT.  Consider the SMEFT term
\begin{align}
\mathcal{L} &= C_{\substack{ledq \\ p r s t}}\ (\bar l_p^j e_r)(\bar d_s q_{jt})\,,
\end{align}
where $p,r,s,t$ are weak-eigenstate indices, summed over the values $1,2,3$.  In  LEFT, this operator breaks up into
\begin{align}
\mathcal{L} &= C_{\substack{ledq \\ p r s t}}\left[ (\bar \nu_{Lp} e_{Rr})(\bar d_{Rs} u_{Lt}) + (\bar e_{Lp} e_{Rr})(\bar d_{Rs} d_{Lt})\right] \,,
\end{align}
where $p,r,s,t$ are still weak-eigenstate indices.  Switching to mass-eigenstate indices for the left-handed $d$ quarks yields
\begin{align}
\mathcal{L} &= C_{\substack{ledq \\ p r s t}} (\bar \nu_{Lp} e_{Rr})(\bar d_{Rs} u_{Lt}) + C_{\substack{ledq \\ p r s t}}   V_{tx} (\bar e_{Lp} e_{Rr})(\bar d_{Rs} d_{Lx}) \,,
\label{2.14}
\end{align}
where $p,r,s,t$ are weak-eigenstate indices, but $x$ is a mass-eigenstate index.  Weak-eigenstate and mass-eigenstate indices are the same for 
$e_{L,R}$, $\nu_L$, $u_{L,R}$, and $d_R$, so we can use Eq.~(\ref{2.14}) with $p=1,2,3=e,\mu,\tau$; $r=1,2,3=e,\mu,\tau$; and $s=1,2,3=d,s,b$ for both terms.  For the sum on $t$, however, the sum for the first term is over $t=1,2=u,c$, whereas the sum for the second term is over $t=1,2,3$ and $x=d,s,b$, since all three left-handed $d$-type quarks remain in the low-energy theory.  Since we will often wish to focus on specific mass-eigenstate operators, we also use the notation
\begin{align}
C_{\substack{ledq \\ e \mu s b}}  (\bar e_{L} \mu_{R})(\bar s_{R} b_{L}) 
\end{align}
for the $p=1$, $r=2$, $s=2$, $x=b$ component of the second term in Eq.(\ref{2.14}), where
\begin{align}
C_{\substack{ledq \\ e \mu s b}}  \equiv  \sum_{r=1}^3 C_{\substack{ledq \\ 1 2 2 r}} V_{rb} .
\end{align}
Notice that care is required only for left-handed $d$-quark indices given our conventions.

\subsection{Gauge-Boson Masses and Couplings}
\label{sec:Gmass} 

The gauge bosons of the spontaneously broken SMEFT need to be redefined so that the gauge kinetic terms have canonical normalization.
Non-canonical normalization arises because dimension-six class-4 Higgs--gauge-boson operators $X^2 H^2$, such as $H^\dagger H W_{\mu \nu}^I W^{I\mu \nu}$, contribute to the gauge kinetic terms in the spontaneously broken theory. In addition, the neutral gauge-boson mass matrix needs to be diagonalized to obtain the gauge-boson mass eigenstates of the spontaneously broken SMEFT.  The gauge-field redefinitions that are needed to rewrite the SMEFT Lagrangian in terms of properly normalized gauge-boson mass eigenstates have been given in detail before.  We summarize the required equations here.

The gauge-field and -coupling redefinitions needed to yield gauge kinetic energy and mass terms that are properly normalized and diagonal are~\cite{Grinstein:1991cd,Alonso:2013hga}
\begin{align}
G_\mu^A &= \mathcal{G}_\mu^A \left(1 + C_{HG} v_T^2 \right), &
W^I_\mu  &=  \mathcal{W}^I_\mu \left(1 + C_{HW} v_T^2 \right), &
B_\mu  &=  \mathcal{B}_\mu \left(1 + C_{HB} v_T^2 \right),
\label{5.16a}
\end{align}
\begin{align}
\gcg &= g_3 \left(1 + C_{HG} \, v_T^2 \right), & \gcw &= g_2 \left(1 + C_{HW} \, v_T^2 \right), & \gcb &= g_1 \left(1 + C_{HB} \, v_T^2 \right),
\label{5.16b}
\end{align}
and
\begin{align}
\begin{pmatrix}
{\cal Z}^\mu \\ {\cal A}^\mu \\
\end{pmatrix} &= 
\begin{pmatrix}
\bar c- {\epsilon  \over 2} \bar s &\qquad -\bar s +  {\epsilon  \over 2} \bar c \\
\bar s+ {\epsilon  \over 2} \bar c &\qquad \bar c +  {\epsilon  \over 2} \bar s \\
\end{pmatrix}
\begin{pmatrix}
{\cal W}_3^\mu \\ {\cal B}^\mu \\
\end{pmatrix}, \quad \epsilon \equiv C_{HWB} v_T^2 .
\end{align}
The neutral gauge-boson mass eigenstates ${\cal Z}^\mu$ and ${\cal A}^\mu$ in the above equation depend on the weak mixing angle $\tc$ through 
\begin{align}
\cos \tc &\equiv \bar c = {{\bar g_2} \over \sqrt{ \bar g_1^2 + \bar g_2^2} }\left[ 1 - {\epsilon \over 2} \ {{\bar g_1} \over {\bar g_2}} \left( {{\bar g_2^2 - \bar g_1^2} \over {\bar g_1^2 + \bar g_2^2}} \right)\right], \nn
\sin \tc &\equiv \bar s = {{\bar g_1} \over \sqrt{ \bar g_1^2 + \bar g_2^2} }\left[ 1 + {\epsilon \over 2} \ {{\bar g_2} \over {\bar g_1}} \left( {{\bar g_2^2 - \bar g_1^2} \over {\bar g_1^2 + \bar g_2^2}} \right)   \right].
\label{2.23}
\end{align}
The massive gauge bosons of the spontaneously broken SMEFT are the ${\cal W}^{\pm \mu}$ and ${\cal Z}^\mu$ with masses
\begin{align}
M_{\cal W}^2 &= \frac 14 \bar g_2^2 v_T^2, \nn
M_{\cal Z}^2 &= \frac 14 \left( \bar g_2^2 + \bar g_1^2 \right) v_T^2 \left( 1 + \frac 12 C_{HD} v_T^2 \right)
+ {\epsilon \over 2} \bar g_1 \bar g_2 v_T^2 .
\end{align}
In the above equations, $G^A_\mu$, $W_\mu^I$, and $B_\mu$ and $g_3$, $g_2$, and $g_1$ are the gauge fields and coupling constants in the unbroken SMEFT, and 
${\cal G}^A_\mu$, ${\cal W}_\mu^I$, and ${\cal B}_\mu$ and $\gcg$, $\gcw$, and $\gcb$ are the gauge fields and coupling constants in the spontaneously broken SMEFT.
Note that products of gauge couplings and gauge fields $g_3 G_\mu^A=\gcg \mathcal{G}_\mu^A$, $g_2 W_\mu^A=\bar g_2 \mathcal{W}_\mu^I$, and $g_1 B_\mu=\bar g_1 \mathcal{B}_\mu$ are unchanged by the above redefinitions.

With these redefinitions, the gauge-covariant derivative in the spontaneously broken SMEFT is given in terms of the gauge-boson mass eigenstates by
\begin{align}
D_{\mu} = \partial_\mu + i \, \gcg \mathcal{G}^A_\mu T^A + i \, \frac{\gcw}{\sqrt{2}} \left[\mathcal{W}_{\mu}^+ T^+ + \mathcal{W}_{\mu}^- T^- \right] + i \,
\gcZ \left[T_3 - \sc^2 Q \right] \mathcal{Z}_\mu + 
i \, \ec \, Q \, \mathcal{A}_\mu,
\label{14.4}
\end{align}
where $Q=T_3+Y$,  and the effective couplings $\ec$ and $\gcZ$ are given by
\begin{align}
\ec &= \gcw \, \sin \tc - \frac{1}{2} \, \cos \tc  \, \gcw \, v_T^2 \, C_{HWB}, \nn
\gcZ &= \frac{\ec}{\sin \tc \cos \tc}  \left[1 +  \frac{\gcb^2+\gcw^2}{2 \gcb \gcw} v_T^2C_{HWB}\right] .
\end{align}

In contrast to the SM, the couplings of the massive gauge bosons $\mathcal{W}^\pm_\mu$ and $\mathcal{Z}_\mu$ to fermions are not completely determined in SMEFT by the gauge-covariant derivative in the fermion kinetic energy terms.  There is an additional contribution to fermion couplings arising from the dimension-six $\psi^2 H^2 D$ operators, such as 
$Q_{Hl}^{(1)}=(H^\dag i\overleftrightarrow{D}_\mu H)(\bar l_p \gamma^\mu l_r)$,  listed in Table~\ref{tab:smeft6ops}, which are the product of Higgs currents times  fermion currents.  The  Higgs currents  
evaluated in unitary gauge using Eq.~\eqref{eq:Hvev} (with $\ckin=0$ to leading order) are 
\begin{align}
\langle H^\dag i\overleftrightarrow{D}_\mu H \rangle &= \frac{\gcZ}{2} \mathcal{Z}_\mu (v_T+h)^2 ,\nn
\langle H^\dag i\overleftrightarrow{D}_\mu^I H \rangle &= 
\begin{cases}
-\frac{\gcw}{2} \mathcal{W}^1_\mu (v_T+h)^2, & I=1 ,\\
-\frac{\gcw}{2} \mathcal{W}^2_\mu (v_T+h)^2,  & I=2 ,\\
-\frac{\gcZ}{2} \mathcal{Z}_\mu (v_T+h)^2,  & I=3,
\end{cases} \nn
\langle i \widetilde H^\dag {D}_\mu H \rangle &=-\frac12  \frac{\gcw}{\sqrt 2} \mathcal{W}_\mu^+ (v_T+h)^2 .
\label{14.5}
\end{align}
Using Eq.~\eqref{14.5}, the additional contribution of the $\psi^2 H^2 D$ operators to the couplings of the massive weak gauge bosons can be easily evaluated. 

The fermion couplings to the massive gauge bosons $\mathcal{W}^\pm_\mu$ and $\mathcal{Z}_\mu$ in the spontaneously broken SMEFT take the usual form 
\begin{align}
\mathcal{L} &= -\frac{\gcw}{\sqrt{2}} \left\{ \mathcal{W}^+_\mu j_{\mathcal{W}}^\mu  + h.c. \right\} - \gcZ \mathcal{Z}_\mu j^\mu_{\mathcal{Z}}
\label{14.9}
\end{align}
with the modified weak charged and neutral currents  
\begin{align}
j_{\mathcal{W}}^\mu  &=    [W_l]_{pr} \overline \nu_{Lp}  \gamma^\mu e_{Lr}  +  [W_q]_{pr}  \overline u_{Lp}  \gamma^\mu d_{Lr}+ [W_R]_{pr}\overline u_{Rp}  \gamma^\mu d_{Rr}, \nn
j^\mu_{\mathcal{Z}} &=[Z_{\nu_L}]_{pr} \overline \nu_{Lp}  \gamma^\mu \nu_{Lr} +  [Z_{e_L}]_{pr}  \overline e_{Lp}  \gamma^\mu e_{Lr} + [Z_{e_R}]_{pr} \overline e_{Rp}  \gamma^\mu e_{Rr} \nn
&+ [Z_{u_L}]_{pr}  \overline u_{Lp}  \gamma^\mu u_{Lr} +  [Z_{u_R}]_{pr} \overline u_{Rp}  \gamma^\mu u_{Rr} +  [Z_{d_L}]_{pr}  \overline d_{Lp}  \gamma^\mu d_{Lr} +  [Z_{d_R}]_{pr} \overline d_{Rp}  \gamma^\mu d_{Rr} ,
\end{align}
where
{\small
\begin{align}
[W_l]_{pr} &= \mathrlap{\left[\delta_{pr} + v_T^2  C^{(3)}_{\substack {Hl \\  pr}} \right], \qquad [W_q]_{pr} = \left[\delta_{pr} +  v_T^2  C^{(3)}_{\substack {Hq \\  pr}} \right], \qquad
[W_R]_{pr} = \left[ \frac12 v_T^2  C_{\substack {Hud \\  pr}} \right], } \nn
[Z_{\nu_L}]_{pr} &= \left[\delta_{pr}\left(\frac 12\right) - \frac12 v_T^2  C^{(1)}_{\substack {Hl \\  pr}} + \frac12 v_T^2  C^{(3)}_{\substack {Hl \\  pr}} \right] ,
\nn
[Z_{e_L}]_{pr} &= \left[\delta_{pr}\left(-\frac 12+\sc^2 \right) - \frac12 v_T^2  C^{(1)}_{\substack {Hl \\  pr}} - \frac12 v_T^2  C^{(3)}_{\substack {Hl \\  pr}} \right],  &
[Z_{e_R}]_{pr} &= \left[\delta_{pr}\left(+\sc^2 \right) - \frac12 v_T^2  C_{\substack {He \\  pr}}  \right], \nn
[Z_{u_L}]_{pr} &=  \left[\delta_{pr}\left(\frac 12-\frac 23 \sc^2 \right) - \frac12 v_T^2  C^{(1)}_{\substack {Hq \\  pr}} + \frac12 v_T^2  C^{(3)}_{\substack {Hq \\  pr}} \right],  &
[Z_{u_R}]_{pr} &=  \left[\delta_{pr}\left(-\frac 23 \sc^2 \right) - \frac12 v_T^2  C_{\substack {Hu \\  pr}}  \right],  \nn
[Z_{d_L}]_{pr} &=  \left[\delta_{pr}\left(-\frac 12+ \frac 13 \sc^2 \right) - \frac12 v_T^2  C^{(1)}_{\substack {Hq \\  pr}} - \frac12 v_T^2  C^{(3)}_{\substack {Hq \\  pr}}  \right], &
[Z_{d_R}]_{pr} &=  \left[\delta_{pr}\left(+\frac13\sc^2 \right) - \frac12 v_T^2  C_{\substack {Hd \\  pr}}  \right] .
\label{14.10b}
\end{align}}%
Here, $[W_l]_{pr}$, $[W_q]_{pr}$, and $[W_R]_{pr}$ are the couplings of  $\mathcal{W}^+_\mu$ to $(\overline \nu_{Lp} \gamma^\mu e_{Lr})$, 
$(\overline u_{Lp} \gamma^\mu d_{Lr})$, and $(\overline u_{Rp} \gamma^\mu d_{Rr})$, respectively, and $[Z_\psi]_{pr}$ are the couplings of 
$\mathcal{Z}_\mu$ to $(\overline \psi_p \gamma^\mu \psi_r)$ for $\psi=\nu_L, e_L, e_R, u_L, u_R, d_L$, and $d_R$.  Note that the couplings Eq.~\eqref{14.10b} are written in the weak-eigenstate basis, so the SM contribution is proportional to 
Kronecker-delta symbols.  The dimension-six $\psi^2 H^2 D$ operators in spontaneously broken SMEFT give the $1/\Lambda^2$ contributions proportional to coefficients $C$ in Eq.~(\ref{14.10b}).  Interestingly, in spontaneously broken SMEFT, $\mathcal{W}^+_\mu$ couples to the right-handed charged current $(\overline u_{Rp} \gamma^\mu d_{Rr})$ with coupling $\left[ W_R \right]_{pr}$ due to the dimension-six operator $Q_{Hud}$~\cite{Alioli:2017ces}.

The couplings Eq.~\eqref{14.10b} are in the weak-eigenstate basis, where the mass matrices have the form Eq.~\eqref{YukExpl}.  Performing a flavor rotation on the left-handed $d$ quarks by $V$ diagonalizes the quark mass matrices.  $V$ is a unitary matrix.  The $\mathcal{W}^\pm$ couplings in Eq.~\eqref{14.10b} are not unitary, however, because of the dimension-six operator contributions~\cite{Dedes:2017zog}. The effective quark-mixing matrix in the left-handed quark sector is given by
\begin{align}
K_{px} &= V_{px} + v_T^2  C^{(3)}_{\substack {Hq \\  ps}}V_{sx},
\label{14.11a}
\end{align}
which satisfies
\begin{align}
\left[K K^\dagger\right]_{pr} &= \delta_{pr} + v_T^2  \left[ C^{(3)}_{\substack {Hq \\  pr}} + C^{(3)*}_{\substack {Hq \\  rp}} \right]
+ \mathcal{O}\left( \frac{v^4}{\Lambda^4} \right)
\label{14.11b}
\end{align}
and
\begin{align}
\left[K^\dagger K \right]_{xy} &= \delta_{xy} + v_T^2  \left[ V^\dagger_{xp} C^{(3)}_{\substack {Hq \\  ps}} V_{sy} + V^\dagger_{xs} C^{(3)*}_{\substack {Hq \\  sp}} V_{py} \right]
+ \mathcal{O}\left( \frac{v^4}{\Lambda^4} \right)\,.
\label{14.11c}
\end{align}
Flavor physics experiments at low energies measure coefficients of the dimension-six flavor-changing terms in the Lagrangian, and thus determine matrix elements of $K$, which is the effective quark-mixing matrix in the low-energy theory.  The two matrices $V$ and $K$ are equal in the SM, but differ in SMEFT due to the presence of dimension-six operators in the weak charged currents. $V$ is unitary, but $K$ is not.  In SMEFT, flavor-changing neutral currents also can be present at order $1/\Lambda^2$ due to dimension-six operators.

Non-unitarity of the effective lepton-mixing matrix was studied previously in the context of neutrino physics in Refs.~\cite{Broncano:2003fq,Broncano:2002rw,Broncano:2004tz}, which considered the 
operator
\begin{align}
\mathcal{O}&=C_{rs} (\overline l_{r i} H^\dagger_j) i \overleftrightarrow{\slashed{\partial}}( l_{s k} H_\ell) \epsilon_{ij} \epsilon_{k\ell}\,.
\label{7.1}
\end{align}
Using the equations of motion converts Eq.~(\ref{7.1}) to the SMEFT operator
\begin{align}
\mathcal{O}&=\frac12 C_{rs} Q^{(1)}_{\substack{ H l \\ r s}}  - C_{rs} \frac12 Q^{(3)}_{\substack{ H l \\ r s}} 
\label{14.8}
\end{align}
which shows that unitarity violation of the effective mixing matrix in the lepton sector also is given by the $\psi^2 H^2D$ operators in spontaneously broken SMEFT.


\section{Power Counting in LEFT}
\label{sec:PowerCounting}

The SM below the electroweak scale can be described by an EFT with an expansion in powers of the inverse electroweak scale $1/v$. The expansion parameter is usually written as $G_F=1/(\sqrt 2\, v^2)$.  The dimensionless small parameter that controls the EFT expansion is $p/v$ or $m/v$, where $p$ and $m$ are the momenta and masses of particles in the EFT. For example, in meson weak decays, such as $K$ and $B$ decays, the low-energy scales are of order $m_K$ and $m_b$.
The low-energy theory only contains particles with masses $m \ll v$.  The gauge bosons remaining in the low-energy theory, the gluons and the photon, are all massless, and there are no massless or massive scalar particles remaining since all components of the single fundamental Higgs scalar doublet are integrated out.  The massive particles in the low-energy theory consist of all SM fermions with the exception of the $t$ quark. 
In the SM, fermions get their mass due to electroweak symmetry breaking, and so have a mass $m \sim y v$ proportional to $v$ times a Yukawa coupling. By assumption, there is a low-energy EFT below $v$ with dynamical fermions. This assumption means that the light fermions have a mass $m \ll v$, so their Yukawa couplings are of order $y \sim m/v \ll 1$, and are parametrically suppressed by the power-counting parameter of the low-energy EFT.

The presence of a low-energy mass in LEFT means that renormalization produces running of the Wilson coefficients of lower-dimensional operators proportional to 
the coefficients of higher-dimensional ones. For example, a dimension-four term in the Lagrangian can have an anomalous dimension proportional to $m^2$ times a dimension-six term, $m^4$ times a dimension-eight term, etc.\footnote{Analogous effects in spontaneously broken SMEFT were computed in Ref.~\cite{Alonso:2013hga}.} The dimension-six contribution to a dimension-four term is of order $m^2/v^2$, and is the same order in power counting as a direct dimension-six contribution to a scattering amplitude, which is order $m^2/v^2$ or $p^2/v^2$.

In the low-energy theory starting with SMEFT, we have an additional expansion in powers of $1/\Lambda$. This additional expansion can be easily included in the power counting by writing $1/\Lambda = 1/v \times (v/\Lambda)$, i.e.\  LEFT uses the low-energy power counting in powers of $1/v$ with additional suppression factors in powers of $v/\Lambda$. There are no particles with masses of order $\Lambda$ in SMEFT, so one cannot get positive powers of $\Lambda$. A term in LEFT of dimension $d$ has the power counting
\begin{align}
L &=  \frac{O_d}{v^{d-4}} \left( \frac{v}{\Lambda} \right)^{\sum_i (D_i-4)}
\end{align}
if it arises from a graph with insertions of SMEFT operators of dimension $D_i$.

In addition to the double expansion in powers of $m/v$ and $v/\Lambda$, the low-energy EFT has a loop expansion in powers of $(\alpha,\alpha_s)/(4\pi)$ since it is weakly coupled. In applications such as low-energy weak interactions, the heaviest mass in the problem is the $b$-quark mass, so
$m/v \sim 1/50$.  We do not know the scale $\Lambda$ of the SMEFT expansion. Current experimental data indicate that it is above a few TeV, so that $v/\Lambda \lesssim 1/5$. However, it is possible that $\Lambda$ is much higher than a few TeV, of the order of the seesaw scale or the GUT scale, in which case $v/\Lambda$ could be as small as $10^{-12}$. We already know that $\Lambda_{\slashed{L}}$ for the dimension-five $\Delta L=2$ operator $Q_5$ and $\Lambda_{\slashed{B}}$ for the dimension-six baryon-number violating operators is this large.

In this paper, we compute the tree-level SMEFT contributions to LEFT operators up to dimension-six. These are the leading BSM contributions to the low-energy amplitudes, and experimental constraints on them provide information about BSM physics. The one-loop corrections to these amplitudes in the SMEFT is beyond the scope of this paper.
Higher order matching corrections have been computed in the SM (for a review, see Ref.~\cite{Buchalla:1995vs}).

We stress that LEFT is the correct low-energy theory even in the case where the high-energy EFT is not given by SMEFT but by HEFT~\cite{Feruglio:1992wf,Grinstein:2007iv}, which relaxes the assumption that the Higgs particle is part of a fundamental electroweak doublet. In this case, the dimension-five LEFT operators come with a suppression factor of $1/\Lambda$ rather than $v/\Lambda^2$ as in the case of SMEFT. Therefore, when systematically considering effects up to dimension six in the LEFT power counting one also has to include effects quadratic in the dimension-five LEFT coefficients. In~\cite{Jenkins:2017dyc}, we present the complete one-loop RGE up to dimension six in the LEFT power counting.

\section{Integrating out  Weak-Scale Particles in SMEFT}
\label{sec:Matching}

In this section, we derive the power counting rules for integrating out a heavy particle in an EFT.  We start with a high-energy EFT, which in this paper is the SMEFT, with a power counting scale $\Lambda$ that suppresses higher-dimension operators. The high-energy theory also contains heavy particles with a mass $M$ of order $v\ll \Lambda$, and we want to construct the low-energy EFT below $v$ in which these heavy particles are integrated out. In the SMEFT, the heavy particles are the $W$ and $Z$ gauge bosons, the $t$ quark, and the Higgs boson $h$. The light particles are those with masses $m$ parametrically smaller than $v$, namely the quarks $u,d,s,c,b$, charged leptons $e$, $\mu$, $\tau$, the left-handed neutrinos $\nu_e, \nu_\mu, \nu_\tau$, the photon, and the gluons.

The power-counting rule in the high-energy EFT is that an arbitrary graph with vertices of operator dimension $D_i$ produces an operator with dimension
\begin{align}
D-4 = \sum_i (D_i-4)\,.
\label{10.1}
\end{align}
Eq.~(\ref{10.1}) is a well-known result, which follows simply from counting powers of $\Lambda$.  Operators of dimension $D_i$  
have coefficients of order $1/\Lambda^{D_i-4}$ in the EFT Lagrangian.  Graphs in the theory cannot generate positive powers of $\Lambda$ since all particles have masses parametrically smaller than $\Lambda$, and loop integrals do not generate powers of $\Lambda$ in dimensional regularization. Comparing powers of $\Lambda$ gives Eq.~(\ref{10.1}).

Now, consider the subset of EFT graphs in which all external particles are light particles with masses and momenta much smaller than $v$.  These graphs include graphs with internal heavy particles depending on heavy masses $M \sim v$.   By expanding the internal heavy propagators in $1/M \sim 1/v$,  one obtains that an operator of dimension $D$ in the low-energy EFT has a coefficient of order
\begin{align}
\frac{1}{\Lambda^a} \frac{1}{v^b}, \qquad a+b=D-4, \qquad a \ge 0
\label{10.2}
\end{align}
by dimensional analysis.  Note that $a$ is non-negative because it is not possible to generate positive powers of $\Lambda$ as discussed above.  For tree graphs, $b>0$, but it is possible to obtain $b<0$ via heavy mass insertions in loop graphs.  For example, a loop graph with two insertions of
\begin{align}
\frac{1}{\Lambda^2} (\overline b \gamma^\mu P_L t)( \overline t \gamma_\mu P_L b )
\end{align}
gives a contribution to $(\overline b \gamma^\mu P_L b)( \overline b \gamma_\mu P_L b )$ with a coefficient of order $M_t^2/\Lambda^4$, i.e.\ $a=4$, $b=-2$.  Thus, in loop graphs, powers of $M$ in the numerator can cancel powers of $v$ in the denominator. In the SM, $M/v$ is a gauge coupling, a Yukawa coupling, or $\sqrt{\lambda}$, so $M/v$ corrections are comparable to radiative corrections. This result is well-known from the matching calculation for the weak interactions in the SM.  

In this paper, we compute tree-level matching when particles of mass $M \sim v$ are integrated out. Each internal heavy fermion line starts at order $1/M$ plus terms with additional factors of $p/M$, and each internal heavy boson line starts at order $1/M^2$ plus terms with additional factors of $p^2/M^2$.  Additional factors of $1/M$ in the denominator are compensated by the dimensions of external fields, i.e.\ a tree graph with heavy internal particle lines generates operators with dimension
\begin{align}
D-4 &= \sum_i (D_i-4) + I_F + 2 I_B ,
\label{10.3}
\end{align}
where $I_F$ is the number of internal heavy fermion lines and $I_B$ is the number of internal heavy boson lines.
Another way to derive Eq.~\eqref{10.3} is to count dimensions.  A graph (tree or loop) with vertices of dimension $D_i$ gives an operator with dimension
\begin{align}
D &= \sum_i D_i  - 3 I_F - 2 I_B ,
\label{10.4}
\end{align}
since each internal heavy line removes two heavy fields, and fermions fields have dimension $3/2$ and boson fields have dimension $1$.  For a tree diagram with only heavy internal lines, one also has the relations 
\begin{align}
V-I=1, \qquad I=I_F+I_B,
\label{10.5}
\end{align}
where $V$ is the number of vertices and $I$ is the number of  internal lines.
Eq.~\eqref{10.5}, when combined with Eq.~\eqref{10.4}, gives back Eq.~\eqref{10.3} for tree graphs with only internal heavy lines.

For tree graphs, we also have the relations
\begin{align}
2 I_F &= \sum_i F_i & 2 I_B &= \sum_i B_i
\label{10.6}
\end{align}
where $F_i$ and $B_i$ are the number of heavy fermion and boson fields at each vertex, since there are no external heavy fields in the low-energy EFT, and each internal heavy line removes two fields.
Eq.~\eqref{10.6} combined with Eq.~\eqref{10.3} gives
\begin{align}
D-4 &= \sum_i \left(D_i+\frac12 F_i + B_i-4\right) = \sum_i w_i, \qquad w_i \equiv D_i+\frac12 F_i + B_i-4\,.
\label{10.7}
\end{align}
$w_i$ can take on integral or half-integral values. There must be an even number of half-integral values of $w_i$ since $\sum_i F_i$ is even.

We construct the EFT below $v$ up to dimension six, so we need $\sum_i w_i \le 2$. Furthermore, since we only compute tree-level matching, every operator vertex must have at least one heavy field, and each graph must have at least two such vertices. Operators with no heavy fields match directly to the low-energy theory, i.e.\ the operator survives in the low-energy theory.
\begin{table}
\small
\begin{align*}
\begin{array}{c|cccccccc}
w & \multicolumn{8}{c}{\text{Operators}} \\
\hline
\frac12 & \psi t A_\mu \\
1 & Z_{\mu \nu} A^{\mu \nu} &  h \psi^2 & \psi^2 Z_\mu & t^2 A_{\mu}  \\
\frac32 & h \psi t & \psi t Z_\mu & \psi t A_{\mu \nu} \\
2 & h^3 & h Z_\mu Z^\mu & h A_{\mu \nu} A^{\mu \nu}  & Z_{\mu \nu} Z^{\mu \nu} & ht^2 & t^2 Z_\mu  &\psi^2 Z_{\mu \nu} 
& t^2 A_{\mu \nu} \\
\end{array}
\end{align*}
\caption{\label{w}Table of operators in SMEFT containing at least one heavy field with weights $w\le2$.   Operators are denoted by their field content, where $Z_\mu$ and $Z_{\mu \nu}$ are the heavy $Z$ (or $W$) field and field-strength tensor, respectively; $A_\mu$ and $A_{\mu \nu}$ are the light photon (or gluon) field and field-strength tensor, respectively; $\psi$ is a light fermion; $t$ is the heavy top quark and $h$ is the heavy Higgs boson.}
\end{table}
The SMEFT vertices with at least one heavy field and $w \le 2$ are given in Table~\ref{w}. The notation is schematic: $\psi$ represents a light fermion, $t$ is the heavy top quark, $h$ is the heavy Higgs boson, $A_\mu$ is a light gauge boson and $A_{\mu \nu}$ is its field-strength tensor, and $Z_\mu$ is a heavy gauge boson and $Z_{\mu \nu}$ is its field-strength tensor.  For SMEFT, the light gauge bosons are photons and gluons, and the heavy ones are the $W$ and $Z$. The smallest weight in Table~\ref{w} is $w=1/2$ for the $\psi t A_\mu$ vertex. This interaction is a $\psi \to t$ interaction due to a photon or gluon, and is not present since QED and QCD gauge couplings are flavor diagonal. The remaining vertices all have $w \ge 1$. Since a tree graph has $V \ge 2$, and from Table~\ref{w}, we see that all vertices have $w \ge 1$, we only need to consider graphs with two insertions of the $w=1$ vertices
\begin{align}
Z_{\mu \nu} A^{\mu \nu} ,\ h \psi^2 ,\ \psi^2 Z_\mu,\ t^2 A_\mu \,.
\label{10.8}
\end{align}

The $t^2 A_\mu$ vertex is the interaction of the $t$ quark with light gauge bosons. It is not possible to draw a tree-graph without external $t$ quarks using this interaction and other $w=1$ vertices in Eq.~\eqref{10.8}, so this interaction can be dropped.

$Z_{\mu \nu}A^{\mu \nu}$ is kinetic mixing~\cite{Galison:1983pa} between a heavy and light gauge boson. Such vertices are produced by operators such as
\begin{align}
Q_{HB} &= H^\dagger H \, B_{\mu \nu} B^{\mu \nu}
\label{10.9}
\end{align}
in the SMEFT when $H$ is replaced by its VEV, and it has a coefficient of order $v^2/\Lambda^2$. Terms of this type are included in rediagonalization of the gauge-boson kinetic energy terms~\cite{Alonso:2013hga}, as discussed in Sec.~\ref{sec:Gmass}.

The $h \psi^2$ interactions are the Yukawa couplings of light fermions to the Higgs boson, and the $\psi^2 Z_\mu$ interactions are the couplings of light fermions to the heavy gauge bosons
$W$ and $Z$. They contribute to the matching via tree graphs with single $h$, $W$, or $Z$ exchange.

The fermion mass matrices and Yukawa couplings in the SMEFT are given in Eqs.~(\ref{eq:MassMatrices}) and~(\ref{eq:Yukawas}) for Dirac fermions and in Eqs. (\ref{eq:nuMass}) and~(\ref{eq:nuYuk}) for Majorana neutrinos, respectively.
Higgs exchange gives four-fermion operators with a coefficient of order
$\mathcal{Y}^2/m_h^2$.  From Eq.~\eqref{eq:Yukawas}, we see that $\mathcal{Y}^2$ has terms of order $(m/v)^2$, $mv/\Lambda^2$, and
$v^4/\Lambda^4$, where $m$ is a light-fermion mass. Thus, Higgs exchange contributions, which are $1/v^2$ times this, are parametrically of the same order as dimension-eight terms, which we have neglected, and can be dropped. The fact that $\mathcal{Y}$ is first order in power counting is a special feature of the SM. If there are additional scalars at the electroweak scale, as in two-Higgs-doublet models, then their Yukawa couplings are in general not related to fermion masses, and tree graphs with the exchange of these scalars must be included in the matching.  

The only tree graphs which we need to include are tree-level $W$ and $Z$ exchange, which give dimension-six operators. This result should be familiar from the Fermi theory of weak interactions. In SMEFT, we need to include $W$ and $Z$ vertices including $1/\Lambda^2$ corrections, since these lead to dimension-six interactions with coefficients of order $1/v^2 \times v^2/\Lambda^2 = 1/\Lambda^2$, which are included in our results. The gauge-boson propagator in unitary gauge is
\begin{align}
-\left(g_{\mu \nu} - \frac{k_\mu k_\nu}{M^2} \right) \frac{1}{k^2-M^2}\,.
\end{align}
where $M$ is the gauge-boson mass. The $k^\mu k^\nu/M^2$ part of the propagator gives terms of order $(m/M)^2$, which are the same order as Higgs exchange contributions, and can be neglected. This result is not an accident, since the two terms are related by gauge invariance. The propagator denominator can be expanded in powers of $k^2/M^2$, and to dimension six, we only need the first term.

In summary, the only contributions we need to keep are those from tree exchange of a single $W$ or $Z$ boson including $1/\Lambda^2$ corrections to the vertices, where the gauge boson propagator can be taken to be $g_{\mu \nu}/M^2$. These contributions are included in the tables of Appendix~\ref{sec:MatchingConditions} using the gauge couplings in Eq.~(\ref{14.10b}). The matching coefficients depend on the product of two gauge couplings, and so have terms of order $1/M^2$, $(1/M^2) (v^2/\Lambda^2)$ and $(1/M^2)(v^4/\Lambda^4)$, with $M \sim v$. The last term, from the product of two dimension-six corrections to the gauge coupling should formally be dropped as it is of higher order in the power counting.


\section{LEFT Operators}
\label{sec:LEFT}

The EFT below the electroweak scale consists of QCD and QED, with $n_u=2$ $u$-type quarks, $n_d=3$ $d$-type quarks, $n_e=3$ charged leptons and 
$n_\nu=3$ left-handed neutrino flavors.  
\begin{table}
	\begin{align*}
		\setlength\arraycolsep{0.3cm}
		\begin{array}{c|c|c|c}
		\toprule
		d & \text{quantum numbers} & n_g=1  & n_g=3 \\
		\midrule
		3 & (\Delta L =2) + {\rm h.c.} & 1 + 1 & 6 + 6 \\
		5 & \Delta B = \Delta L=0 & 5+5 & 35+35 \\
		5 &(\Delta L=2) + {\rm h.c.} & {0+0} & 3+3 \\
		6 & \Delta B = \Delta L=0 & 80= 57_+ + 23_- & 3631=1933_+ + 1698_- \\
		6 & (\Delta L=2)+ {\rm h.c.} &  {11+11}   & 600+600 \\
		6 & (\Delta L=4) + {\rm h.c.} &  {0+0} &  6+6\\
		6 & (\Delta B = \Delta L=1)+ {\rm h.c.} & {6+6}  & 288+288  \\
		6 & (\Delta B = -\Delta L=1)+ {\rm h.c.} & {2+2}  &  228+228  \\
		\bottomrule
		\end{array}
	\end{align*}
	\caption{Number and quantum numbers of operators in LEFT of dimensions three, five, and six.  The first column gives the operator dimension $d$, and the second column gives the $\Delta B$ and $\Delta L$ quantum numbers.  The third and fourth columns list the number of Hermitian operators in LEFT for $n_g=1$ and $n_g=3$ generations of fermions, split according to their sign under $CP$.
	\label{tab:nleft} 
}
\end{table}
The operators in LEFT are built out of fermion fields $u_{Lr}$, $u_{Rr}$ $d_{Lr}$, $d_{Rr}$, $e_{Lr}$, $e_{Rr}$, and $\nu_{Lr}$, where $r$ is a weak-eigenstate 
index\footnote{Since the weak-eigenstate index is equal to the mass-eigenstate index for all fermions except the left-handed $d$ quarks, converting to mass-eigenstate indices only involves conversion of $d_{Lr}$ to $V_{rx} d_{Lx}$, as discussed in Sec.~\ref{sec:indices}.}, the gauge-covariant derivative $D_\mu = \partial_\mu +i g T^A G^A_\mu + ie Q A_\mu$, and gauge field strengths: the photon field-strength $F_{\mu \nu}$ and the gluon field-strengths $G_{\mu \nu}^A$. 

The QCD and QED Lagrangian is 
\begin{align}
\mathcal{L}_{\rm QCD + QED} &= -\frac14 F_{\mu \nu} F^{\mu\nu} - \frac14 G_{\mu \nu}^A G^{A \mu \nu} + \theta_{\rm QCD} \frac{g^2}{32 \pi^2} G_{\mu \nu}^A \widetilde G^{A \mu \nu} +  \theta_{\rm QED} \frac{e^2}{32 \pi^2} F_{\mu \nu} \widetilde F^{\mu \nu}\nn
&+ \sum_{\psi=u,d,e,\nu_L}\overline \psi i \slashed{D} \psi   - \left[ \sum_{\psi=u,d,e}  \overline \psi_{Rr} [M_\psi]_{rs} \psi_{Ls} + \text{h.c.} \right],
\label{qcd}
\end{align}
which contains QCD gauge interactions of $n_u=2$ $u$-type quarks and $n_d=3$ $d$-type quarks and QED gauge interactions of the $u$ quarks, $d$ quarks, and $n_e=3$ charged leptons
at dimension four, and Dirac-fermion mass terms for $u$, $d$, and $e$ at dimension three.  The $n_\nu =3$ left-handed neutrinos are gauge singlets with no mass term.  Theta terms for QCD and QED are included as well.  

The LEFT Lagrangian is the QCD and QED Lagrangian (\ref{qcd}) plus additional $SU(3) \times U(1)_Q$ gauge-invariant operators at dimension three and higher dimension $d>4$, beginning at $d=5$.  The number and quantum numbers of operators in LEFT at each dimension can be obtained by counting invariants~\cite{Jenkins:2009dy,Hanany:2010vu,Henning:2015alf,Henning:2017fpj,Kobach:fr}.  In this paper, we consider operators in LEFT up to dimension six.  Table~\ref{tab:nleft} gives the number and quantum numbers of LEFT operators at dimension three, five, and six for $n_g=1$ and $n_g=3$ generations.  A complete and independent LEFT operator basis up to dimension six is constructed and presented in 
Tables~\ref{tab:oplist1} and \ref{tab:oplist2} of Appendix~\ref{sec:LEFTBasis}. Table~\ref{tab:oplist1} contains the baryon- and lepton-number-conserving operators of dimension five and six, as well as
the dimension-three and dimension-five $\Delta L =\pm 2$ operators that correspond to Majorana-neutrino mass  and dipole operators, respectively.  
Table~\ref{tab:oplist2} contains the dimension-six operators that violate lepton number and/or baryon number.  
LEFT operators are denoted by $\mathcal{O}$ and LEFT operator coefficients are denoted by $L$ to distinguish them from the SMEFT operators $Q$ and coefficients $C$, since some operators, such as the SMEFT operator $Q_G$ and the LEFT operator $\mathcal{O}_G$ look identical.
Appendix~\ref{sec:MatchingConditions} contains more detailed tables, Tables~\ref{dim3}--\ref{dim6bml}, 
listing the LEFT operators in each operator sector, their number for arbitrary numbers of fermion flavors $n_\nu$, $n_e$, $n_u$, and $n_d$ and for the values 
$n_\nu=3$, $n_e=3$, $n_u=2$, and $n_d=3$ of the SM, and the tree-level matching conditions from SMEFT including operators up to dimension six.   For example, the tree-level matching for the operator $\mathcal{O}_G$ in Table~\ref{dim6X3} is simply $L_G = C_G$.  Appendix~\ref{sec:MatchingConditions} also contains Table~\ref{dim6cp}, which divides the LEFT operators 
of Tables~\ref{dim3}--\ref{dim6bml} into $CP$-even and $CP$-odd operators.  
The leading LEFT operators of Table~\ref{tab:nleft} are described below according to operator dimension.

\subsection{Dimension-Three Operators}
\label{subsec:Maj}

The $\Delta L=2 + {\rm h.c.}$ Majorana-neutrino mass operators given in Table~\ref{dim3} arise at dimension three.  The Majorana-neutrino mass matrix $M_\nu$ is symmetric in weak-eigenstate indices, and the Majorana-neutrino mass term in the LEFT Lagrangian is
\begin{align}
\mathcal{L}^{(3)}_{\slashed{L}} &= - \frac12 [M_\nu]_{rs} \mathcal{O}_{\substack{\nu \\ rs}} + \hc = -\frac12 [M_\nu]_{rs} (\nu_{Lr} C \nu_{Ls}) + \hc ,
\end{align}
with $[M_\nu]_{rs} = -C_{\substack{5 \\ rs}} v_T^2 $ if the mass arises from the dimension-five $\Delta L=2$ operator in SMEFT.  Consequently, $M_\nu$ is of order $v^2/\Lambda_{\slashed{L}}$.  There are six $\Delta L =2$ operators for $n_\nu=3$, plus their conjugates.  Note that the $\mathcal{O}_\nu$ coefficient is $L_\nu \equiv - \frac 12 M_\nu$.

\subsection{Dimension-Five Operators}

All of the dimension-five LEFT operators are dipole operators.  

The $\Delta L=2+{\rm h.c.}$ Majorana-neutrino dipole operators are given in Table~\ref{dim5n}.
These operators are antisymmetric in the neutrino weak-eigenstate indices, so there are three $\Delta L=2$ operators plus their Hermitian conjugates.  The tree-level matching to $L_{\nu \gamma}$ from SMEFT up to dimension-six contributions vanishes.
The first non-vanishing contribution in spontaneously broken SMEFT arises at dimension seven,  $L_{\nu \gamma} \sim v^2/(\Lambda_{\slashed{L}} \Lambda^2)$.

The $\Delta B=\Delta L=0$ dipole operators $(\overline L R)X$ are given in Table~\ref{dim5mag}.  The flavor-changing operators in this table lead to interesting processes such as $\mu \to e \gamma$ and $b \to s \gamma$, as well as magnetic and electric dipole moments.  There are 35 $(\overline L R) X$ operators, plus 35 Hermitian conjugate $(\overline R L)X$ operators.
Tree-level matching in spontaneously broken SMEFT generates these dimension-five dipole operators at order $v/\Lambda$ from the dimension-six class-6 dipole operators $\psi^2 X H$. 

Kobach has shown~\cite{Kobach:2016ami} that in the SMEFT, there are no odd-dimensional $SU(3) \times SU(2) \times U(1)$ gauge-invariant operators that preserve both baryon and lepton number.  LEFT has dimension-five dipole operators which preserve baryon number and lepton number, since the gauge group is now $SU(3) \times U(1)_Q$.

\subsection{Dimension-Six Operators}

The dimension-six LEFT operators divide into the baryon- and lepton-number-conserving operators given in Table~\ref{tab:oplist1}, and $\Delta L = \pm 4$, $\Delta L = \pm 2$, $\Delta B = \Delta L = \pm 1$, and $\Delta B = - \Delta L = \pm 1$ operators given in Table~\ref{tab:oplist2}. 

Since there are numerous $\psi^4$ operators, it is convenient to further divide the $\psi^4$ operators
into subclasses, according to chirality $L$ and $R$ of the fermion bilinears, as was done in SMEFT~\cite{Grzadkowski:2010es}, and according to scalar, vector, and tensor Dirac structure $S$, $V$, and $T$.  Fierz identities can be used to convert operators between the different subclasses, so the choice of independent LEFT operators is not unique.  We have constructed the LEFT operator basis from fermion bilinears in the form $(\overline \psi \Gamma \chi)$ or $(\psi^T \Gamma \chi)$ that contain either two lepton or two quark fields, avoiding leptoquark bilinears.  In addition, we have eliminated tensor Dirac matrices $\Gamma=\sigma^{\mu\nu}$ as far as possible from the operator basis.
It is a non-trivial exercise to make sure there is no double counting of operators due to the Fierz identities. 

\begin{enumerate}
\item {\bf $\Delta B= \Delta L =0$ operators:}
These operators preserve baryon and lepton number, and they are the only dimension-six operators that could be present at the TeV scale. The operators divide into  $X^3$ and $\psi^4$ operators.  There are two triple-gluon $X^3$ operators and 78 four-fermion $\psi^4$ operators, neglecting flavor.
\begin{enumerate}
\item $X^3$: There are two pure gauge operators in Table~\ref{dim6X3} constructed from three gluon field strengths.  These two operators also exist in SMEFT, and the tree-level matching between SMEFT and LEFT is $L_{G}= C_G$ and $L_{\widetilde G} = C_{\widetilde G}$.
\item $(\bar L L)(\bar LL)$: The only fermion bilinears of the form $(\bar L \Gamma L)$ are left-handed vector currents $(\bar L \gamma^\mu L)$. $(\bar L L)(\bar LL)$ operators are the product of two left-handed currents. Integrating out the $W$ and $Z$ bosons in the SM to get the Fermi theory of weak interactions produces these operators.  Four-fermion dimension-six operators in unbroken SMEFT also give a tree-level contribution.  
Table~\ref{dim6ll} lists the independent operators in the LEFT operator basis.  The operators are divided into purely leptonic, semileptonic, and non-leptonic operators.
\item $(\bar RR)(\bar RR)$: These operators are products of two right-handed vector currents.  They are produced by $Z$ exchange in the SM.  Four-fermion dimension-six
operators in unbroken SMEFT give a tree-level contribution.  There also is a contribution from $\mathcal{W}$ exchange in spontaneously broken SMEFT since $\mathcal{W}$ couples to the right-handed charged current $(\overline u_{Rp} \gamma^\mu d_{Rr})$ due to the dimension-six operator $Q_{Hud}$.
Table~\ref{dim6rr} lists the independent operators in the LEFT operator basis.  The operators are divided into purely leptonic, semileptonic, and non-leptonic operators.
\item $(\bar LL)(\bar RR)$:  These operators are products of a left-handed vector current and a right-handed vector current.  
Table~\ref{dim6lr} lists the independent operators in the LEFT operator basis.  The operators are divided into purely leptonic, semileptonic, and non-leptonic operators.  Most operators are produced in the SM from $Z$ exchange.  Four-fermion dimension-six operators in unbroken SMEFT give a tree-level contribution in most cases.  In addition, $\mathcal{W}$ exchange in spontaneously broken SMEFT produces a number of operators due to the coupling of $\mathcal{W}$ to right-handed charged quark currents.  The operator $\op{uddu}{V8}{LR}$ and its Hermitian conjugate are not produced in SMEFT at this level.
\item $(\bar LR)(\bar RL)+\hc$: The fermion bilinear $(\bar L R)$ can be either a scalar $(\bar L R)$ or a tensor $(\bar L \sigma^{\mu \nu} R)$.  Only products of scalar fermion bilinears exist in the LEFT operator basis due to the identity $(\bar L \sigma^{\mu \nu} R)(\bar R \sigma_{\mu \nu} L)=0$.  This identity contracts a self-dual tensor with an anti-self-dual one. 
$(\bar L \sigma^{\mu \nu} R)$ transforms as $(0,1)$ under the Lorentz group, and $(\bar R \sigma^{\mu \nu} L)$ as $(1,0)$.  It is not possible to combine the two tensor bilinears into a Lorentz singlet $(0,0)$, which is why the (apparently) Lorentz singlet combination $(\bar L \sigma^{\mu \nu} R)(\bar R \sigma_{\mu \nu} L)$ vanishes.  Table~\ref{dim6lrrl} lists the three semileptonic scalar operators of the LEFT operator basis in this category.  Two of the operators receive a tree-level matching from dimension-six operators in unbroken SMEFT, whereas the operator 
$\op{eu}{S}{RL}$ is not produced in SMEFT at tree level.
\item $(\bar LR)(\bar LR)+\hc$: In this case, both scalar and tensor operators are possible. 
Table~\ref{dim6lrlr} lists the independent operators in the LEFT operator basis.  The operators are divided into purely leptonic, semileptonic, and non-leptonic operators.  Some of the operators are present in unbroken SMEFT, but seven operators in this category have no SMEFT matching at tree level.
\end{enumerate}
\item {\bf $\Delta L =\pm4$ operators:}  These operators are the square of the Majorana-neutrino mass term.   
Table~\ref{dim6l4} lists the single $\Delta L = 4$ LEFT operator. 
The operator transforms as the $\Yboxdim{8pt}\yng(2,2)$ representation under the neutrino-flavor symmetry group.  This operator receives a tree-level contribution in spontaneously broken SMEFT from Higgs and gauge-boson exchange that is not included in the table,  since we are dropping such terms, as discussed in Sec.~\ref{sec:Matching}. These operators lead e.g.\ to neutrinoless quadruple $\beta$ decay~\cite{Heeck:2013rpa,Arnold:2017bnh}.
\item {\bf $\Delta L =\pm2$ operators:} Table~\ref{dim6l2} lists  LEFT operators in this category.  Some of the operators are produced in spontaneously broken SMEFT from Higgs exchange with one Higgs Yukawa coupling in the SM and one Higgs coupling to the $\Delta L=\pm 2$ Majorana-neutrino bilinear, as well as from gauge-boson exchange.
The operators $\op{\nu edu}{S}{LR}$, $\op{\nu edu}{V}{RL}$, $\op{\nu edu}{V}{RR}$ lead to $\beta$ decay with an emitted neutrino rather than an antineutrino.

\item {\bf $\Delta B= \Delta L =\pm1$ operators:} Table~\ref{dim6bl} lists the LEFT operators in this category.  Many of the operators receive a tree-level matching from the $\Delta B = \Delta L = \pm 1$ operators in SMEFT.  A number of the operators are not produced in SMEFT at tree level.
\item {\bf $\Delta B= -\Delta L =\pm1$ operators:} Table~\ref{dim6bml} lists  LEFT operators in this category.  Operators with these quantum numbers do not exist in SMEFT.
\end{enumerate}

The tables give the tree-level matching coefficients up to order $1/\Lambda^2$ in SMEFT. SMEFT gives non-trivial correlations between the coefficients, as can be seen, for example, by looking at Table~\ref{dim6lrlr}.  Consequently, one can test whether BSM physics arises via  SMEFT, i.e. whether it respects the SM electroweak gauge symmetry breaking $SU(3) \times SU(2) \times U(1) \to SU(3) \times U(1)_Q$, by seeing whether the SMEFT correlations are satisfied.  As an example, in the case of $(\bar LR)(\bar LR)$ operators, SMEFT predictions are
\begin{align}
& \wc{\substack{ee \\ prst}}{S}{RR} = 0, &
& \wc{\substack{ed \\ prst}}{S}{RR} = 0, &
& \wc{\substack{ed \\ prst}}{T}{RR} = 0, \nn
& \wc{\substack{uu \\ prst}}{S1}{RR} = 0, &
& \wc{\substack{uu \\ prst}}{S8}{RR} = 0, &
& \wc{\substack{dd \\ prst}}{S1}{RR} = 0, \nn
& \wc{\substack{dd \\ prst}}{S8}{RR} = 0, &
& \wc{\substack{eu \\ prst}}{S}{RR} + \wc{\substack{\nu e d u \\ prst}}{S}{RR} = 0, &
& \wc{\substack{eu \\ prst}}{T}{RR} + \wc{\substack{\nu e d u \\ prst}}{T}{RR} = 0, \nn
& \wc{\substack{ud \\ prst}}{S1}{RR} + \wc{\substack{uddu \\ stpr}}{S1}{RR}  = 0, &
& \wc{\substack{ud \\ prst}}{S8}{RR} + \wc{\substack{uddu \\ stpr}}{S8}{RR}  = 0,
\end{align}
so low-energy constraints on these operators provide important information on whether electroweak symmetry breaking results from the VEV of a single fundamental scalar doublet as assumed in SMEFT.
Many additional examples are provided by operators with vanishing matching coefficients in SMEFT in the tables.


\section{Flavor Physics and $B$ anomalies}
\label{sec:FlavorPhysics}

In this section, we study the implications of the RGE for several low-energy flavor-changing weak decays.  We start with $\mu$ decay,
which is used to extract the value of $G_F$, and then discuss lepton non-universality in $B$ decays, which has received a lot of attention due to recent results from LHCb. In the following discussion, we assume that deviations in LEFT coefficients from their SM values are small, i.e.\ they are suppressed by $1/\Lambda^2$, where $\Lambda$ is the scale of new physics, and then we restrict to the special case where the LEFT operators arise from matching to the SMEFT, which imposes the restriction that the new physics is invariant under $SU(3) \times SU(2) \times U(1)$ gauge symmetry. We work to order $1/\Lambda^2$, since we have neglected operators with dimension greater than six. We do not perform a detailed fit to the experimental results. The aim of this section is to discuss which LEFT coefficients contribute to the decay amplitudes, and their RGE evolution including mixing with other operators. We use results for the RGE of LEFT operators from a subsequent paper~\cite{Jenkins:2017dyc}.  Another important and interesting example not discussed here is flavor-changing $\mu \to e$ transitions. This application has been studied in detail, including the LEFT renormalization considered here, in recent publications~\cite{Davidson:2016edt,Crivellin:2017rmk,Cirigliano:2017azj}. An analysis of $B$ anomalies within an EFT based on flavor symmetries has been presented in~\cite{Bordone:2017anc}.

\subsection{$\mu$ Decay and $G_F$}

The value of the VEV in the SM is obtained from the measurement of  $G_F$ in $\mu$ decay, $\mu^- \rightarrow e^- + \bar{\nu}_e + \nu_\mu$.
The terms in the LEFT Lagrangian which contribute to $\mu$ decay are
\begin{align}
\label{eq:MuDecayLEFT}
\mathcal{L} &= \lcc{}{V}{LL}  \, (\bar \nu_{L \mu} \gamma^\mu \nu_{Le}) (\bar e_{L} \gamma_\mu \mu_{L}) 
+ \lcc{}{V}{LR} \, (\bar \nu_{L \mu} \gamma^\mu \nu_{Le}) (\bar e_{R} \gamma_\mu \mu_{R}) \,,
\end{align}
where the coefficients are evaluated at the scale of the muon mass $\mu = m_\mu$, and we have used $L^X \equiv L_{\substack{\nu e\\ \mu e e \mu} }^X$ to simplify the notation.
The muon decay rate computed from Eq.~(\ref{eq:MuDecayLEFT}) has contributions proportional to $\abs{\lcc{}{V}{LL} }^2$,
$\abs{ \lcc{}{V}{LR} }^2$ and an interference contribution proportional to $\text{Re} \left( \lcc{}{V}{LL}\lcc{}{V}{LR*}\right)  $. The SM only has left-handed charged currents, and the right-handed current coefficient
$\lcc{}{V}{LR} $ is of order $v^2/\Lambda^2$, so the $\abs{ \lcc{}{V}{LR} }^2$  contribution to the decay rate is of order $v^4/\Lambda^4$ and can be dropped. The interference term is helicity suppressed, of order $m_e/m_\mu  \times v^2/\Lambda^2$, since the electrons have opposite chirality in the two operators. While $m_e/m_\mu$ is the ratio of two low-energy scales, numerically $m_e/m_\mu \sim 1/200$, and the interference term can also be dropped. Thus, the $\mu$ decay rate is obtained
from the left-handed current operator.  Comparing with the usual Fermi theory gives
\begin{align}
-\frac{4 \mathcal{G_F} }{\sqrt 2}  &=\lcc{}{V}{LL}=\wc{\substack{\nu e\\ \mu e e \mu} }{V}{LL} \,.
\label{7.0}
\end{align}
The coefficients $\lcc{}{V}{LL} $ and $\lcc{}{V}{LR} $ are evolved down to $\mu=m_\mu$ using the RGE given in Ref.~\cite{Jenkins:2017dyc}. The only contributions to the RGEs of $\lcc{}{V}{LL}$ and $\lcc{}{V}{LR}$ are from penguin and box graphs, and they vanish for the off-diagonal terms needed, so the coefficients at $\mu=m_\mu$ are the same as those at $\mu=M_Z$.  Thus, Eq.~(\ref{7.0}) with the r.h.s.\ evaluated at $\mu=M_Z$ is fixed by $\mathcal{G_F}$.

Tree-level matching of the LEFT Lagrangian to the SMEFT gives
\begin{align}
\lcc{}{V}{LL} & = - \frac{2}{v_T^2}+C_{\substack{ll \\ \mu e e \mu}} + C_{\substack{ll \\ e \mu \mu e}}
- 2C_{\substack{Hl \\ \mu \mu}}^{(3)} - 2C_{\substack{Hl \\\ e e}}^{(3)}\,,  &
\lcc{}{V}{LR}  & = C_{\substack{le \\ \mu ee\mu}}  \,,
\label{7.2}
\end{align}
at $\mu=M_Z$. Combining with Eq.~(\ref{7.0}), $\mathcal{G_F}$ in SMEFT~\cite{Alonso:2013hga} is
\begin{align}
\frac{4 \mathcal{G_F}}{\sqrt 2} &= \frac{2}{v_T^2} - C_{\substack{ll \\ \mu e e \mu}} - C_{\substack{ll \\ e \mu \mu e}}
+ 2C_{\substack{Hl \\ \mu \mu}}^{(3)} + 2C_{\substack{Hl \\\ e e}}^{(3)}\,,
\label{7.3}
\end{align}
evaluated at $\mu=M_Z$,
and is fixed by the experimentally measured $\mu$ decay. The $\tau \to \nu_\tau \ell \overline \nu_\ell$, $\ell=e,\mu$  rate depends on the linear combination in Eq.~(\ref{7.3}) with the subscripts $\mu \to \tau$ and $e \to \ell$, which we denote by $\mathcal{G_F}(\tau \to \ell)$. Precision tests on lepton universality in $\tau$ decay~(for a review, see Ref.~\cite{Pich:2013lsa}) give
\begin{align}
\frac{\mathcal{G_F}(\tau \to \mu)}{\mathcal{G_F}(\tau \to e)} &= 1.0018 \pm 0.0014 \,, &
\frac{\mathcal{G_F}(\tau \to e)}{\mathcal{G_F}} &= 1.00011 \pm 0.0015 \,.
\label{6.6}
\end{align}
There is a small (but not significant) deviation from unity in ${\mathcal{G_F}(\tau \to \mu)}/{\mathcal{G_F}(\tau \to e)}$. The precision of $0.001$ in the ratios Eq.~(\ref{6.6}) means that they are sensitive to new physics scales of order 7~TeV.

\subsection{$\overline b \to \overline c \tau \nu$ Decays}

There are possible deviations from the SM in $B$ semileptonic decay ratios~\cite{Lees:2013uzd,Aaij:2015yra,Hirose:2016wfn}

\begin{align}
R_D &= \frac{ \Gamma(B \to D \tau^+ \nu_\tau)}{ \Gamma(B \to D \ell^+ \nu_\ell)}, & R_{D^*} &= \frac{ \Gamma(B \to D^* \tau^+ \nu_\tau)}{ \Gamma(B \to D^* \ell^+ \nu_\ell)},
\end{align}
where $\ell=e,\mu$. The semileptonic $B$ decay rates are roughly equal to their SM values, so we will assume that the LEFT coefficients have only small deviations from SM values, and the deviations arise from interference with the SM amplitude.
The terms in the LEFT Lagrangian that contribute to semileptonic $b \to c$ decays are
\begin{align}
\mathcal{L} &= \lcc{r}{V}{LL}\, (\bar \nu_{Lr} \gamma^\mu e_{Lr}) (\bar b_{L} \gamma_\mu c_{L})  +
\lcc{r}{V}{LR}\, (\bar \nu_{Lr} \gamma^\mu e_{Lr})(\bar b_{R} \gamma_\mu c_{R})  +
\lcc{r}{S}{RL}  \, (\bar \nu_{Lr} e_{Rr}) (\bar b_{R}  c_{L})   \nn
&+ \lcc{r}{S}{RR}\, (\bar   \nu_{Lr} e_{Rr})  (\bar b_{L} c_{R} )  +
\lcc{r}{T}{RR} \, (\bar  \nu_{Lr}  \sigma^{\mu \nu} e_{Rr} )  (\bar  b_{L}  \sigma_{\mu \nu} c_{R} ) \,,
\label{7.6}
\end{align}
where we have used $L_r^X \equiv L_{\substack{\nu e d u \\ rrbc}}^X$ to simplify the notation, and
switched to the quark mass-eigenstate basis, as discussed in Sec.~\ref{sec:indices}.

In the SMEFT at $\mu=M_Z$, one obtains the coefficients
\begin{align}
\lcc{r}{V}{LL} 
& \mathrlap{ = - \frac{2}{v_T^2} V_{cb}^*   + 2 V_{ib}^*\, C_{\substack{lq \\ rr i c}}^{(3)} -2 V_{cb}^*\, C_{\substack{Hl \\  rr }}^{(3)}-2 V_{ib}^*\,   C_{\substack{Hq \\ ci}}^{(3)*}\,,}  \nonumber \\[5pt]
\lcc{r}{V}{LR} 
&= -   V_{ib}^*\, C_{\substack{Hud\\ ci}}^* \,,  &
\lcc{r}{S}{RL} 
&=  V_{ib}^* \,C_{\substack{ledq \\ rr i c}} \,,   \nn
\lcc{r}{S}{RR} 
&=  V_{ib}^*\, C_{\substack{lequ \\ rr i c}}^{(1)} \,,  &
\lcc{r}{T}{RR} 
&=  V_{ib}^*\, C_{\substack{lequ \\ rr i c}}^{(3)} \,.
\end{align}
In the SM at tree level, only $\lcc{r}{V}{LL}$ is non-zero, so the other coefficients are assumed to be of order $1/\Lambda^2$.

The dominant contribution to the semileptonic $b$ decay rate is from $\abs{\lcc{r}{V}{LL}}^2$, since that is the only coefficient that exists in the SM. The only other contributions up to order $v^2/\Lambda^2$ are the interference terms of $\lcc{r}{V}{LL}$ with the other terms. The interference of $\lcc{r}{V}{LL}$ with $S$ and $T$ operators is of order $(m_\ell/m_b)(v^2/\Lambda^2)$ because they are helicity suppressed, and can  be dropped.  

Normalizing to $\mathcal{G_F}$ in Eq~(\ref{7.3}), we see that the two terms that contribute to the semi\-leptonic $b$ decay amplitudes in SMEFT are
{\small
\begin{align}
\lcc{r}{V}{LL} &= -\frac{4 \mathcal{G_F}}{\sqrt 2} 
V_{cb}^*\left[1+v_T^2 \left(\frac12  C_{\substack{ll \\ \mu e e \mu}} + \frac12  C_{\substack{ll \\ e \mu \mu e}}
-  C_{\substack{Hl \\ \mu \mu}}^{(3)} -  C_{\substack{Hl \\\ e e}}^{(3)} + C_{\substack{Hl \\ r r }}^{(3)}
-  \frac{V_{ib}^*}{V_{cb}^*} C_{\substack{lq \\ r r i c}}^{(3)}  + \frac{V_{ib}^*}{V_{cb}^*}  C_{\substack{Hq \\ ci}}^{(3)*}\right)  \right] \,, \nn
\lcc{r}{V}{LR} &= -\frac{4 \mathcal{G_F}}{\sqrt 2}  \frac12 V_{ib}^*\, v_T^2 C_{\substack{Hud \\ ci}}^*\,.
\label{7.7}
\end{align}}%
Note that in SMEFT to order $v^2/\Lambda^2$, $\lcc{r}{V}{LR}$ does not depend on  lepton flavor, and cannot be responsible for the $R_{D,D^*}$ anomaly.

The $B \to D$ decay amplitude is proportional to the vector-current matrix element with coefficient $(\lcc{r}{V}{LL}+\lcc{r}{V}{LR})/2$, whereas the $B \to D^*$ decay amplitude depends on the vector current, as well as the axial current with coefficient 
$(-\lcc{r}{V}{LL}+\lcc{r}{V}{LR})/2$. The vector and axial current do not interfere in the total rate.
The RGE for $\lcc{r}{V}{LL}$, $\lcc{r}{V}{LR}$ are given in Ref.~\cite{Jenkins:2017dyc},
\begin{align}
\dlcc{r}{V}{LL} &= -4 e^2 \lcc{r}{V}{LL} \,,&
\dlcc{r}{V}{LR} &= -2e^2 \lcc{r}{V}{LR}\,.
\label{7.6b}
\end{align}
Thus, in  LEFT, $\lcc{r}{V}{LL}$, $\lcc{r}{V}{LR}$ are multiplicatively renormalized by QED corrections, and do not depend on  other LEFT coefficients. The renormalization is small,  increasing $\lcc{r}{V}{LL}$ and $\lcc{r}{V}{LR}$ by about 2\% and 1\% respectively between $M_Z$ and $m_b$. The ratio of the coefficients for $\tau$ and $\ell=e,\mu$ is 
\begin{align}
\frac{ \lcc{\tau}{V}{LL} \pm \lcc{\tau}{V}{LR}  }{ \lcc{\ell}{V}{LL} \pm \lcc{\ell}{V}{LR}  } & =1+v_T^2 \left( C_{\substack{Hl \\ \tau \tau }}^{(3)} -C_{\substack{Hl \\ \ell \ell }}^{(3)}
-  \frac{V_{ib}^*}{V_{cb}^*} C_{\substack{lq \\\tau \tau i c}}^{(3)}
+  \frac{V_{ib}^*}{V_{cb}^*} C_{\substack{lq \\ \ell \ell i c}}^{(3)}  \right)
\label{7.8}
\end{align}
and the RGE factor cancels in the ratio.

The $R_{D,D^*}$ anomalies are usually assumed to arise from deviations in the $\tau$ decay rate from the SM values, with $\ell=e,\mu$ rates close to their SM values. In the SMEFT, a simple way to do this is to have SMEFT coefficients for $\ell=e,\mu$ be small, and
\begin{align}
C_{\substack{Hl \\ \tau \tau }}^{(3)} 
-  \frac{V_{ib}^*}{V_{cb}^*} C_{\substack{lq \\\tau \tau i c}}^{(3)} > 0\,.
\label{7.9}
\end{align}

\subsection{$b \to s \ell^+ \ell^-$ Decays}

LHCb has also measured anomalies in $B \to K^* \ell^+ \ell^-$  and $B \to K \ell^+ \ell^-$ decays~\cite{Aaij:2014ora,Aaij:2017vbb},
\begin{align}
R_K &= \frac{ \Gamma(B \to K \mu^+ \mu^-)}{ \Gamma(B \to K e^+ e^-)}
= 0.745^{+0.090}_{-0.074}\mathrm{\,(stat)}\,\pm0.036\mathrm{\,(syst)} \,, \nn
R_{K^*}  &= \frac{ \Gamma(B \to K^* \mu^+ \mu^-)}{ \Gamma(B \to K^* e^+ e^-)} = \begin{cases}
0.66~^{+~0.11}_{-~0.07}\mathrm{\,(stat)}  \pm 0.03\mathrm{\,(syst)}	& \textrm{for } 0.045 < q^2 < 1.1~\text{GeV}^2 \, , \\
0.69~^{+~0.11}_{-~0.07}\mathrm{\,(stat)} \pm 0.05\mathrm{\,(syst)}	& \textrm{for } 1.1 < q^2 < 6.0~\text{GeV}^2 \, .
\end{cases}
\label{RK}
\end{align}
Most explanations of these anomalies have focused on new physics contributions to the electromagnetic and semileptonic operators in the $b \to s$ weak Hamiltonian~\cite{Alonso:2014csa}. The relevant terms in the LEFT Lagrangian are
\begin{align}
\mathcal{L} &=
c_7  \bar s_{L}  \sigma^{\mu \nu} b_{R}\, F_{\mu \nu} + c_7^\prime \, \bar s_{R}  \sigma^{\mu \nu} b_{L}\, F_{\mu \nu} +
\wcc{\ell}{V}{LL}\, (\bar l_{L}  \gamma^\mu l_{L})(\bar s_{L} \gamma_\mu b_{L})  \nn
& +
\wcc{\ell}{V}{RR}\, (\bar l_{R}  \gamma^\mu l_{R})(\bar s_{R} \gamma_\mu b_{R})  +
\wcc{\ell}{\prime V}{LR}\, (\bar l_{R} \gamma^\mu l_{R})  (\bar s_{L} \gamma_\mu b_{L})    +
\wcc{\ell}{V}{LR}\, (\bar l_{L}  \gamma^\mu l_{L})(\bar s_{R} \gamma_\mu b_{R})  \nn
& +
\wcc{\ell}{S}{RL}\, (\bar l_{L} l_{R}) (\bar s_{R} b_{L})+\wcc{\ell}{\prime S}{RL}\, (\bar l_{R} l_{L}) (\bar s_{L} b_{R})   +
\wcc{\ell}{S}{RR}\, (\bar l_{L} l_{R})(\bar s_{L} b_{R}) \nn
& + \wcc{\ell}{\prime S}{RR*}\, (\bar l_{R} l_{L})(\bar s_{R} b_{L})  +
\wcc{\ell}{T}{RR}\, (\bar l_{L} \sigma^{\mu \nu} l_{R}) (\bar s_{L} \sigma_{\mu \nu} b_{R})
+\wcc{\ell}{\prime T}{RR*}\, (\bar l_{R} \sigma^{\mu \nu} l_{L}) (\bar s_{R} \sigma_{\mu \nu} b_{L})  \,,
\label{7.11}
\end{align}
where, to simplify the notation, we use
\begin{align}
c_7 &= L_{\substack{d \gamma \\ sb} }\,, &
c_7^\prime &= L_{\substack{d \gamma \\ bs}}^*\,, &
\wcc{\ell}{V}{LL} &= \wc{\substack{ed \\  llsb}}{V}{LL}\,, &
\wcc{\ell}{V}{RR} &= \wc{\substack{ed \\  llsb}}{V}{RR}\,, \nn
\wcc{\ell}{\prime V}{LR} &= \wc{\substack{de\\sbll}}{V}{LR}\,, &
\wcc{\ell}{V}{LR} &= \wc{\substack{ed \\ llsb}}{V}{LR}\,, &
\wcc{\ell}{S}{RL} &= \wc{\substack{ed \\ llsb}}{S}{RL}\,, &
\wcc{\ell}{\prime S}{RL} &= \wc{\substack{ed \\ llbs}}{S}{RL*}\,,  \nn
\wcc{\ell}{S}{RR} &= \wc{\substack{ed \\ llsb} }{S}{RR}\,, &
\wcc{\ell}{\prime S}{RR} &= \wc{\substack{ed \\ llbs} }{S}{RR*}\,, &
\wcc{\ell}{T}{RR} &=\wc{ \substack{ed \\ llsb} }{T}{RR}\,, &
\wcc{\ell}{\prime T}{RR} &= \wc{ \substack{ed \\ llbs} }{T}{RR*}\,.
\label{7.11a}
\end{align}
In addition to the operators shown explicitly in Eq.~(\ref{7.11}), there are also four-quark operators $O_1-O_6$ and the $b\to s$ chromomagnetic operator $O_8$ that mix with the above operators under RGE, and are included in the usual analysis of $B$ decay.

Our operator basis for LEFT is in terms of fields with definite chiral properties. Traditionally, in weak decays, a basis of operators with definite parity has been used (i.e.\ scalar, pseudoscalar, etc.). The conversion to the basis used in Ref.~\cite{Alonso:2014csa} is
\begin{align}
\lambda_1 C_7 &= c_7\,, &
\lambda_1 C_7^\prime &= c_7^\prime\,, \nn
\lambda_2 C_9 & = \wcc{\ell}{V}{LL} + \wcc{\ell}{\prime V}{LR}\,, &
\lambda_2 C_{10} &=-\wcc{\ell}{V}{LL} + \wcc{\ell}{\prime V}{LR}\,,  \nn
\lambda_2 C_9^\prime &=\wcc{\ell}{V}{LR} + \wcc{\ell}{ V}{RR}\,,  &
\lambda_2 C_{10}^\prime &=-\wcc{\ell}{V}{LR} + \wcc{\ell}{ V}{RR}\,,\nn
\lambda_2 C_S & =\wcc{\ell}{S}{RR} + \wcc{\ell}{\prime S}{RL}\,, &
\lambda_2 C_P  & =\wcc{\ell}{S}{RR} - \wcc{\ell}{\prime S}{RL}\,, \nn
\lambda_2 C_S^\prime &=\wcc{\ell}{S}{RL} + \wcc{\ell}{\prime S}{RR}\,, &
\lambda_2 C_P^\prime  &=\wcc{\ell}{S}{RL} - \wcc{\ell}{\prime S}{RR}\,,  \nn
\lambda_2 C_T  &=\wcc{\ell}{\prime T}{RR} + \wcc{\ell}{ T}{RR}\,, &
\lambda_2 C_{T5}  &=-\wcc{\ell}{\prime T}{RR} + \wcc{\ell}{ T}{RR}\,,
\label{7.13}
\end{align}
where
\begin{align}
\lambda_1 &\equiv -\frac{4 \mathcal{G_F}}{\sqrt 2}\frac{e\,  m_b}{16 \pi^2}\,, & \lambda_2 &\equiv -\frac{8 \mathcal{G_F}}{\sqrt 2}\frac{e^2}{16\pi^2}\,. 
\label{7.13a}
\end{align}
%

Matching at tree-level from the SMEFT at $\mu=M_Z$ gives
\begin{align}
c_7 &=  \frac{1}{\sqrt 2}  \left( -C_{\substack{dW \\ sb}} s_W + C_{\substack{dB \\ sb}} c_W \right) v_T \,, &
\wcc{\ell}{S}{RR} &= 0 \,,  \nn
c_7^\prime &=  \frac{1}{\sqrt 2}  \left( -C_{\substack{dW \\ bs}}^* s_W + C_{\substack{dB \\ bs}}^* c_W \right) v_T \,, &
\wcc{\ell}{\prime S}{RL} &= C_{\substack{ ledq \\ llbs}}^*  \,, \nn
\wcc{\ell}{V}{LR} 
& =  C_{\substack{ qe \\ sbll}} 
-\frac{\gcZ^2 v_T^2}{ 2 M_Z^2}  \sc^2 \q_e \left( C_{\substack{Hq \\ sb}}^{(1)} + C_{\substack{Hq \\ sb}}^{(3)}  \right) \,,  &
\wcc{\ell}{S}{RL} &= C_{\substack{ ledq \\ llsb}} \,, \nn
\wcc{\ell}{V}{LL} 
&= C^{(1)}_{\substack{ lq \\ llsb}} + C^{(3)}_{\substack{ lq \\ llsb}} 
-\frac{\gcZ^2 v_T^2}{ 2 M_Z^2}  \left(\frac12+\sc^2 \q_e \right) \left(C_{\substack{Hq \\ sb}}^{(1)} +  C_{\substack{Hq \\ sb}}^{(3)}  \right) \,, &
\wcc{\ell}{\prime S}{ RR} &=0 \,,  \nn
\wcc{\ell}{V}{RR} &= C_{\substack{ ed \\ llsb}} -\frac{\gcZ^2 v_T^2}{2 M_Z^2}  \sc^2 \q_e  C_{\substack{Hd \\ sb}} \,,  &
\wcc{\ell}{T}{RR}  &=0 \,, \nn
\wcc{\ell}{V}{LR} 
&=C_{\substack{ ld \\ llsb}} 
-\frac{\gcZ^2 v_T^2}{2 M_Z^2}   \left(\frac12 +\sc^2 \q_e\right)  C_{\substack{Hd \\ sb}} \,,   &
\wcc{\ell}{\prime T}{ RR} &=0 \,.
\label{7.12}
\end{align}
Assuming that all BSM physics is via the SMEFT, i.e.\ it respects the SM electroweak symmetry-breaking mechanism, Eq.~(\ref{7.12}) leads to the relations $\wcc{\ell}{S}{RR}=\wcc{\ell}{T}{RR}=\wcc{\ell}{\prime S}{RR}=\wcc{\ell}{\prime T}{RR}=0$, which are 
equivalent to the relations found in Ref.~\cite{Alonso:2014csa},
\begin{align}
C_S + C_P &= 0 \,, & C_S^\prime - C_P^\prime &=0 \,, & C_T&=0 \,, & C_{T5} &=0 \,.
\label{7.14}
\end{align}

To obtain $R_{K,K^*} \not =1$ requires a violation of $e-\mu$ universality in $b \to s \ell^+ \ell^-$ decays. The operators $O_1-O_6$ and $O_8$ do not involve leptons, but can generate the operators involving leptons in Eq.~(\ref{7.11}) via RGE. This contribution is the same for all lepton flavors since the operators and anomalous dimensions are flavor blind. Similarly, the contributions of the photon penguin operators $c_7$ and $c_7^\prime$ cancel in $R_{K,K^*}$, since the photon coupling to leptons is flavor blind. Flavor-dependent gauge couplings due to Higgs operators, such as those in Eq.~(\ref{14.10b}) for the $W$ and $Z$, do not exist for the photon.

The other operators in Eq.~(\ref{7.11}) can depend on lepton flavor and contribute to $R_{K,K^*}-1$ by an amount proportional to the difference of the operators for $\mu$ and $e$. 
The RGE for the leptonic operators $\wcc{\ell}{V}{LL}$, etc.\ is rather involved and contains mixing via four-quark operators through penguin diagrams, as well as non-linear terms involving squares of dipole coefficients~\cite{Jenkins:2017dyc}. However, these cancel in the RGE for the differences between $\mu$ and $e$, which reduce to the simple form
\begin{align}
	\dwcc{\Delta}{V}{LL} &= 4e^2  \wcc{\Delta}{V}{LL} \,, &
	\dwcc{\Delta}{V}{RR} &= 4e^2  \wcc{\Delta}{V}{RR} \,, \nn
	\dwcc{\Delta}{V}{LR} &= -4e^2  \wcc{\Delta}{V}{LR} \,, &
	\dwcc{\Delta}{S}{RL} &= -\left(\frac{20}{3}e^2+8g^2 \right)  \wcc{\Delta}{S}{RL}  \,,\nonumber
\end{align}
\begin{align}
\left[ \begin{array}{cc} \dwcc{\Delta}{S}{RR}  \\[5pt] \dwcc{\Delta}{T}{RR}  \end{array}\right]=
\left[ \begin{array}{cc} -\frac{20}{3}e^2-8g^2 & -32 e^2 \\[5pt] -\frac23 e^2 & \frac{20}{9}e^2+\frac83 g^2  \end{array}\right]
\left[ \begin{array}{cc} \wcc{\Delta}{S}{RR}  \\[5pt] \wcc{\Delta}{T}{RR}  \end{array}\right]
+ \left[ \begin{array}{cc} -64 e^2 L_{ \substack{d\gamma \\  sb }} \left(L_{\substack{e\gamma \\ \mu\mu}}-L_{\substack{e\gamma \\ ee}}\right) \\[5pt] 0  \end{array}\right],
\label{7.16}
\end{align}
where $c_{\Delta}^X \equiv c_{\mu}^X - c_e^X$, and the primed coefficients have the same RGE as the unprimed ones. Using $c_\Delta^X$ for computing lepton universality violation is valid as long as the RG correction is treated in perturbation theory. A more accurate analysis requires the full RGE for $c^X_\mu$ and $c^X_e$ separately~\cite{Jenkins:2017dyc}, which are considerably more complicated.

The LEFT RGE is non-linear, and dimension-six operator coefficients have terms that depend on the square of dimension-five dipole coefficients. The non-linear dipole term in Eq.~(\ref{7.16}) depends on the $e$ and $\mu$ dipole operators, which are strongly constrained by the electric and magnetic moments of the electron and muon~\cite{Baron:2013eja,Bennett:2008dy,Hanneke:2008tm,Bennett:2006fi,Mohr:2015ccw},
\begin{align}
\abs{\text{Re}\, L_{\substack{e\gamma \\ ee}} } &\le 3.85 \times 10^{-11}\ \text{GeV}^{-1} \,, &
\abs{\text{Im}\, L_{\substack{e\gamma \\ ee}} }  &\le 6.7 \times 10^{-16}\ \text{GeV}^{-1} \, (90\% \text{ C.L.}) \,, \nn
\abs{\text{Re}\, L_{\substack{e\gamma \\ \mu\mu}} } &\le 4.5 \times 10^{-10}\ \text{GeV}^{-1} \,, &
\abs{\text{Im}\, L_{\substack{e\gamma \\ \mu\mu}} } &\le 1.5 \times 10^{-6}\ \text{GeV}^{-1} \, (95\% \text{ C.L.}) \,,
\label{6.21}
\end{align}
and so the non-linear terms are negligible. In Eq.~(\ref{6.21}), we have used the experimental uncertainty on muon $g-2$ as the limit on $\text{Re}\, L_{\substack{e\gamma \\ \mu\mu}} $. The current discrepancy between experiment and the SM prediction~\cite{Blum:2013xva} corresponds to $\text{Re}\, L_{\substack{e\gamma \\ \mu\mu}} \approx (20\pm 6) \times 10^{-10}\, \text{GeV}^{-1}$.

The Lagrangian coefficients at $\mu=m_b$ are related to those at $\mu=M_Z$ by
\renewcommand{\arraystretch}{1.25}
\begin{align}
\left[ \begin{array}{cc}  \wcc{\Delta}{V}{LL} \\ \wcc{\Delta}{V}{RR} \\ \wcc{\Delta}{V}{LR} \\ \wcc{\Delta}{S}{RL} \\
\wcc{\Delta}{S}{RR}  \\ \wcc{\Delta}{T}{RR}  \end{array}\right]_{\mu=m_b} &=
\left[ \begin{array}{cccccc} 
0.99 & 0 & 0 & 0 & 0 & 0 \\
0 & 0.99 & 0 & 0 & 0 & 0 \\
0 & 0 & 1.01  & 0 & 0  & 0\\
 0 & 0 & 0 & 1.38 & 0 & 0 \\
 0 & 0 & 0 & 0 & 1.38 & 0.07 \\
 0 & 0 & 0 & 0 & 0.001 & 0.90 \\
\end{array}\right]
 \left[ \begin{array}{cc}  \wcc{\Delta}{V}{LL} \\ \wcc{\Delta}{V}{RR} \\ \wcc{\Delta}{V}{LR} \\  \wcc{\Delta}{S}{RL} \\
\wcc{\Delta}{S}{RR}  \\ \wcc{\Delta}{T}{RR}  \end{array}\right]_{\mu=m_Z}.
\label{7.17}
\end{align}
The electromagnetic correction is 7\% from the $32 e^2$ term in Eq.~(\ref{7.16}).

The inclusive $b \to s \ell^+ \ell^-$ decay rate is quadratic in the coefficients appearing in Eq.~(\ref{7.11}). The dipole and vector operators produce leptons of the same chirality, whereas the scalar and tensor operators produce leptons of opposite chirality. The interference terms between the two are helicity suppressed by $m_\ell/m_b$, and can be neglected. Since the scalar and tensor operators have no SM contribution, they only contribute quadratically (i.e.\ at order $v^4/\Lambda^4$) to the rate, and can be neglected. 

The most likely way to obtain the $R_{K,K^*}$ ratios Eq.~(\ref{RK}) is through lepton universality violation in the coefficients 
$c^{V,LL},c^{V,RR},c^{V,LR},c^{\prime V, LR}$. $B \to K$ decays depend on the hadronic matrix elements of $(\overline s \gamma^\mu b)$ and 
$(\overline s \sigma^{\mu \nu} b)$, whereas $B \to K^*$ decays depend on these, as well as the matrix elements of $(\overline s \gamma^\mu \gamma_5 b)$. Note that $(\overline s  \sigma^{\mu \nu} \gamma_5 b )= -(i/2) \epsilon^{\mu \nu \alpha \beta} (\overline s  \sigma_{\alpha \beta}b) $, and is not an independent operator. The general matrix elements can be obtained from the results of Ref.~\cite{Ligeti:2016npd}.
A global fit to $B$-decay experiments~\cite{Alonso:2014csa} indicates that the simplest explanation for $R_{K^*}$ is due to a deviation in the direction $\delta C_9=- \delta C_{10}$ from the SM values for the Wilson coefficients, i.e.\ $\wcc{\Delta}{V}{LL} \not = 0 $ is the source of the discrepancy in $R_{K^*}$, with $\delta C_9 + \delta C_{10} \approx 1$. Furthermore, Ref.~\cite{Alonso:2014csa} also concludes that explaining $R_K$ in Eq.~(\ref{RK}) requires non-zero $\wcc{\Delta}{V}{LR}, \wcc{\Delta}{ V}{RR}$. The lepton universality-violating coefficients needed in Ref.~\cite{Alonso:2014csa} to explain the $R_{K,K^*}$ anomalies only have small electromagnetic running as given in Eq.~(\ref{7.17}), so the results of Ref.~\cite{Alonso:2014csa} at $\mu=m_b$ can be taken to be unchanged at $\mu=M_Z$.


\section{Conclusions}
\label{sec:Conclusions}

In this paper, we have classified all the operators up to dimension six that can appear in an $SU(3) \times U(1)_Q$ invariant low-energy effective field theory Lagrangian below 
the electroweak scale to order $G_F \sim 1/v^2$, and we have constructed a complete operator basis for this low-energy EFT up to dimension six.   The LEFT Lagrangian contains
70 Hermitian operators of dimension five, and 3631 Hermitian operators of dimension six that do not violate baryon or lepton number, as well as baryon- and lepton-number-violating operators. At dimension three, LEFT contains $\Delta L = \pm 2$ Majorana-neutrino mass operators, and at dimension five, it contains $\Delta L = \pm 2$ Majorana-neutrino dipole operators.  At dimension six, there are numerous additional LEFT operators: there are $\Delta L = \pm 4$, $\Delta L= \pm 2$, $\Delta B = \Delta L = \pm 1$ and $\Delta B = -\Delta L =\pm 1$ operator sectors.  

The tree-level matching to the LEFT operator basis at the electroweak scale has been computed from SMEFT up to dimension six operators in this paper.  The complete one-loop RGE of the LEFT Lagrangian is computed in a companion paper~\cite{Jenkins:2017dyc}. 
While parts of the RGE for flavor-violating processes are known to higher order~\cite{Buchalla:1995vs},
our calculation gives the complete renormalization of the entire set of LEFT operators up to dimension six, including non-linear terms and including corrections to the RGE of the QCD and QED gauge couplings and fermion masses due to higher-dimension operators in LEFT.
The RGE results presented in~\cite{Jenkins:2017dyc} show that some contributions to four-fermion operator coefficients that are quadratic in dimension-five coefficients come with large numerical prefactors of 96 or 192. These terms contribute e.g.\ to processes that change flavor by two units, such as $K$-$\bar K$ mixing or $\tau^-\to \mu^+e^-e^-$. Phenomenological implications of these terms will be the subject of further study.

The results obtained here together with the one-loop RGE in LEFT~\cite{Jenkins:2017dyc}, combined with previous results on SMEFT~\cite{Grojean:2013kd,Jenkins:2013zja,Jenkins:2013wua,Alonso:2013hga,Alonso:2014zka}, allow one to compute the low-energy Lagrangian starting from the SMEFT at a scale $\Lambda$ far above the electroweak scale to leading-log order, i.e.\ using tree-level matching and one-loop running. The low-energy Lagrangian then can be used to compute experimental observables without large logarithms, e.g.\ by using $\mu=m_b$ to compute $B$ decays, etc.

Determining LEFT parameters from low-energy experimental data provides a model-independent way to constrain BSM physics via low-energy observables. Observables measured above the electroweak scale can be used to constrain the parameters of SMEFT. By comparing these determinations of LEFT and SMEFT parameters with the matching relations between the two theories, one can test whether the SMEFT is a good description of physics below the TeV scale. 
SMEFT assumes that electroweak gauge symmetry is broken by a single fundamental Higgs doublet that acquires a vacuum expectation value.
Consistency of the SMEFT and LEFT parameters thus tests whether the electroweak gauge symmetry breaking mechanism of the SM and SMEFT is correct.

\section*{Acknowledgements}
\addcontentsline{toc}{section}{Acknowledgements}

We thank V.~Cirigliano for useful discussions
and the authors of Ref.~\cite{Fuentes-Martin:2020zaz} for pointing out a mistake in a previous version of the paper.
Financial support by
the DOE (Grant No.\ DE-SC0009919)
is gratefully acknowledged.
P.S.\ is supported by a grant of the Swiss National Science Foundation (Project No.\
P300P2\_167751).


\begin{appendix}


\section{SMEFT Operator Basis}
\label{sec:SMEFTBasis}

This appendix lists the SMEFT operators up to dimension six. The operators were listed in Ref.~\cite{Grzadkowski:2010es}. They are reproduced here since we make extensive use of them in this paper.

\vspace{2cm}


\begin{table}[H]
\begin{center}
\small
\begin{minipage}[t]{5.2cm}
\renewcommand{\arraystretch}{1.5}
\begin{tabular}[t]{c|c}
\multicolumn{2}{c}{\boldmath{$\Delta L = 2 \qquad (LL)HH+\hc$}} \\
\hline
$Q_{5}$      & $\epsilon^{ij} \epsilon^{k\ell} (l_{ip}^T C l_{kr} ) H_j H_\ell  $  \\
\end{tabular}
\end{minipage}
\end{center}
\caption{Dimension-five $\Delta L=2$ operator $Q_5$ in SMEFT.  There is also the Hermitian conjugate $\Delta L = -2$ operator $Q_5^\dagger$, as indicated by ${}+\hc$ in the table heading.
Subscripts $p$ and $r$ are weak-eigenstate indices.}
\label{tab:smeft5ops}
\end{table}


\begin{table}[H]
\hspace{-0.5cm}
\begin{center}
\begin{adjustbox}{width=1\textwidth,center}
\small
\begin{minipage}[t]{4.45cm}
\renewcommand{\arraystretch}{1.5}
\begin{tabular}[t]{c|c}
\multicolumn{2}{c}{\boldmath$1:X^3$} \\
\hline
$Q_G$                & $f^{ABC} G_\mu^{A\nu} G_\nu^{B\rho} G_\rho^{C\mu} $ \\
$Q_{\widetilde G}$          & $f^{ABC} \widetilde G_\mu^{A\nu} G_\nu^{B\rho} G_\rho^{C\mu} $ \\
$Q_W$                & $\epsilon^{IJK} W_\mu^{I\nu} W_\nu^{J\rho} W_\rho^{K\mu}$ \\ 
$Q_{\widetilde W}$          & $\epsilon^{IJK} \widetilde W_\mu^{I\nu} W_\nu^{J\rho} W_\rho^{K\mu}$ \\
\end{tabular}
\end{minipage}
\begin{minipage}[t]{2.7cm}
\renewcommand{\arraystretch}{1.5}
\begin{tabular}[t]{c|c}
\multicolumn{2}{c}{\boldmath$2:H^6$} \\
\hline
$Q_H$       & $(H^\dag H)^3$ 
\end{tabular}
\end{minipage}
\begin{minipage}[t]{5.1cm}
\renewcommand{\arraystretch}{1.5}
\begin{tabular}[t]{c|c}
\multicolumn{2}{c}{\boldmath$3:H^4 D^2$} \\
\hline
$Q_{H\Box}$ & $(H^\dag H)\Box(H^\dag H)$ \\
$Q_{H D}$   & $\ \left(H^\dag D^\mu H\right)^* \left(H^\dag D_\mu H\right)$ 
\end{tabular}
\end{minipage}
\begin{minipage}[t]{2.7cm}
\renewcommand{\arraystretch}{1.5}
\begin{tabular}[t]{c|c}
\multicolumn{2}{c}{\boldmath$5: \psi^2H^3 + \hc$} \\
\hline
$Q_{eH}$           & $(H^\dag H)(\bar l_p e_r H)$ \\
$Q_{uH}$          & $(H^\dag H)(\bar q_p u_r \widetilde H )$ \\
$Q_{dH}$           & $(H^\dag H)(\bar q_p d_r H)$\\
\end{tabular}
\end{minipage}

\vspace{0.25cm}

\end{adjustbox}

\begin{adjustbox}{width=0.9\textwidth,center}

\begin{minipage}[t]{4.7cm}
\renewcommand{\arraystretch}{1.5}
\begin{tabular}[t]{c|c}
\multicolumn{2}{c}{\boldmath$4:X^2H^2$} \\
\hline
$Q_{H G}$     & $H^\dag H\, G^A_{\mu\nu} G^{A\mu\nu}$ \\
$Q_{H\widetilde G}$         & $H^\dag H\, \widetilde G^A_{\mu\nu} G^{A\mu\nu}$ \\
$Q_{H W}$     & $H^\dag H\, W^I_{\mu\nu} W^{I\mu\nu}$ \\
$Q_{H\widetilde W}$         & $H^\dag H\, \widetilde W^I_{\mu\nu} W^{I\mu\nu}$ \\
$Q_{H B}$     & $ H^\dag H\, B_{\mu\nu} B^{\mu\nu}$ \\
$Q_{H\widetilde B}$         & $H^\dag H\, \widetilde B_{\mu\nu} B^{\mu\nu}$ \\
$Q_{H WB}$     & $ H^\dag \tau^I H\, W^I_{\mu\nu} B^{\mu\nu}$ \\
$Q_{H\widetilde W B}$         & $H^\dag \tau^I H\, \widetilde W^I_{\mu\nu} B^{\mu\nu}$ 
\end{tabular}
\end{minipage}
\begin{minipage}[t]{5.2cm}
\renewcommand{\arraystretch}{1.5}
\begin{tabular}[t]{c|c}
\multicolumn{2}{c}{\boldmath$6:\psi^2 XH+\hc$} \\
\hline
$Q_{eW}$      & $(\bar l_p \sigma^{\mu\nu} \tau^I e_r) H W_{\mu\nu}^I$ \\
$Q_{eB}$        & $(\bar l_p \sigma^{\mu\nu} e_r) H B_{\mu\nu}$ \\
$Q_{uG}$        & $(\bar q_p \sigma^{\mu\nu} T^A u_r) \widetilde H \, G_{\mu\nu}^A$ \\
$Q_{uW}$        & $(\bar q_p \sigma^{\mu\nu} \tau^I u_r) \widetilde H \, W_{\mu\nu}^I$ \\
$Q_{uB}$        & $(\bar q_p \sigma^{\mu\nu} u_r) \widetilde H \, B_{\mu\nu}$ \\
$Q_{dG}$        & $(\bar q_p \sigma^{\mu\nu} T^A d_r) H\, G_{\mu\nu}^A$ \\
$Q_{dW}$         & $(\bar q_p \sigma^{\mu\nu} \tau^I d_r) H\, W_{\mu\nu}^I$ \\
$Q_{dB}$        & $(\bar q_p \sigma^{\mu\nu} d_r) H\, B_{\mu\nu}$ 
\end{tabular}
\end{minipage}
\begin{minipage}[t]{5.4cm}
\renewcommand{\arraystretch}{1.5}
\begin{tabular}[t]{c|c}
\multicolumn{2}{c}{\boldmath$7:\psi^2H^2 D$} \\
\hline
$Q_{H l}^{(1)}$      & $(H^\dag i\overleftrightarrow{D}_\mu H)(\bar l_p \gamma^\mu l_r)$\\
$Q_{H l}^{(3)}$      & $(H^\dag i\overleftrightarrow{D}^I_\mu H)(\bar l_p \tau^I \gamma^\mu l_r)$\\
$Q_{H e}$            & $(H^\dag i\overleftrightarrow{D}_\mu H)(\bar e_p \gamma^\mu e_r)$\\
$Q_{H q}^{(1)}$      & $(H^\dag i\overleftrightarrow{D}_\mu H)(\bar q_p \gamma^\mu q_r)$\\
$Q_{H q}^{(3)}$      & $(H^\dag i\overleftrightarrow{D}^I_\mu H)(\bar q_p \tau^I \gamma^\mu q_r)$\\
$Q_{H u}$            & $(H^\dag i\overleftrightarrow{D}_\mu H)(\bar u_p \gamma^\mu u_r)$\\
$Q_{H d}$            & $(H^\dag i\overleftrightarrow{D}_\mu H)(\bar d_p \gamma^\mu d_r)$\\
$Q_{H u d}$ + h.c.   & $i(\widetilde H ^\dag D_\mu H)(\bar u_p \gamma^\mu d_r)$\\
\end{tabular}
\end{minipage}
\end{adjustbox}
\end{center}

\begin{center}
\begin{adjustbox}{width=0.9\textwidth,center}
\begin{minipage}[t]{4.75cm}
\renewcommand{\arraystretch}{1.5}
\begin{tabular}[t]{c|c}
\multicolumn{2}{c}{\boldmath$8:(\bar L L)(\bar L L)$} \\
\hline
$Q_{ll}$        & $(\bar l_p \gamma^\mu l_r)(\bar l_s \gamma_\mu l_t)$ \\
$Q_{qq}^{(1)}$  & $(\bar q_p \gamma^\mu q_r)(\bar q_s \gamma_\mu q_t)$ \\
$Q_{qq}^{(3)}$  & $(\bar q_p \gamma^\mu \tau^I q_r)(\bar q_s \gamma_\mu \tau^I q_t)$ \\
$Q_{lq}^{(1)}$                & $(\bar l_p \gamma^\mu l_r)(\bar q_s \gamma_\mu q_t)$ \\
$Q_{lq}^{(3)}$                & $(\bar l_p \gamma^\mu \tau^I l_r)(\bar q_s \gamma_\mu \tau^I q_t)$ 
\end{tabular}
\end{minipage}
\hspace{0.2cm}
\begin{minipage}[t]{5.25cm}
\renewcommand{\arraystretch}{1.5}
\begin{tabular}[t]{c|c}
\multicolumn{2}{c}{\boldmath$8:(\bar R R)(\bar R R)$} \\
\hline
$Q_{ee}$               & $(\bar e_p \gamma^\mu e_r)(\bar e_s \gamma_\mu e_t)$ \\
$Q_{uu}$        & $(\bar u_p \gamma^\mu u_r)(\bar u_s \gamma_\mu u_t)$ \\
$Q_{dd}$        & $(\bar d_p \gamma^\mu d_r)(\bar d_s \gamma_\mu d_t)$ \\
$Q_{eu}$                      & $(\bar e_p \gamma^\mu e_r)(\bar u_s \gamma_\mu u_t)$ \\
$Q_{ed}$                      & $(\bar e_p \gamma^\mu e_r)(\bar d_s\gamma_\mu d_t)$ \\
$Q_{ud}^{(1)}$                & $(\bar u_p \gamma^\mu u_r)(\bar d_s \gamma_\mu d_t)$ \\
$Q_{ud}^{(8)}$                & $(\bar u_p \gamma^\mu T^A u_r)(\bar d_s \gamma_\mu T^A d_t)$ \\
\end{tabular}
\end{minipage}
\hspace{0.2cm}
\begin{minipage}[t]{4.75cm}
\renewcommand{\arraystretch}{1.5}
\begin{tabular}[t]{c|c}
\multicolumn{2}{c}{\boldmath$8:(\bar L L)(\bar R R)$} \\
\hline
$Q_{le}$               & $(\bar l_p \gamma^\mu l_r)(\bar e_s \gamma_\mu e_t)$ \\
$Q_{lu}$               & $(\bar l_p \gamma^\mu l_r)(\bar u_s \gamma_\mu u_t)$ \\
$Q_{ld}$               & $(\bar l_p \gamma^\mu l_r)(\bar d_s \gamma_\mu d_t)$ \\
$Q_{qe}$               & $(\bar q_p \gamma^\mu q_r)(\bar e_s \gamma_\mu e_t)$ \\
$Q_{qu}^{(1)}$         & $(\bar q_p \gamma^\mu q_r)(\bar u_s \gamma_\mu u_t)$ \\ 
$Q_{qu}^{(8)}$         & $(\bar q_p \gamma^\mu T^A q_r)(\bar u_s \gamma_\mu T^A u_t)$ \\ 
$Q_{qd}^{(1)}$ & $(\bar q_p \gamma^\mu q_r)(\bar d_s \gamma_\mu d_t)$ \\
$Q_{qd}^{(8)}$ & $(\bar q_p \gamma^\mu T^A q_r)(\bar d_s \gamma_\mu T^A d_t)$\\
\end{tabular}
\end{minipage}

\end{adjustbox}

\vspace{0.25cm}

\begin{adjustbox}{width=0.57\textwidth,center}

\begin{minipage}[t]{3.75cm}
\renewcommand{\arraystretch}{1.5}
\begin{tabular}[t]{c|c}
\multicolumn{2}{c}{\boldmath$8:(\bar LR)(\bar RL)+\hc$} \\
\hline
$Q_{ledq}$ & $(\bar l_p^j e_r)(\bar d_s q_{tj})$ 
\end{tabular}
\end{minipage}
\hspace{0.4cm}
\begin{minipage}[t]{5.5cm}
\renewcommand{\arraystretch}{1.5}
\begin{tabular}[t]{c|c}
\multicolumn{2}{c}{\boldmath$8:(\bar LR)(\bar L R)+\hc$} \\
\hline
$Q_{quqd}^{(1)}$ & $(\bar q_p^j u_r) \epsilon_{jk} (\bar q_s^k d_t)$ \\
$Q_{quqd}^{(8)}$ & $(\bar q_p^j T^A u_r) \epsilon_{jk} (\bar q_s^k T^A d_t)$ \\
$Q_{lequ}^{(1)}$ & $(\bar l_p^j e_r) \epsilon_{jk} (\bar q_s^k u_t)$ \\
$Q_{lequ}^{(3)}$ & $(\bar l_p^j \sigma_{\mu\nu} e_r) \epsilon_{jk} (\bar q_s^k \sigma^{\mu\nu} u_t)$
\end{tabular}
\end{minipage}
\end{adjustbox}
\end{center}
\caption{The 76 dimension-six operators that conserve baryon and lepton number in SMEFT. The operators are divided into eight classes according to their field content. The class-8 $\psi^4$ four-fermion operators are further divided into subclasses according to their chiral properties. Operators with ${}+\hc$ have Hermitian conjugates, as does the $\psi^2 H^2 D$ operator $Q_{Hud}$. The subscripts $p, r, s, t$ are weak-eigenstate indices.}
\label{tab:smeft6ops}
\end{table}


\begin{table}[H]
\begin{center}
\small
\begin{minipage}[t]{5.2cm}
\renewcommand{\arraystretch}{1.5}
\begin{tabular}[t]{c|c}
\multicolumn{2}{c}{\boldmath{$\Delta B = \Delta L = 1 +\hc$}} \\
\hline
$Q_{duql}$      & $\epsilon^{\alpha \beta \gamma} \epsilon^{ij} (d^T_{\alpha p} C u_{\beta r} ) (q^T_{\gamma i s} C l_{jt})  $  \\
$Q_{qque}$      & $\epsilon^{\alpha \beta \gamma} \epsilon^{ij} (q^T_{\alpha i p} C q_{\beta j r} ) (u^T_{\gamma s} C e_{t})  $  \\
$Q_{qqql}$      & $\epsilon^{\alpha \beta \gamma} \epsilon^{i\ell} \epsilon^{jk} (q^T_{\alpha i p} C q_{\beta j r} ) (q^T_{\gamma k s} C l_{\ell t})  $  \\
$Q_{duue}$      & $\epsilon^{\alpha \beta \gamma} (d^T_{\alpha p} C u_{\beta r} ) (u^T_{\gamma s} C e_{t})  $  \\
\end{tabular}
\end{minipage}
\end{center}
\caption{Dimension-six $\Delta B= \Delta L=1$ operators in SMEFT.  There are also Hermitian conjugate $\Delta B = \Delta L = -1$ operators, as indicated by ${}+\hc$ in the table heading.
Subscripts $p$, $r$, $s$ and $t$ are weak-eigenstate indices.}
\label{tab:smeft6baryonops}
\end{table}


\clearpage

\section{LEFT Operator Basis}
\label{sec:LEFTBasis}

This appendix lists the LEFT operators up to dimension six.  Weak-eigenstate indices of the operators are not shown---e.g.\ $\op{ee}{V}{LL}$ with the weak-eigenstate indices 
included is $\op{\substack{ ee \\ prst } }{V}{LL}$.

\begin{table}[H]
\vspace{-0.75cm}
\begin{adjustbox}{width=0.85\textwidth,center}
\begin{minipage}[t]{3cm}
\renewcommand{\arraystretch}{1.51}
\small
\begin{align*}
\begin{array}[t]{c|c}
\multicolumn{2}{c}{\boldsymbol{\nu \nu+\hc}} \\
\hline
\O_{\nu} & (\nu_{Lp}^T C \nu_{Lr})  \\
\end{array}
\end{align*}
\end{minipage}
%
\begin{minipage}[t]{3cm}
\renewcommand{\arraystretch}{1.51}
\small
\begin{align*}
\begin{array}[t]{c|c}
\multicolumn{2}{c}{\boldsymbol{(\nu \nu) X+\hc}} \\
\hline
\O_{\nu \gamma} & (\nu_{Lp}^T C   \sigma^{\mu \nu}  \nu_{Lr})  F_{\mu \nu}  \\
\end{array}
\end{align*}
\end{minipage}
\begin{minipage}[t]{3cm}
\renewcommand{\arraystretch}{1.51}
\small
\begin{align*}
\begin{array}[t]{c|c}
\multicolumn{2}{c}{\boldsymbol{(\overline L R ) X+\hc}} \\
\hline
\O_{e \gamma} & \bar e_{Lp}   \sigma^{\mu \nu} e_{Rr}\, F_{\mu \nu}  \\
\O_{u \gamma} & \bar u_{Lp}   \sigma^{\mu \nu}  u_{Rr}\, F_{\mu \nu}   \\
\O_{d \gamma} & \bar d_{Lp}  \sigma^{\mu \nu} d_{Rr}\, F_{\mu \nu}  \\
\O_{u G} & \bar u_{Lp}   \sigma^{\mu \nu}  T^A u_{Rr}\,  G_{\mu \nu}^A  \\
\O_{d G} & \bar d_{Lp}   \sigma^{\mu \nu} T^A d_{Rr}\,  G_{\mu \nu}^A \\
\end{array}
\end{align*}
\end{minipage}
\begin{minipage}[t]{3cm}
\renewcommand{\arraystretch}{1.51}
\small
\begin{align*}
\begin{array}[t]{c|c}
\multicolumn{2}{c}{\boldsymbol{X^3}} \\
\hline
\O_G     & f^{ABC} G_\mu^{A\nu} G_\nu^{B\rho} G_\rho^{C\mu}  \\
\O_{\widetilde G} & f^{ABC} \widetilde G_\mu^{A\nu} G_\nu^{B\rho} G_\rho^{C\mu}   \\
\end{array}
\end{align*}
\end{minipage}
\end{adjustbox}
%

%
\mbox{}\\[-1.25cm]

\begin{adjustbox}{width=1.05\textwidth,center}
\begin{minipage}[t]{3cm}
\renewcommand{\arraystretch}{1.51}
\small
\begin{align*}
\begin{array}[t]{c|c}
\multicolumn{2}{c}{\boldsymbol{(\overline L L)(\overline L  L)}} \\
\hline
\op{\nu\nu}{V}{LL} & (\bar \nu_{Lp}  \gamma^\mu \nu_{Lr} )(\bar \nu_{Ls} \gamma_\mu \nu_{Lt})   \\
\op{ee}{V}{LL}       & (\bar e_{Lp}  \gamma^\mu e_{Lr})(\bar e_{Ls} \gamma_\mu e_{Lt})   \\
\op{\nu e}{V}{LL}       & (\bar \nu_{Lp} \gamma^\mu \nu_{Lr})(\bar e_{Ls}  \gamma_\mu e_{Lt})  \\
\op{\nu u}{V}{LL}       & (\bar \nu_{Lp} \gamma^\mu \nu_{Lr}) (\bar u_{Ls}  \gamma_\mu u_{Lt})  \\
\op{\nu d}{V}{LL}       & (\bar \nu_{Lp} \gamma^\mu \nu_{Lr})(\bar d_{Ls} \gamma_\mu d_{Lt})     \\
\op{eu}{V}{LL}      & (\bar e_{Lp}  \gamma^\mu e_{Lr})(\bar u_{Ls} \gamma_\mu u_{Lt})   \\
\op{ed}{V}{LL}       & (\bar e_{Lp}  \gamma^\mu e_{Lr})(\bar d_{Ls} \gamma_\mu d_{Lt})  \\
\op{\nu edu}{V}{LL}      & (\bar \nu_{Lp} \gamma^\mu e_{Lr}) (\bar d_{Ls} \gamma_\mu u_{Lt})  + \hc   \\
\op{uu}{V}{LL}        & (\bar u_{Lp} \gamma^\mu u_{Lr})(\bar u_{Ls} \gamma_\mu u_{Lt})    \\
\op{dd}{V}{LL}   & (\bar d_{Lp} \gamma^\mu d_{Lr})(\bar d_{Ls} \gamma_\mu d_{Lt})    \\
\op{ud}{V1}{LL}     & (\bar u_{Lp} \gamma^\mu u_{Lr}) (\bar d_{Ls} \gamma_\mu d_{Lt})  \\
\op{ud}{V8}{LL}     & (\bar u_{Lp} \gamma^\mu T^A u_{Lr}) (\bar d_{Ls} \gamma_\mu T^A d_{Lt})   \\[-0.5cm]
\end{array}
\end{align*}
\renewcommand{\arraystretch}{1.51}
\small
\begin{align*}
\begin{array}[t]{c|c}
\multicolumn{2}{c}{\boldsymbol{(\overline R  R)(\overline R R)}} \\
\hline
\op{ee}{V}{RR}     & (\bar e_{Rp} \gamma^\mu e_{Rr})(\bar e_{Rs} \gamma_\mu e_{Rt})  \\
\op{eu}{V}{RR}       & (\bar e_{Rp}  \gamma^\mu e_{Rr})(\bar u_{Rs} \gamma_\mu u_{Rt})   \\
\op{ed}{V}{RR}     & (\bar e_{Rp} \gamma^\mu e_{Rr})  (\bar d_{Rs} \gamma_\mu d_{Rt})   \\
\op{uu}{V}{RR}      & (\bar u_{Rp} \gamma^\mu u_{Rr})(\bar u_{Rs} \gamma_\mu u_{Rt})  \\
\op{dd}{V}{RR}      & (\bar d_{Rp} \gamma^\mu d_{Rr})(\bar d_{Rs} \gamma_\mu d_{Rt})    \\
\op{ud}{V1}{RR}       & (\bar u_{Rp} \gamma^\mu u_{Rr}) (\bar d_{Rs} \gamma_\mu d_{Rt})  \\
\op{ud}{V8}{RR}    & (\bar u_{Rp} \gamma^\mu T^A u_{Rr}) (\bar d_{Rs} \gamma_\mu T^A d_{Rt})  \\
\end{array}
\end{align*}
\end{minipage}
%
%
\begin{minipage}[t]{3cm}
\renewcommand{\arraystretch}{1.51}
\small
\begin{align*}
\begin{array}[t]{c|c}
\multicolumn{2}{c}{\boldsymbol{(\overline L  L)(\overline R  R)}} \\
\hline
\op{\nu e}{V}{LR}     & (\bar \nu_{Lp} \gamma^\mu \nu_{Lr})(\bar e_{Rs}  \gamma_\mu e_{Rt})  \\
\op{ee}{V}{LR}       & (\bar e_{Lp}  \gamma^\mu e_{Lr})(\bar e_{Rs} \gamma_\mu e_{Rt}) \\
\op{\nu u}{V}{LR}         & (\bar \nu_{Lp} \gamma^\mu \nu_{Lr})(\bar u_{Rs}  \gamma_\mu u_{Rt})    \\
\op{\nu d}{V}{LR}         & (\bar \nu_{Lp} \gamma^\mu \nu_{Lr})(\bar d_{Rs} \gamma_\mu d_{Rt})   \\
\op{eu}{V}{LR}        & (\bar e_{Lp}  \gamma^\mu e_{Lr})(\bar u_{Rs} \gamma_\mu u_{Rt})   \\
\op{ed}{V}{LR}        & (\bar e_{Lp}  \gamma^\mu e_{Lr})(\bar d_{Rs} \gamma_\mu d_{Rt})   \\
\op{ue}{V}{LR}        & (\bar u_{Lp} \gamma^\mu u_{Lr})(\bar e_{Rs}  \gamma_\mu e_{Rt})   \\
\op{de}{V}{LR}         & (\bar d_{Lp} \gamma^\mu d_{Lr}) (\bar e_{Rs} \gamma_\mu e_{Rt})   \\
\op{\nu edu}{V}{LR}        & (\bar \nu_{Lp} \gamma^\mu e_{Lr})(\bar d_{Rs} \gamma_\mu u_{Rt})  +\hc \\
\op{uu}{V1}{LR}        & (\bar u_{Lp} \gamma^\mu u_{Lr})(\bar u_{Rs} \gamma_\mu u_{Rt})   \\
\op{uu}{V8}{LR}       & (\bar u_{Lp} \gamma^\mu T^A u_{Lr})(\bar u_{Rs} \gamma_\mu T^A u_{Rt})    \\ 
\op{ud}{V1}{LR}       & (\bar u_{Lp} \gamma^\mu u_{Lr}) (\bar d_{Rs} \gamma_\mu d_{Rt})  \\
\op{ud}{V8}{LR}       & (\bar u_{Lp} \gamma^\mu T^A u_{Lr})  (\bar d_{Rs} \gamma_\mu T^A d_{Rt})  \\
\op{du}{V1}{LR}       & (\bar d_{Lp} \gamma^\mu d_{Lr})(\bar u_{Rs} \gamma_\mu u_{Rt})   \\
\op{du}{V8}{LR}       & (\bar d_{Lp} \gamma^\mu T^A d_{Lr})(\bar u_{Rs} \gamma_\mu T^A u_{Rt}) \\
\op{dd}{V1}{LR}      & (\bar d_{Lp} \gamma^\mu d_{Lr})(\bar d_{Rs} \gamma_\mu d_{Rt})  \\
\op{dd}{V8}{LR}   & (\bar d_{Lp} \gamma^\mu T^A d_{Lr})(\bar d_{Rs} \gamma_\mu T^A d_{Rt}) \\
\op{uddu}{V1}{LR}   & (\bar u_{Lp} \gamma^\mu d_{Lr})(\bar d_{Rs} \gamma_\mu u_{Rt})  + \hc  \\
\op{uddu}{V8}{LR}      & (\bar u_{Lp} \gamma^\mu T^A d_{Lr})(\bar d_{Rs} \gamma_\mu T^A  u_{Rt})  + \hc \\
\end{array}
\end{align*}
\end{minipage}

\begin{minipage}[t]{3cm}
\renewcommand{\arraystretch}{1.51}
\small
\begin{align*}
\begin{array}[t]{c|c}
\multicolumn{2}{c}{\boldsymbol{(\overline L R)(\overline L R)+\hc}} \\
\hline
\op{ee}{S}{RR} 		& (\bar e_{Lp}   e_{Rr}) (\bar e_{Ls} e_{Rt})   \\
\op{eu}{S}{RR}  & (\bar e_{Lp}   e_{Rr}) (\bar u_{Ls} u_{Rt})   \\
\op{eu}{T}{RR} & (\bar e_{Lp}   \sigma^{\mu \nu}   e_{Rr}) (\bar u_{Ls}  \sigma_{\mu \nu}  u_{Rt})  \\
\op{ed}{S}{RR}  & (\bar e_{Lp} e_{Rr})(\bar d_{Ls} d_{Rt})  \\
\op{ed}{T}{RR} & (\bar e_{Lp} \sigma^{\mu \nu} e_{Rr}) (\bar d_{Ls} \sigma_{\mu \nu} d_{Rt})   \\
\op{\nu edu}{S}{RR} & (\bar   \nu_{Lp} e_{Rr})  (\bar d_{Ls} u_{Rt} ) \\
\op{\nu edu}{T}{RR} &  (\bar  \nu_{Lp}  \sigma^{\mu \nu} e_{Rr} )  (\bar  d_{Ls}  \sigma_{\mu \nu} u_{Rt} )   \\
\op{uu}{S1}{RR}  & (\bar u_{Lp}   u_{Rr}) (\bar u_{Ls} u_{Rt})  \\
\op{uu}{S8}{RR}   & (\bar u_{Lp}   T^A u_{Rr}) (\bar u_{Ls} T^A u_{Rt})  \\
\op{ud}{S1}{RR}   & (\bar u_{Lp} u_{Rr})  (\bar d_{Ls} d_{Rt})   \\
\op{ud}{S8}{RR}  & (\bar u_{Lp} T^A u_{Rr})  (\bar d_{Ls} T^A d_{Rt})  \\
\op{dd}{S1}{RR}   & (\bar d_{Lp} d_{Rr}) (\bar d_{Ls} d_{Rt}) \\
\op{dd}{S8}{RR}  & (\bar d_{Lp} T^A d_{Rr}) (\bar d_{Ls} T^A d_{Rt})  \\
\op{uddu}{S1}{RR} &  (\bar u_{Lp} d_{Rr}) (\bar d_{Ls}  u_{Rt})   \\
\op{uddu}{S8}{RR}  &  (\bar u_{Lp} T^A d_{Rr}) (\bar d_{Ls}  T^A u_{Rt})  \\[-0.5cm]
\end{array}
\end{align*}
\renewcommand{\arraystretch}{1.51}
\small
\begin{align*}
\begin{array}[t]{c|c}
\multicolumn{2}{c}{\boldsymbol{(\overline L R)(\overline R L) +\hc}} \\
\hline
\op{eu}{S}{RL}  & (\bar e_{Lp} e_{Rr}) (\bar u_{Rs}  u_{Lt})  \\
\op{ed}{S}{RL} & (\bar e_{Lp} e_{Rr}) (\bar d_{Rs} d_{Lt}) \\
\op{\nu edu}{S}{RL}  & (\bar \nu_{Lp} e_{Rr}) (\bar d_{Rs}  u_{Lt})  \\
\end{array}
\end{align*}
\end{minipage}
\end{adjustbox}
\setlength{\abovecaptionskip}{0.15cm}
\setlength{\belowcaptionskip}{-3cm}
\caption{The operators for LEFT of dimension three, five, and six that conserve baryon and lepton number, and the dimension-three and dimension-five $\Delta L=\pm 2$ operators.  The dimension-three $\Delta L=2$ operator $\O_\nu$ is the Majorana-neutrino mass operator, while the dimension-five $\Delta L=2$ operator $\O_{\nu\gamma}$ is the Majorana-neutrino dipole operator.  There are 5 additional dimension-five dipole operators $(\bar L R)X$.  The 80 dimension-six operators consist of 2 pure gauge operators $X^3$ and 78 four-fermion operators $\psi^4$, which are further divided by their chiral structure.  The $\psi^4$ operator superscripts $V$, $S$, $T$ refer to products of vector, scalar, and tensor fermion bilinears, and the additional two labels $L$ or $R$ refer to the chiral projectors in the bilinears.  Operators with ${}+\hc$ have Hermitian conjugates.  The subscripts $p, r, s, t$ are weak-eigenstate indices.}
\label{tab:oplist1}
\end{table}

\begin{table}
%
\centering
\begin{minipage}[t]{3cm}
\renewcommand{\arraystretch}{1.5}
\small
\begin{align*}
\begin{array}[t]{c|c}
\multicolumn{2}{c}{\boldsymbol{\Delta L = 4 + \hc}}  \\
\hline
\op{\nu\nu}{S}{LL} &  (\nu_{Lp}^T C \nu_{Lr}^{}) (\nu_{Ls}^T C \nu_{Lt}^{} )  \\
\end{array}
\end{align*}
\end{minipage}

\begin{adjustbox}{width=\textwidth,center}
\begin{minipage}[t]{3cm}
\renewcommand{\arraystretch}{1.5}
\small
\begin{align*}
\begin{array}[t]{c|c}
\multicolumn{2}{c}{\boldsymbol{\Delta L =2 + \hc}}  \\
\hline
\op{\nu e}{S}{LL}  &  (\nu_{Lp}^T C \nu_{Lr}) (\bar e_{Rs} e_{Lt})   \\
\op{\nu e}{T}{LL} &  (\nu_{Lp}^T C \sigma^{\mu \nu} \nu_{Lr}) (\bar e_{Rs}\sigma_{\mu \nu} e_{Lt} )  \\
\op{\nu e}{S}{LR} &  (\nu_{Lp}^T C \nu_{Lr}) (\bar e_{Ls} e_{Rt} )  \\
\op{\nu u}{S}{LL}  &  (\nu_{Lp}^T C \nu_{Lr}) (\bar u_{Rs} u_{Lt} )  \\
\op{\nu u}{T}{LL}  &  (\nu_{Lp}^T C \sigma^{\mu \nu} \nu_{Lr}) (\bar u_{Rs} \sigma_{\mu \nu} u_{Lt} ) \\
\op{\nu u}{S}{LR}  &  (\nu_{Lp}^T C \nu_{Lr}) (\bar u_{Ls} u_{Rt} )  \\
\op{\nu d}{S}{LL}   &  (\nu_{Lp}^T C \nu_{Lr}) (\bar d_{Rs} d_{Lt} ) \\
\op{\nu d}{T}{LL}   &  (\nu_{Lp}^T C \sigma^{\mu \nu}  \nu_{Lr}) (\bar d_{Rs} \sigma_{\mu \nu} d_{Lt} ) \\
\op{\nu d}{S}{LR}  &  (\nu_{Lp}^T C \nu_{Lr}) (\bar d_{Ls} d_{Rt} ) \\
\op{\nu edu}{S}{LL} &  (\nu_{Lp}^T C e_{Lr}) (\bar d_{Rs} u_{Lt} )  \\
\op{\nu edu}{T}{LL}  & (\nu_{Lp}^T C  \sigma^{\mu \nu} e_{Lr}) (\bar d_{Rs}  \sigma_{\mu \nu} u_{Lt} ) \\
\op{\nu edu}{S}{LR}   & (\nu_{Lp}^T C e_{Lr}) (\bar d_{Ls} u_{Rt} ) \\
\op{\nu edu}{V}{RL}   & (\nu_{Lp}^T C \gamma^\mu e_{Rr}) (\bar d_{Ls} \gamma_\mu u_{Lt} )  \\
\op{\nu edu}{V}{RR}   & (\nu_{Lp}^T C \gamma^\mu e_{Rr}) (\bar d_{Rs} \gamma_\mu u_{Rt} )  \\
\end{array}
\end{align*}
\end{minipage}
%
\begin{minipage}[t]{3cm}
\renewcommand{\arraystretch}{1.5}
\small
\begin{align*}
\begin{array}[t]{c|c}
\multicolumn{2}{c}{\boldsymbol{\Delta B = \Delta L = 1 + \hc}} \\
\hline
\op{udd}{S}{LL} &  \epsilon_{\alpha\beta\gamma}  (u_{Lp}^{\alpha T} C d_{Lr}^{\beta}) (d_{Ls}^{\gamma T} C \nu_{Lt}^{})   \\
\op{duu}{S}{LL} & \epsilon_{\alpha\beta\gamma}  (d_{Lp}^{\alpha T} C u_{Lr}^{\beta}) (u_{Ls}^{\gamma T} C e_{Lt}^{})  \\
\op{uud}{S}{LR} & \epsilon_{\alpha\beta\gamma}  (u_{Lp}^{\alpha T} C u_{Lr}^{\beta}) (d_{Rs}^{\gamma T} C e_{Rt}^{})  \\
\op{duu}{S}{LR} & \epsilon_{\alpha\beta\gamma}  (d_{Lp}^{\alpha T} C u_{Lr}^{\beta}) (u_{Rs}^{\gamma T} C e_{Rt}^{})   \\
\op{uud}{S}{RL} & \epsilon_{\alpha\beta\gamma}  (u_{Rp}^{\alpha T} C u_{Rr}^{\beta}) (d_{Ls}^{\gamma T} C e_{Lt}^{})   \\
\op{duu}{S}{RL} & \epsilon_{\alpha\beta\gamma}  (d_{Rp}^{\alpha T} C u_{Rr}^{\beta}) (u_{Ls}^{\gamma T} C e_{Lt}^{})   \\
\op{dud}{S}{RL} & \epsilon_{\alpha\beta\gamma}  (d_{Rp}^{\alpha T} C u_{Rr}^{\beta}) (d_{Ls}^{\gamma T} C \nu_{Lt}^{})   \\
\op{ddu}{S}{RL} & \epsilon_{\alpha\beta\gamma}  (d_{Rp}^{\alpha T} C d_{Rr}^{\beta}) (u_{Ls}^{\gamma T} C \nu_{Lt}^{})   \\
\op{duu}{S}{RR}  & \epsilon_{\alpha\beta\gamma}  (d_{Rp}^{\alpha T} C u_{Rr}^{\beta}) (u_{Rs}^{\gamma T} C e_{Rt}^{})  \\
\end{array}
\end{align*}
\end{minipage}
%
\begin{minipage}[t]{3cm}
\renewcommand{\arraystretch}{1.5}
\small
\begin{align*}
\begin{array}[t]{c|c}
\multicolumn{2}{c}{\boldsymbol{\Delta B = - \Delta L = 1 + \hc}}  \\
\hline
\op{ddd}{S}{LL} & \epsilon_{\alpha\beta\gamma}  (d_{Lp}^{\alpha T} C d_{Lr}^{\beta}) (\bar e_{Rs}^{} d_{Lt}^\gamma )  \\
\op{udd}{S}{LR}  & \epsilon_{\alpha\beta\gamma}  (u_{Lp}^{\alpha T} C d_{Lr}^{\beta}) (\bar \nu_{Ls}^{} d_{Rt}^\gamma )  \\
\op{ddu}{S}{LR} & \epsilon_{\alpha\beta\gamma}  (d_{Lp}^{\alpha T} C d_{Lr}^{\beta})  (\bar \nu_{Ls}^{} u_{Rt}^\gamma )  \\
\op{ddd}{S}{LR} & \epsilon_{\alpha\beta\gamma}  (d_{Lp}^{\alpha T} C d_{Lr}^{\beta}) (\bar e_{Ls}^{} d_{Rt}^\gamma ) \\
\op{ddd}{S}{RL}  & \epsilon_{\alpha\beta\gamma}  (d_{Rp}^{\alpha T} C d_{Rr}^{\beta}) (\bar e_{Rs}^{} d_{Lt}^\gamma )  \\
\op{udd}{S}{RR}  & \epsilon_{\alpha\beta\gamma}  (u_{Rp}^{\alpha T} C d_{Rr}^{\beta}) (\bar \nu_{Ls}^{} d_{Rt}^\gamma )  \\
\op{ddd}{S}{RR}  & \epsilon_{\alpha\beta\gamma}  (d_{Rp}^{\alpha T} C d_{Rr}^{\beta}) (\bar e_{Ls}^{} d_{Rt}^\gamma )  \\
\end{array}
\end{align*}
\end{minipage}
\end{adjustbox}
%
\caption{The LEFT dimension-six four-fermion operators that violate baryon and/or lepton number.  All operators have Hermitian conjugates.  The operator superscripts $V$, $S$, $T$ refer to products of vector, scalar, and tensor fermion bilinears, and the additional two labels $L$ or $R$ refer to the chiral projectors in the bilinears.  The subscripts $p, r, s, t$ are weak-eigenstate indices.}
\label{tab:oplist2}
\end{table}


\section{Matching Conditions}
\label{sec:MatchingConditions}

This appendix gives the number of operators of each Lorentz type, broken up into leptonic, semileptonic, and nonleptonic categories, and the  tree-level matching conditions in SMEFT up to dimension six. Table~\ref{dim6cp} gives the number of $CP$-even and $CP$-odd operators.

\begin{table}[h]
\renewcommand{\arraystretch}{1.2}
\small
\begin{align*}
\begin{array}[t]{c|c|c|c}
\multicolumn{4}{c}{\boldsymbol{\Delta L=2 \qquad \nu \nu+{\rm h.c.}}} \\
\toprule
 & \text{Number} & \text{SM} &   \text{Matching} \\
\midrule\midrule
\O_{\nu} & \frac12 n_\nu (n_\nu+1) & 6  & \frac{1}{2} C_{\substack{5 \\ pr}} v_T^2 \\
\bottomrule
\end{array}
\end{align*}
\setlength{\belowcaptionskip}{-2cm}
\caption{Dimension-three $\Delta L=2$ Majorana neutrino mass operators in LEFT.  There are also Hermitian conjugate $\Delta L=-2$ operators $\O_{\nu}^\dagger$, as indicated in the table heading.  The second column is the number of operators for an arbitrary number of neutrino flavors $n_\nu$, and the third column is the number in the SM LEFT with $n_\nu =3$.  The last column is the tree-level matching coefficient in SMEFT. }
\label{dim3}
\end{table}

\begin{table}[h]
\renewcommand{\arraystretch}{1.2}
\small
\begin{align*}
\begin{array}[t]{c|c|c|c}
\multicolumn{4}{c}{\boldsymbol{\Delta L =2 \qquad (\nu \nu) X+{\rm h.c.}}} \\
\toprule
 & \text{Number} & \text{SM} &   \text{Matching} \\
\midrule\midrule
\O_{\nu \gamma} & \frac12 n_\nu (n_\nu-1) & 3  & 0 \\
\bottomrule
\end{array}
\end{align*}
\setlength{\abovecaptionskip}{0cm}
\caption{Dimension-five $\Delta L=2$ Majorana neutrino dipole operators in LEFT.  There are also Hermitian conjugate $\Delta L=-2$ operators $\O_{\nu \gamma}^\dagger$, as indicated in the table heading.
The second column is the number of operators for an arbitrary number of neutrino flavors $n_\nu$, and the third column is number in the SM LEFT with $n_\nu=3$.  The last column is the tree-level matching coefficient in SMEFT, which vanishes. }
\label{dim5n}
\end{table}

\begin{table}[h]
\renewcommand{\arraystretch}{1.2}
\small
\begin{align*}
\begin{array}[t]{c|c|c|c}
\multicolumn{4}{c}{\boldsymbol{(\bar L R ) X+{\rm h.c.}}} \\
\toprule
 & \text{Number} & \text{SM} &   \text{Matching} \\
\midrule\midrule
\multicolumn{4}{c}{ \text{Leptonic} } \\
\hline
\O_{e \gamma} & n_e^2 & 9  & \frac{1}{\sqrt 2}\left( -C_{\substack{eW \\ pr}} \bar s + C_{\substack{eB \\ pr}}  \bar c  \right) v_T \\
\midrule
\multicolumn{4}{c}{ \text{Nonleptonic} } \\
\hline
\O_{u \gamma}   & n_u^2 & 4  &  \frac{1}{\sqrt 2} \left( C_{\substack{uW \\ pr}} \bar s + C_{\substack{uB \\ pr}} \bar c \right) v_T \\
\O_{d \gamma}   & n_d^2 & 9  &  \frac{1}{\sqrt 2}  \left( -C_{\substack{dW \\ pr}} \bar s + C_{\substack{dB \\ pr}} \bar c \right) v_T \\
\O_{u G}  & n_u^2 & 4  &  \frac{1}{\sqrt 2} C_{\substack{uG \\ pr}} v_T \\
\O_{d G}  & n_d^2 & 9  & \frac{1}{\sqrt 2} C_{\substack{dG \\ pr}} v_T \\
\hline
\text{Total}  & 2 n_u^2+ 2n_d^2 & 26 & \\
\bottomrule
%
\end{array}
\end{align*}
\setlength{\abovecaptionskip}{0cm}
\caption{Dimension-five $(\bar L R)X$ dipole operators in LEFT.  There are also Hermitian conjugate dipole operators $(\bar R L)X$, as indicated in the table heading.  The operators are divided into the leptonic and nonleptonic operators.  The second column is the number of operators for an arbitrary number of charged lepton flavors $n_e$, $u$-type quark flavors $n_u$, and $d$-type quark flavors $n_d$, and the third column is the number in the SM LEFT with $n_e=3$, $n_u=2$ and $n_d=3$.  The last column is the tree-level matching coefficient in SMEFT. $\bar s$ and $\bar c$ are defined in Eq.~(\ref{2.23}).}
\label{dim5mag}
\end{table}

\begin{table}[h]
\renewcommand{\arraystretch}{1.2}
\small
\begin{align*}
\begin{array}[t]{c|c|c|c}
\multicolumn{4}{c}{\boldsymbol{X^3}} \\
\toprule
 & \text{Number} & \text{SM} &   \text{Matching} \\
\midrule\midrule
\O_G      & 1 & 1 & C_G \\
\O_{\widetilde G}  & 1 & 1  & C_{\widetilde G} \\
\hline
\text{Total}  & 2 & 2 \\
\bottomrule
\end{array}
\end{align*}
\setlength{\abovecaptionskip}{0cm}
\caption{Dimension-six triple-gauge-boson operators in LEFT.  The tree-level matching coefficient of each operator is equal to the coefficient of the corresponding operator in SMEFT. }
\label{dim6X3}
\end{table}

\begin{table}[h]
\renewcommand{\arraystretch}{1.2}
\small
\begin{align*}
\begin{adjustbox}{center}
\begin{array}[t]{c|c|c|c}
\multicolumn{4}{c}{\boldsymbol{(\bar L  L)(\bar L  L)}} \\
\toprule
& \text{Number} & \text{SM} &   \text{Matching} \\
\midrule\midrule
\multicolumn{4}{c}{ \text{Leptonic} } \\
\hline
\op{\nu\nu}{V}{LL}  & \frac14 n_\nu^2 (n_\nu+1)^2 & 36   & C_{\substack{ ll \\ prst}} -\frac{\gcZ^2}{4 M_Z^2} \left[Z_\nu \right]_{pr} \left[Z_\nu \right]_{st} -\frac{\gcZ^2}{4 M_Z^2} \left[Z_\nu \right]_{pt} \left[Z_\nu \right]_{sr}  \\
\op{ee}{V}{LL}      & \frac14 n_e^2 (n_e+1)^2 & 36   & C_{\substack{ ll \\ prst}} -\frac{\gcZ^2}{4 M_Z^2}   \left[Z_{e_L} \right]_{pr} \left[Z_{e_L} \right]_{st} -\frac{\gcZ^2}{4 M_Z^2}   \left[Z_{e_L} \right]_{pt} \left[Z_{e_L} \right]_{sr} \\
\op{\nu e}{V}{LL}    & n_e^2 n_\nu^2 & 81   &  C_{\substack{ ll \\ prst}} + C_{\substack{ ll \\ stpr}} -\frac{\gcw^2}{2 M_W^2} 
\left[W_l\right]_{pt} \left[W_l\right]_{rs}^*  -\frac{\gcZ^2}{M_Z^2} 
\left[Z_\nu \right]_{pr} \left[Z_{e_L} \right]_{st} \\
\hline
\text{Total}  & n_e^2 n_\nu^2 + \frac{1}{4} n_e^2(n_e+1)^2 & \\
	 & {} + \frac{1}{4} n_\nu^2(n_\nu+1)^2  & 153 & \\
\midrule
\multicolumn{4}{c}{ \text{Semileptonic} } \\
\hline
\op{\nu u}{V}{LL}       & n_\nu^2 n_u^2  & 36   &  C^{(1)}_{\substack{ lq \\ prst}} + C^{(3)}_{\substack{ lq \\ prst}} 
-\frac{\gcZ^2}{ M_Z^2}   \left[Z_\nu \right]_{pr} \left[Z_{u_L} \right]_{st} \\
\op{\nu d}{V}{LL}          & n_\nu^2 n_d^2  & 81    & C^{(1)}_{\substack{ lq \\ prst}} -  C^{(3)}_{\substack{ lq \\ prst}}
-\frac{\gcZ^2}{ M_Z^2} 
 \left[Z_\nu \right]_{pr} \left[Z_{d_L} \right]_{st}    \\
\op{eu}{V}{LL}     & n_e^2 n_u^2  & 36    & C^{(1)}_{\substack{ lq \\ prst}} -  C^{(3)}_{\substack{ lq \\ prst}}
-\frac{\gcZ^2}{ M_Z^2}   \left[Z_{e_L} \right]_{pr} \left[Z_{u_L} \right]_{st}  \\
\op{ed}{V}{LL}      & n_e^2 n_d^2  & 81   & C^{(1)}_{\substack{ lq \\ prst}} + C^{(3)}_{\substack{ lq \\ prst}} 
-\frac{\gcZ^2}{ M_Z^2}   \left[Z_{e_L} \right]_{pr} \left[Z_{d_L} \right]_{st}  \\
\op{\nu e d u}{V}{LL}   + {\rm h.c.}  & 2 \times n_e n_\nu n_u n_d   & 2\times 54    &  2 C^{(3)}_{\substack{ lq \\ prst}} -\frac{\gcw^2}{2 M_W^2} 
\left[W_l\right]_{pr} \left[W_q\right]_{ts}^* \\
\hline
\text{Total}  & (n_e^2 + n_\nu^2)(n_u^2+n_d^2) & & \\
	 & {}+ 2 n_e n_\nu n_u n_d & 342 & \\
\midrule
\multicolumn{4}{c}{ \text{Nonleptonic} } \\
\hline
\op{uu}{V}{LL}       & \frac12 n_u^2 (n_u^2+1)   & 10   &  C^{(1)}_{\substack{ qq \\ prst}} + C^{(3)}_{\substack{ qq \\ prst}}  -\frac{\gcZ^2}{2 M_Z^2}   \left[Z_{u_L} \right]_{pr} \left[Z_{u_L} \right]_{st} \\
\op{dd}{V}{LL}     & \frac12 n_d^2 (n_d^2+1)  & 45   & C^{(1)}_{\substack{ qq \\ prst}} + C^{(3)}_{\substack{ qq \\ prst}}  -\frac{\gcZ^2}{2 M_Z^2}   \left[Z_{d_L} \right]_{pr} \left[Z_{d_L} \right]_{st} \\
\op{ud}{V1}{LL}    & n_u^2 n_d^2  & 36     &  C^{(1)}_{\substack{ qq \\ prst}} + C^{(1)}_{\substack{ qq \\ stpr}} - C^{(3)}_{\substack{ qq \\ prst}} - C^{(3)}_{\substack{ qq \\ stpr}}  +\frac{2}{N_c} C^{(3)}_{\substack{ qq \\ ptsr}} +\frac{2}{N_c} C^{(3)}_{\substack{ qq \\ srpt}} \\
& &   & -\frac{\gcw^2}{2 M_W^2} 
\left[W_q\right]_{pt} \left[W_q\right]_{rs}^* \frac{1}{N_c}   -\frac{\gcZ^2}{ M_Z^2}    \left[Z_{u_L} \right]_{pr} \left[Z_{d_L} \right]_{st} \\
\op{ud}{V8}{LL}      & n_u^2 n_d^2  & 36   &  4 C^{(3)}_{\substack{ qq \\ ptsr}} +4 C^{(3)}_{\substack{ qq \\ srpt}} -\frac{\gcw^2}{M_W^2} 
\left[W_q\right]_{pt} \left[W_q\right]_{rs}^*   \\
\hline 
\text{Total} &  2 n_u^2 n_d^2 + \frac12 n_u^2 (n_u^2+1)  & &  \\
	 &  {}+\frac12 n_d^2 (n_d^2+1)   & 127 &  \\
\bottomrule
\end{array}
\end{adjustbox}
\end{align*}
\caption{Dimension-six four-fermion operators: two left-handed currents in LEFT.  The $(\bar L  L)(\bar L  L)$ operators are divided into leptonic, semileptonic, and nonleptonic operators.  The semileptonic operator 
$\op{\nu e d u}{V}{LL}$ and its Hermitian conjugate ${\op{\nu e d u}{V}{LL}}^\dagger$ are both present.  All other operators are Hermitian.
The second column is the number of operators for an arbitrary number of neutrino flavors $n_\nu$, charged lepton flavors $n_e$, $u$-type quark flavors $n_u$, and $d$-type quark flavors $n_d$, and the third column is the number in the SM LEFT with $n_\nu=3$, $n_e=3$, $n_u=2$, and $n_d=3$.  The last column is the tree-level matching coefficient in SMEFT.}
\label{dim6ll}
\end{table}

\begin{table}[h]
\renewcommand{\arraystretch}{1.2}
\small
\begin{align*}
\begin{adjustbox}{center}
\begin{array}[t]{c|c|c|c}
\multicolumn{4}{c}{\boldsymbol{(\bar R  R)(\bar R  R)}} \\
\toprule
 & \text{Number} & \text{SM} &   \text{Matching} \\
\midrule\midrule
\multicolumn{4}{c}{ \text{Leptonic} } \\
\hline
\op{ee}{V}{RR}  & \frac14 n_e^2 (n_e+1)^2 & 36   &  C_{\substack{ ee \\ prst}} -\frac{\gcZ^2}{4 M_Z^2}   \left[Z_{e_R} \right]_{pr} \left[Z_{e_R} \right]_{st} -\frac{\gcZ^2}{4 M_Z^2}   \left[Z_{e_R} \right]_{pt} \left[Z_{e_R} \right]_{sr}  \\
\midrule
%
%
\multicolumn{4}{c}{ \text{Semileptonic} } \\
\hline
\op{eu}{V}{RR}   & n_e^2 n_u^2 & 36  & C_{\substack{ eu \\ prst}} -\frac{\gcZ^2}{ M_Z^2}   \left[Z_{e_R} \right]_{pr} \left[Z_{u_R} \right]_{st} \\
\op{ed}{V}{RR}   & n_e^2 n_d^2 & 81   &   C_{\substack{ ed \\ prst}} -\frac{\gcZ^2}{M_Z^2}   \left[Z_{e_R} \right]_{pr} \left[Z_{d_R} \right]_{st} \\
\hline
\text{Total}  & n_e^2 (n_u^2+n_d^2)  & 117 & \\
\midrule
\multicolumn{4}{c}{ \text{Nonleptonic} } \\
\hline
\op{uu}{V}{RR}   & \frac12 n_u^2 (n_u^2+1)   & 10   &  C_{\substack{ uu \\ prst}}
-\frac{\gcZ^2}{2 M_Z^2}   \left[Z_{u_R} \right]_{pr} \left[Z_{u_R} \right]_{st}
 \\
\op{dd}{V}{RR}   & \frac12 n_d^2 (n_d^2+1)  & 45   &  C_{\substack{ dd \\ prst}}
-\frac{\gcZ^2}{2 M_Z^2}   \left[Z_{d_R} \right]_{pr} \left[Z_{d_R} \right]_{st} \\
\op{ud}{V1}{RR}    & n_u^2 n_d^2  & 36 &  C^{(1)}_{\substack{ ud \\ prst}}-\frac{\gcw^2}{2 M_W^2} 
\left[W_R\right]_{pt} \left[W_R\right]_{rs}^* \frac{1}{N_c} -\frac{\gcZ^2}{ M_Z^2}   \left[Z_{u_R} \right]_{pr} \left[Z_{d_R} \right]_{st}  \\
\op{ud}{V8}{RR}    & n_u^2 n_d^2  & 36     & C^{(8)}_{\substack{ ud \\ prst}} -\frac{\gcw^2}{M_W^2} 
\left[W_R\right]_{pt} \left[W_R\right]_{rs}^*  \\
\hline
\text{Total} &  2 n_u^2 n_d^2 + \frac12 n_u^2 (n_u^2+1)    & & \\
	& {}  +\frac12 n_d^2 (n_d^2+1) & 127 & \\
\bottomrule
\end{array}
\end{adjustbox}
\end{align*}
\caption{Dimension-six four-fermion operators: two right-handed currents in LEFT.  The $(\bar R  R)(\bar R  R)$ operators are divided into leptonic, semileptonic, and nonleptonic operators.  The second column is the number of operators for an arbitrary number of charged lepton flavors $n_e$, 
$u$-type quark flavors $n_u$, and $d$-type quark flavors $n_d$, and the third column is the number in the SM LEFT with $n_e=3$, $n_u=2$, and $n_d=3$.  The last column is the tree-level matching coefficient in SMEFT.}
\label{dim6rr}
\end{table}
%
\begin{table}[h]
\renewcommand{\arraystretch}{1.2}
\small
\begin{align*}
\begin{adjustbox}{center}
\begin{array}[t]{c|c|c|c}
\multicolumn{4}{c}{} \\[-1cm]
\multicolumn{4}{c}{\boldsymbol{(\bar L L)(\bar R  R)}} \\
\toprule
& \text{Number} & \text{SM} &   \text{Matching} \\
\midrule\midrule
\multicolumn{4}{c}{ \text{Leptonic} } \\
\hline
\op{\nu e}{V}{LR}   & n_e^2 n_\nu^2 & 81   & C_{\substack{ le \\ prst}} 
-\frac{\gcZ^2}{ M_Z^2}  \left[Z_\nu \right]_{pr} \left[Z_{e_R} \right]_{st} 
 \\
\op{ee}{V}{LR}  & n_e^4 & 81   & C_{\substack{ le \\ prst}}
-\frac{\gcZ^2}{ M_Z^2}   \left[Z_{e_L} \right]_{pr} \left[Z_{e_R} \right]_{st} 
\\
\hline
\text{Total}  & n_e^2( n_e^2 + n_\nu^2)  & 162 & \\
\midrule
\multicolumn{4}{c}{ \text{Semileptonic} } \\ 
\hline
\op{\nu u}{V}{LR}  &  n_\nu^2 n_u^2 & 36  &  C_{\substack{ lu \\ prst}}  
-\frac{\gcZ^2}{ M_Z^2}   \left[Z_\nu \right]_{pr} \left[Z_{u_R} \right]_{st}  \\
\op{\nu d}{V}{LR}    & n_\nu^2 n_d^2 &  81  & C_{\substack{ ld \\ prst}}  
-\frac{\gcZ^2}{ M_Z^2}   \left[Z_\nu \right]_{pr} \left[Z_{d_R} \right]_{st}  \\
\op{eu}{V}{LR}   & n_e^2 n_u^2 & 36   & C_{\substack{ lu \\ prst}} 
-\frac{\gcZ^2}{M_Z^2}   \left[Z_{e_L} \right]_{pr} \left[Z_{u_R} \right]_{st} \\
\op{ed}{V}{LR}   & n_e^2 n_d^2 & 81   & C_{\substack{ ld \\ prst}} 
-\frac{\gcZ^2}{M_Z^2}   \left[Z_{e_L} \right]_{pr} \left[Z_{d_R} \right]_{st} \\
\op{ue}{V}{LR}   & n_e^2 n_u^2 & 36  & C_{\substack{ qe \\ prst}}
 -\frac{\gcZ^2}{M_Z^2}   \left[Z_{u_L} \right]_{pr} \left[Z_{e_R} \right]_{st}  \\
\op{de}{V}{LR}     & n_e^2 n_d^2 & 81  &  C_{\substack{ qe \\ prst}} 
-\frac{\gcZ^2}{ M_Z^2}    \left[Z_{d_L} \right]_{pr} \left[Z_{e_R} \right]_{st}  \\
\op{\nu edu}{V}{LR}  +{ \rm h.c.}  & 2\times n_e n_\nu n_u n_d &  2\times 54   & -\frac{\gcw^2}{2 M_W^2} 
\left[W_l\right]_{pr} \left[W_R\right]_{ts}^* \\
\hline
\text{Total} &  (2n_e^2 +n_\nu^2)(n_u^2 + n_d^2)  & & \\
	& {}+ 2 n_e n_\nu n_u n_d & 459 & \\
\midrule
\multicolumn{4}{c}{ \text{Nonleptonic} } \\
\hline
\op{uu}{V1}{LR}     & n_u^4 & 16   &  C^{(1)}_{\substack{ qu \\ prst}} 
 -\frac{\gcZ^2}{ M_Z^2}   \left[Z_{u_L} \right]_{pr} \left[Z_{u_R} \right]_{st}  \nn
\op{uu}{V8}{LR}   & n_u^4 & 16  &  C^{(8)}_{\substack{ qu \\ prst}}  \\
\op{ud}{V1}{LR}    & n_u^2 n_d^2  &  36   & C^{(1)}_{\substack{ qd \\ prst}} 
 -\frac{\gcZ^2}{ M_Z^2}   \left[Z_{u_L} \right]_{pr} \left[Z_{d_R} \right]_{st} \nn
\op{ud}{V8}{LR}   & n_u^2 n_d^2  &  36   & C^{(8)}_{\substack{ qd \\ prst}} \\
\op{du}{V1}{LR}     & n_u^2 n_d^2 & 36   & C^{(1)}_{\substack{ qu \\ prst}} 
-\frac{\gcZ^2}{M_Z^2}  \left[Z_{d_L} \right]_{pr} \left[Z_{u_R} \right]_{st}  \nn
\op{du}{V8}{LR}     & n_u^2 n_d^2  &  36   &  C^{(8)}_{\substack{ qu \\ prst}}
\\
\op{dd}{V1}{LR}   & n_d^4 & 81    & C^{(1)}_{\substack{ qd \\ prst}} 
-\frac{\gcZ^2}{ M_Z^2}   \left[Z_{d_L} \right]_{pr} \left[Z_{d_R} \right]_{st}  \nn
\op{dd}{V8}{LR}   & n_d^4 & 81    & C^{(8)}_{\substack{ qd \\ prst}} 
\\
\op{uddu}{V1}{LR}  + {\rm h.c.}  & 2\times n_u^2 n_d^2 & 2\times 36    &  -\frac{\gcw^2}{2 M_W^2} 
\left[W_q\right]_{pr} \left[W_R\right]_{ts}^* \nn
\op{uddu}{V8}{LR} + {\rm h.c.}  & 2\times n_u^2 n_d^2 & 2\times 36   & 0
\\
\hline
\text{Total}  & 2(n_u^4+n_d^4+4n_u^2n_d^2)  & 482 &  \\
\bottomrule
\end{array}
\end{adjustbox}
\end{align*}
\caption{Dimension-six four-fermion operators: left-handed times right-handed currents in LEFT.  The $(\bar L  L)(\bar R  R)$ operators are divided into leptonic, semileptonic, and nonleptonic operators.  Semileptonic operators $\op{\nu edu}{V}{LR}$ and nonleptonic operators $\op{uddu}{V1}{LR}$
and $\op{uddu}{V8}{LR}$ all come with additional Hermitian conjugate operators.  All other operators are Hermitian.  The second column is the number of operators for an arbitrary number of neutrino flavors $n_\nu$, charged lepton flavors $n_e$, $u$-type quark flavors $n_u$, and $d$-type quark flavors $n_d$, and the third column is the number in the SM LEFT with $n_\nu=3$, $n_e=3$, $n_u=2$, and $n_d=3$.  The last column is the tree-level matching coefficient in SMEFT.}
\label{dim6lr}
\end{table}
%
\begin{table}[h]
\renewcommand{\arraystretch}{1.2}
\small
\begin{align*}
\begin{array}[t]{c|c|c|c}
\multicolumn{4}{c}{\boldsymbol{(\bar L R)(\bar R L) +{\rm h.c.}}} \\
\toprule
& \text{Number} & \text{SM} &   \text{Matching} \\
\midrule\midrule
\multicolumn{4}{c}{ \text{Semileptonic} } \\
\hline
\op{eu}{S}{RL}   & n_e^2 n_u^2  & 36 & 0
 \\
\op{ed}{S}{RL}  & n_e^2 n_d^2  & 81 & C_{\substack{ ledq \\ prst}} 
\\
\op{\nu edu}{S}{RL}  & n_e n_\nu n_u n_d  & 54 &  C_{\substack{ ledq \\ prst}} 
\\
\hline
\text{Total}  & n_e^2 (n_u^2+n_d^2) + n_e n_\nu n_u n_d & 171 \\
\bottomrule
\end{array}
\end{align*}
\caption{Dimension-six four-fermion operators: $(\bar L R)(\bar R L)$ scalar bilinears in LEFT.  There are also Hermitian conjugate operators, as indicated in the table heading.  All of the operators are semileptonic operators.  The second column is the number of operators for an arbitrary number of neutrino flavors $n_\nu$, charged lepton flavors $n_e$, $u$-type quark flavors $n_u$, and $d$-type quark flavors $n_d$, and the third column is the number in the SM LEFT with $n_\nu=3$, $n_e=3$, $n_u=2$, and $n_d=3$.  The last column is the tree-level matching coefficient in SMEFT.}
\label{dim6lrrl}
\end{table}
%
\begin{table}[h]
\renewcommand{\arraystretch}{1.2}
\small
\begin{align*}
\begin{adjustbox}{center}
\begin{array}[t]{c|c|c|c}
\multicolumn{4}{c}{\boldsymbol{(\bar L R)(\bar L R)+{\rm h.c.}}} \\
\toprule
 & \text{Number} & \text{SM} &   \text{Matching} \\
\midrule\midrule
\multicolumn{4}{c}{ \text{Leptonic} } \\
\hline
\op{ee}{S}{RR}   & \frac12 n_e^2(n_e^2+1)  & 45 & 0 \nn
\midrule
\multicolumn{4}{c}{ \text{Semileptonic} } \\
\hline
\op{eu}{S}{RR}  & n_e^2 n_u^2 & 36 & -C^{(1)}_{\substack{ lequ \\ prst}}
 \\
\op{eu}{T}{RR} & n_e^2 n_u^2 & 36 & -C^{(3)}_{\substack{ lequ \\ prst}}  \\
\op{ed}{S}{RR} & n_e^2 n_d^2  & 81 &  0
 \\
\op{ed}{T}{RR} & n_e^2 n_d^2  & 81&  0 \\
\op{\nu edu}{S}{RR} & n_e n_\nu n_u n_d  & 54 &  C^{(1)}_{\substack{ lequ \\ prst}} 
\\
\op{\nu edu}{T}{RR}  & n_e n_\nu n_u n_d  & 54 &  C^{(3)}_{\substack{ lequ \\ prst}}  \\
\hline
\text{Total} &  2 n_e^2(n_u^2+n_d^2) + 2 n_e n_\nu n_u n_d & 342  \\
\midrule
\multicolumn{4}{c}{ \text{Nonleptonic} } \\
\hline
\op{uu}{S1}{RR}  & \frac12 n_u^2(n_u^2+1)  & 10 & 0 \nn
\op{uu}{S8}{RR}   & \frac12 n_u^2(n_u^2+1)  & 10 & 0 \\
\op{ud}{S1}{RR} &  n_u^2 n_d^2  & 36 &   C^{(1)}_{\substack{ quqd \\ prst}} \\
\op{ud}{S8}{RR}  &  n_u^2 n_d^2  & 36 &  C^{(8)}_{\substack{ quqd \\ prst}}  \\
\op{dd}{S1}{RR}  & \frac12 n_d^2(n_d^2+1)  & 45 &   0 \\
\op{dd}{S8}{RR}  & \frac12 n_d^2(n_d^2+1)  & 45 & 0 \\
\op{uddu}{S1}{RR} & n_u^2 n_d^2  & 36 &   -C^{(1)}_{\substack{ quqd \\ stpr}}
 \\
\op{uddu}{S8}{RR}   & n_u^2 n_d^2  & 36 &  -C^{(8)}_{\substack{ quqd \\ stpr}} \\
\hline
\text{Total}  & 4 n_u^2 n_d^2 + n_u^2(n_u^2+1) & & \\
	 & {}+n_d^2(n_d^2+1) & 254 & \\
\bottomrule
\end{array}
\end{adjustbox}
\end{align*}
\caption{Dimension-six four-fermion operators: $(\bar L R)(\bar L R)$ scalar and tensor bilinears in LEFT.  There are also Hermitian conjugate operators, as indicated in the table heading.  The operators are divided into leptonic, semileptonic, and nonleptonic operators.  The second column is the number of operators for an arbitrary number of neutrino flavors $n_\nu$, charged lepton flavors $n_e$, $u$-type quark flavors $n_u$, and $d$-type quark flavors $n_d$, and the third column is the number in the SM LEFT with $n_\nu=3$, $n_e=3$, $n_u=2$, and $n_d=3$.  The last column is the tree-level matching coefficient in SMEFT.}
\label{dim6lrlr}
\end{table}

\begin{table}[h]
\renewcommand{\arraystretch}{1.2}
\small
\begin{align*}
\begin{array}[t]{c|c|c|c}
\multicolumn{4}{c}{\boldsymbol{\Delta L = 4 + {\rm h.c.}}} \\
\toprule
 & \text{Number} &  \text{SM} &   \text{Matching} \\
\midrule\midrule
\op{\nu\nu}{S}{LL} & \frac{1}{12} n_\nu^2(n_\nu^2-1) & 6 & 0 
\\
\bottomrule
\end{array}
\end{align*}
\caption{Dimension-six $\Delta L=4$ operators in LEFT.  There are also Hermitian conjugate operators, as indicated in the table heading.
The second column is the number of operators for an arbitrary number of neutrino flavors $n_\nu$, and the third column is the number in the SM LEFT with $n_\nu=3$.  The last column is the tree-level matching coefficient in SMEFT.}
\label{dim6l4}
\end{table}

%
\begin{table}[h]
\renewcommand{\arraystretch}{1.2}
\small
\begin{align*}
\begin{array}[t]{c|c|c|c}
\multicolumn{4}{c}{\boldsymbol{\Delta L =2 + {\rm h.c.}}} \\
\toprule
& \text{Number} & \text{SM} &   \text{Matching} \\
\midrule\midrule
\multicolumn{4}{c}{ \text{Leptonic} } \\
\hline
\op{\nu e}{S}{LL}  & \frac12 n_\nu(n_\nu+1) n_e^2 & 54 & 
0 \\
\op{\nu e}{T}{LL}  & \frac12 n_\nu(n_\nu-1) n_e^2  & 27 &  0 \\
\op{\nu e}{S}{LR}  & \frac12 n_\nu(n_\nu+1) n_e^2 & 54 & 
0 \\
\hline
\text{Total} &  \frac{1}{2} n_\nu( 3n_\nu + 1 ) n_e^2 & 135 & \\
\midrule
\multicolumn{4}{c}{ \text{Semileptonic} } \\
\hline
\op{\nu u}{S}{LL}  & \frac12 n_\nu(n_\nu+1) n_u^2 & 24 &  
0 \\
\op{\nu u}{T}{LL}  &  \frac12 n_\nu(n_\nu-1) n_u^2  & 12 &  0 \\
\op{\nu u}{S}{LR}  & \frac12 n_\nu(n_\nu+1) n_u^2  & 24 &  
0 \\
\op{\nu d}{S}{LL}  & \frac12 n_\nu(n_\nu+1) n_d^2 & 54 & 
0 \\
\op{\nu d}{T}{LL}  & \frac12 n_\nu(n_\nu-1) n_d^2   & 27 & 0 \\
\op{\nu d}{S}{LR} & \frac12 n_\nu(n_\nu+1) n_d^2  & 54 &  
0 \\
\op{\nu edu}{S}{LL} & n_e n_\nu n_u n_d & 54 & 0 \\
\op{\nu edu}{T}{LL} & n_e n_\nu n_u n_d & 54 & 0 \\
\op{\nu edu}{S}{LR} & n_e n_\nu n_u n_d & 54 &  0\\
\op{\nu edu}{V}{RL}   & n_e n_\nu n_u n_d & 54 &  0 \\
\op{\nu edu}{V}{RR}  & n_e n_\nu n_u n_d & 54 & 0 \\
\hline
\text{Total}  & \frac12 n_\nu (3n_\nu+1)(n_u^2+n_d^2) + 5 n_e n_\nu n_u n_d & 465 \\
\bottomrule
\end{array}
\end{align*}
\caption{Dimension-six $\Delta L=2$ operators in LEFT.  There are also Hermitian conjugate operators, as indicated in the table heading.  The operators are divided into leptonic and semileptonic operators.  The second column is the number of operators for an arbitrary number of neutrino flavors $n_\nu$, charged lepton flavors $n_e$, $u$-type quark flavors $n_u$, and $d$-type quark flavors $n_d$, and the third column is the number in the SM LEFT with $n_\nu=3$, $n_e=3$, $n_u=2$, and $n_d=3$.  The last column is the tree-level matching coefficient in SMEFT.}
\label{dim6l2}
\end{table}
%

%
\begin{table}[h]
\renewcommand{\arraystretch}{1.2}
\small
\begin{align*}
\begin{adjustbox}{center}
\begin{array}[t]{c|c|c|c}
\multicolumn{4}{c}{\boldsymbol{\Delta B = \Delta L = 1 + {\rm h.c.}}} \\
\toprule
& \text{Number} &  \text{SM} &   \text{Matching} \\
\midrule\midrule
\op{udd}{S}{LL} & n_\nu n_u n_d^2 &  54 & { C_{\substack{ qqql \\ srpt}} - C_{\substack{ qqql \\ rspt}} + C_{\substack{ qqql \\ rpst}} } \\
\op{duu}{S}{LL}  & n_e n_d n_u^2  &  36 &  { C_{\substack{ qqql \\ srpt}} - C_{\substack{ qqql \\ rspt}} + C_{\substack{ qqql \\ rpst}} }  \\
%
%
\op{uud}{S}{LR}  & \frac12 n_d n_u (n_u-1) n_e & 9 &  0 \\
\op{duu}{S}{LR}  &  n_e n_u^2 n_d &  36 &  -C_{\substack{ qque \\ prst}} - C_{\substack{ qque \\ rpst}} \\
%
%
%
\op{uud}{S}{RL} & \frac12 n_d n_u (n_u-1) n_e & 9 &  0 \\
\op{duu}{S}{RL}  & n_e n_u^2 n_d & 36 &  C_{\substack{ duql \\ prst}} \\
\op{dud}{S}{RL} & n_\nu n_u n_d^2  & 54 &  -C_{\substack{ duql \\ prst}} \\
\op{ddu}{S}{RL}  & \frac12 n_d (n_d-1)n_u n_\nu  & 18  & 0 \\
%
\op{duu}{S}{RR}  &  n_e n_d n_u^2   &  36 &  C_{\substack{ duue \\ prst}} \\
\hline
\text{Total} &  \frac52 n_d^2 n_\nu n_u+ 5 n_d n_e n_u^2 - n_d n_e n_u - \frac12 n_d n_\nu n_u 
 &   288 \\
\bottomrule
\end{array}
\end{adjustbox}
\end{align*}
\caption{Dimension-six $\Delta B = \Delta L=1$ operators in LEFT.  There are also Hermitian conjugate $\Delta B = \Delta L=-1$ operators, as indicated in the table heading.  The second column is the number of operators for an arbitrary number of neutrino flavors $n_\nu$, charged lepton flavors $n_e$, $u$-type quark flavors $n_u$, and $d$-type quark flavors $n_d$, and the third column is the number in the SM LEFT with $n_\nu=3$, $n_e=3$, $n_u=2$, and $n_d=3$.  The last column is the tree-level matching coefficient in SMEFT.}
\label{dim6bl}
\end{table}

\begin{table}[h]
\renewcommand{\arraystretch}{1.2}
\small
\begin{align*}
\begin{adjustbox}{center}
\begin{array}[t]{c|c|c|c}
\multicolumn{4}{c}{\boldsymbol{\Delta B = - \Delta L = 1 + {\rm h.c.}}} \\
\toprule
 & \text{Number} & \text{SM} &   \text{Matching} \\
\midrule\midrule
\op{ddd}{S}{LL} & \frac13 n_d (n_d^2-1) n_e & 24 & 0 \\
%
%
\op{udd}{S}{LR}  & n_\nu n_u n_d^2  & 54 &  0 \\
\op{ddu}{S}{LR} & \frac12 n_d (n_d-1)n_u n_\nu  & 18 & 0 \\
\op{ddd}{S}{LR} & \frac12 n_d^2 (n_d-1) n_e & 27 &  0 \\
%
%
%
\op{ddd}{S}{RL} & \frac12 n_d^2 (n_d-1) n_e & 27 & 0 \\
%
%
%
\op{udd}{S}{RR}  & n_\nu n_u n_d^2 & 54 & 0 \\
\op{ddd}{S}{RR} & \frac13 n_d (n_d^2-1) n_e & 24 &  0 \\
\hline
\text{Total} &  \frac53 n_d^3 n_e +\frac52  n_\nu n_d^2 n_u - n_d^2 n_e -\frac12 n_d n_\nu n_u - \frac 23 n_d n_e    & 228 \\
\bottomrule
\end{array}
\end{adjustbox}
\end{align*}
\caption{Dimension-six $\Delta B = -\Delta L=1$ operators in LEFT.  There are also Hermitian conjugate $\Delta B = -\Delta L=-1$ operators, as indicated in the table heading.  The second column is the number of operators for an arbitrary number of neutrino flavors $n_\nu$, charged lepton flavors $n_e$, $u$-type quark flavors $n_u$, and $d$-type quark flavors $n_d$, and the third column is the number in the SM LEFT with $n_\nu=3$, $n_e=3$, $n_u=2$, and $n_d=3$.  The last column is the tree-level matching coefficient in SMEFT.}
\label{dim6bml}
\end{table}

%
%
\begin{table}[h]
\renewcommand{\arraystretch}{1}
\small
\begin{align*}
\\[-2cm]
\begin{adjustbox}{width=1.15\textwidth,center}
\begin{array}[t]{l|c|c|c|c}
\toprule
\text{Operator type} & \multicolumn{2}{c|}{\text{$CP$-even}} &  \multicolumn{2}{c}{\text{$CP$-odd}}  \\
& & \text{SM} & & \text{SM} \\
\midrule\midrule
\Delta L=2 + {\rm h.c.} &							\frac12 n_\nu (n_\nu+1) &					6 &			\frac12 n_\nu (n_\nu+1) &				6  \\
\midrule
(\bar L R ) X+ {\rm h.c.} : \text{leptonic}  &		n_e^2 &								9 &			n_e^2 &							9 \\[5pt]
(\bar L R ) X+ {\rm h.c.} : \text{nonleptonic}  &	2( n_u^2 + n_d^2) &						26 &			2(n_u^2+n_d^2) &					26 \\[5pt]
\Delta L = 2 \ \ (\nu \nu) X+ {\rm h.c.} &						\frac12 n_\nu (n_\nu-1) &					3 &			\frac12 n_\nu (n_\nu-1) &				3 \\
\midrule
X^3 & 1 & 1 & 1 & 1 \\[5pt]
(\bar L L)(\bar L L): \text{leptonic}		& 	\frac12 ( n_e^2 n_\nu^2 + n_e n_\nu) + \frac18 n_e (n_e^3+2 n_e^2+3 n_e+2)	&		&		\frac12 ( n_e^2 n_\nu^2 - n_e n_\nu) + \frac18 n_e(n_e^2-1)(n_e+2)	&		\\
										&	{} + \frac18 n_\nu (n_\nu^3+2 n_\nu^2+3 n_\nu+2)						&	87	&		{} + \frac18 n_\nu(n_\nu^2-1)(n_\nu+2) 						&	66	 \\[5pt]
(\bar L L)(\bar L L): \text{semileptonic}	&	\frac12 (n_u^2 + n_d^2) (n_e^2 + n_\nu^2) 							&		&		\frac12 (n_u^2 + n_d^2) (n_e^2 + n_\nu^2) 					&		\\
										&	{}+ \frac12(n_u + n_d) (n_e + n_\nu) + n_e n_\nu n_u n_d					&	186	&		{}- \frac12(n_u + n_d) (n_e + n_\nu) + n_e n_\nu n_u n_d			&	156	\\[5pt]
(\bar L L)(\bar L L): \text{nonleptonic}		&	\frac14( n_u^4+4n_u^2 n_d^2  + n_d^4								&		&		\frac14( n_u^4+4n_u^2 n_d^2  + n_d^4						&		\\
										&	{}+3n_u^2+4 n_u n_d + 3n_d^2)									&	76	&		{}- n_u^2-4 n_u n_d - n_d^2)								&	51	\\[5pt]
(\bar R R)(\bar R R): \text{leptonic}		&	\frac18 n_e(n_e+1)(n_e^2+n_e+2)									&	21	&		\frac18 n_e(n_e^2-1)(n_e+2)								&	15	\\[5pt]
(\bar R R)(\bar R R): \text{semileptonic}	&	\frac12 n_e(n_e (n_u^2 + n_d^2) + n_u + n_d)							&	66	&		\frac12 n_e(n_e (n_u^2 + n_d^2) - n_u - n_d )					&	51	\\[5pt]
(\bar R R)(\bar R R): \text{nonleptonic}	&	\frac14 (n_u^4 +4 n_u^2 n_d^2 + n_d^4								&		&		\frac14 (n_u^4 +4 n_u^2 n_d^2 + n_d^4						&		\\
										&	{}+3 n_u^2 + 4n_u n_d  + 3 n_d^2 )									&	76	&		{}- n_u^2 -4n_u n_d -n_d^2 )								&	51	\\[5pt]
(\bar L L)(\bar R R): \text{leptonic}		&	\frac12 n_e(n_e^3+ n_e(n_\nu^2+1) + n_\nu)							&	90	&		\frac12 n_e(n_e^3+ n_e(n_\nu^2-1) - n_\nu)					&	72	\\[5pt]
(\bar L L)(\bar R R): \text{semileptonic}	&	\frac12(n_u^2+n_d^2)(2n_e^2+n_\nu^2)								&		&		\frac12(n_u^2+n_d^2)(2n_e^2+n_\nu^2)						&		\\ 
										&	{} + \frac12(n_u+n_d)(2n_e+n_\nu) + n_e n_\nu n_u n_d					&	252	&		{} - \frac12(n_u+n_d)(2n_e+n_\nu) + n_e n_\nu n_u n_d			&	207	\\[5pt]
(\bar L L)(\bar R R): \text{nonleptonic}	&	n_u^4+4 n_u^2 n_d^2 +n_d^4	+ n_u^2 + 2 n_u n_d + n_d^2				&	266	&		n_u^4+4 n_u^2 n_d^2 +n_d^4	- n_u^2 - 2 n_u n_d -  n_d^2		&	216	\\[5pt]
(\bar L R)(\bar R L) + {\rm h.c.}				&	n_e^2 (n_u^2+n_d^2) + n_e n_\nu n_u n_d							&	171	&		n_e^2 (n_u^2+n_d^2) + n_e n_\nu n_u n_d					&	171	\\[5pt]
(\bar L R)(\bar L R) + {\rm h.c.} : \text{leptonic}	&	\frac12 n_e^2(n_e^2+1) & 45 & \frac12 n_e^2(n_e^2+1) & 45 \\[5pt]
(\bar L R)(\bar L R) + {\rm h.c.} : \text{semileptonic} & 2 n_e^2(n_u^2+n_d^2) + 2 n_e n_\nu n_u n_d							&	342	&		2 n_e^2(n_u^2+n_d^2) + 2 n_e n_\nu n_u n_d					&	342	\\[5pt]
(\bar L R)(\bar L R) + {\rm h.c.} : \text{nonleptonic} &	n_u^2(n_u^2+1)+n_d^2(n_d^2+1) + 4 n_u^2 n_d^2					&	254	&		n_u^2(n_u^2+1)+n_d^2(n_d^2+1) + 4 n_u^2 n_d^2			&	254	\\[5pt]
\Delta L = 4 + {\rm h.c}							&	\frac{1}{12} n_\nu^2(n_\nu^2-1)									&	6	&		\frac{1}{12} n_\nu^2(n_\nu^2-1)							&	6	\\
\Delta L = 2 + {\rm h.c.} : \text{leptonic}				&	\frac{1}{2} n_\nu( 3n_\nu + 1 ) n_e^2									&	135	&		\frac{1}{2} n_\nu( 3n_\nu + 1 ) n_e^2							&	135	\\[5pt]
\Delta L = 2 + {\rm h.c.} : \text{semileptonic}			&	\frac12 n_\nu (3n_\nu+1)(n_u^2+n_d^2) + 5 n_e n_\nu n_u n_d			&	465	&		\frac12 n_\nu (3n_\nu+1)(n_u^2+n_d^2) + 5 n_e n_\nu n_u n_d	&	465	\\[5pt]
\Delta B = \Delta L = 1 + {\rm h.c.}					&	\frac52 n_d^2 n_\nu n_u+ 5 n_d n_e n_u^2 							&		&		\frac52 n_d^2 n_\nu n_u+ 5 n_d n_e n_u^2 					&		\\
										&	{} - n_d n_e n_u - \frac12 n_d n_\nu n_u								&	288	&		{} - n_d n_e n_u - \frac12 n_d n_\nu n_u						&	288	\\[5pt]
\Delta B = - \Delta L = 1 + {\rm h.c.}					&	\frac53 n_d^3 n_e +\frac52  n_\nu n_d^2 n_u - n_d^2 n_e 					&		&		\frac53 n_d^3 n_e +\frac52  n_\nu n_d^2 n_u - n_d^2 n_e 			&		\\
										&	{}-\frac12 n_d n_\nu n_u - \frac 23 n_d n_e							&	228	& 		{}-\frac12 n_d n_\nu n_u - \frac 23 n_d n_e					&	228	\\[5pt]
\midrule
\text{Total}								&																& 3099	&															& 2864	\\
\bottomrule
\end{array}
\end{adjustbox}
\end{align*}
\caption{Number of operators in LEFT at dimensions three, five, and six, divided into $CP$-even and $CP$-odd operators.  
The number of operators is given for an arbitrary number of neutrino flavors $n_\nu$, charged lepton flavors $n_e$, $u$-type quark flavors $n_u$, and $d$-type quark flavors $n_d$, and for the case of SM LEFT with $n_\nu=3$, $n_e=3$, $n_u=2$, and $n_d=3$.  The LEFT operators in each category are given explicitly in Tables~\ref{dim3}--\ref{dim6bml}. }
\label{dim6cp}
\end{table}

\end{appendix}

\clearpage


\bibliographystyle{JHEP}
\bibliography{RGL}

\providecommand{\href}[2]{#2}\begingroup\raggedright\begin{thebibliography}{10}

\bibitem{Buchmuller:1985jz}
W.~Buchm\"uller and D.~Wyler, {\it {Effective Lagrangian Analysis of New
  Interactions and Flavor Conservation}},  {\em Nucl.Phys.} {\bf B268} (1986)
  621.

\bibitem{Grzadkowski:2010es}
B.~Grzadkowski, M.~Iskrzy\'nski, M.~Misiak, and J.~Rosiek, {\it {Dimension-Six
  Terms in the Standard Model Lagrangian}},  {\em JHEP} {\bf 10} (2010) 085,
  [\href{http://arxiv.org/abs/1008.4884}{{\tt arXiv:1008.4884}}].

\bibitem{Alonso:2013hga}
R.~Alonso, E.~E. Jenkins, A.~V. Manohar, and M.~Trott, {\it {Renormalization
  Group Evolution of the Standard Model Dimension Six Operators III: Gauge
  Coupling Dependence and Phenomenology}},  {\em JHEP} {\bf 04} (2014) 159,
  [\href{http://arxiv.org/abs/1312.2014}{{\tt arXiv:1312.2014}}].

\bibitem{Weinberg:1979sa}
S.~Weinberg, {\it {Baryon and Lepton Nonconserving Processes}},  {\em
  Phys.Rev.Lett.} {\bf 43} (1979) 1566--1570.

\bibitem{Wilczek:1979hc}
F.~Wilczek and A.~Zee, {\it {Operator Analysis of Nucleon Decay}},  {\em
  Phys.Rev.Lett.} {\bf 43} (1979) 1571--1573.

\bibitem{Abbott:1980zj}
L.~Abbott and M.~B. Wise, {\it {The Effective Hamiltonian for Nucleon Decay}},
  {\em Phys.Rev.} {\bf D22} (1980) 2208.

\bibitem{Alonso:2014zka}
R.~Alonso, H.-M. Chang, E.~E. Jenkins, A.~V. Manohar, and B.~Shotwell, {\it
  {Renormalization group evolution of dimension-six baryon number violating
  operators}},  {\em Phys. Lett.} {\bf B734} (2014) 302--307,
  [\href{http://arxiv.org/abs/1405.0486}{{\tt arXiv:1405.0486}}].

\bibitem{Lehman:2014jma}
L.~Lehman, {\it {Extending the Standard Model Effective Field Theory with the
  Complete Set of Dimension-7 Operators}},  {\em Phys. Rev.} {\bf D90} (2014),
  no.~12 125023, [\href{http://arxiv.org/abs/1410.4193}{{\tt
  arXiv:1410.4193}}].

\bibitem{Lehman:2015coa}
L.~Lehman and A.~Martin, {\it {Low-derivative operators of the Standard Model
  effective field theory via Hilbert series methods}},  {\em JHEP} {\bf 02}
  (2016) 081, [\href{http://arxiv.org/abs/1510.00372}{{\tt arXiv:1510.00372}}].

\bibitem{Henning:2015alf}
B.~Henning, X.~Lu, T.~Melia, and H.~Murayama, {\it {2, 84, 30, 993, 560, 15456,
  11962, 261485, ...: Higher dimension operators in the SM EFT}},  {\em JHEP}
  {\bf 08} (2017) 016, [\href{http://arxiv.org/abs/1512.03433}{{\tt
  arXiv:1512.03433}}].

\bibitem{Kobach:2016ami}
A.~Kobach, {\it {Baryon Number, Lepton Number, and Operator Dimension in the
  Standard Model}},  {\em Phys. Lett.} {\bf B758} (2016) 455--457,
  [\href{http://arxiv.org/abs/1604.05726}{{\tt arXiv:1604.05726}}].

\bibitem{Liao:2016hru}
Y.~Liao and X.-D. Ma, {\it {Renormalization Group Evolution of Dimension-seven
  Baryon- and Lepton-number-violating Operators}},  {\em JHEP} {\bf 11} (2016)
  043, [\href{http://arxiv.org/abs/1607.07309}{{\tt arXiv:1607.07309}}].

\bibitem{Brivio:2017vri}
I.~Brivio and M.~Trott, {\it {The Standard Model as an Effective Field
  Theory}},  \href{http://arxiv.org/abs/1706.08945}{{\tt arXiv:1706.08945}}.

\bibitem{PDG}
{\bf Particle Data Group} Collaboration, C.~Patrignani et~al., {\it {Review of
  Particle Physics}},  {\em Chin. Phys.} {\bf C40} (2016), no.~10 100001.

\bibitem{Buchalla:1995vs}
G.~Buchalla, A.~J. Buras, and M.~E. Lautenbacher, {\it {Weak decays beyond
  leading logarithms}},  {\em Rev. Mod. Phys.} {\bf 68} (1996) 1125--1144,
  [\href{http://arxiv.org/abs/hep-ph/9512380}{{\tt hep-ph/9512380}}].

\bibitem{Aebischer:2015fzz}
J.~Aebischer, A.~Crivellin, M.~Fael, and C.~Greub, {\it {Matching of gauge
  invariant dimension-six operators for $b\to s$ and $b\to c$ transitions}},
  {\em JHEP} {\bf 05} (2016) 037, [\href{http://arxiv.org/abs/1512.02830}{{\tt
  arXiv:1512.02830}}].

\bibitem{Grojean:2013kd}
C.~Grojean, E.~E. Jenkins, A.~V. Manohar, and M.~Trott, {\it {Renormalization
  Group Scaling of Higgs Operators and $h \to \gamma \gamma$ Decay }},  {\em
  JHEP} {\bf 04} (2013) 016, [\href{http://arxiv.org/abs/1301.2588}{{\tt
  arXiv:1301.2588}}].

\bibitem{Jenkins:2013zja}
E.~E. Jenkins, A.~V. Manohar, and M.~Trott, {\it {Renormalization Group
  Evolution of the Standard Model Dimension Six Operators I: Formalism and
  lambda Dependence}},  {\em JHEP} {\bf 10} (2013) 087,
  [\href{http://arxiv.org/abs/1308.2627}{{\tt arXiv:1308.2627}}].

\bibitem{Jenkins:2013wua}
E.~E. Jenkins, A.~V. Manohar, and M.~Trott, {\it {Renormalization Group
  Evolution of the Standard Model Dimension Six Operators II: Yukawa
  Dependence}},  {\em JHEP} {\bf 01} (2014) 035,
  [\href{http://arxiv.org/abs/1310.4838}{{\tt arXiv:1310.4838}}].

\bibitem{Elias-Miro:2013gya}
J.~Elias-Mir{\'o}, J.~Espinosa, E.~Masso, and A.~Pomarol, {\it {Renormalization
  of Dimension-Six Operators Relevant for the Higgs Decays $h\rightarrow
  \gamma\gamma,\gamma Z$}},  {\em JHEP} {\bf 08} (2013) 033,
  [\href{http://arxiv.org/abs/1302.5661}{{\tt arXiv:1302.5661}}].

\bibitem{Elias-Miro:2013mua}
J.~Elias-Mir\'o, J.~Espinosa, E.~Masso, and A.~Pomarol, {\it {Higgs Windows to
  New Physics Through D = 6 Operators: Constraints and One-Loop Anomalous
  Dimensions}},  {\em JHEP} {\bf 11} (2013) 066,
  [\href{http://arxiv.org/abs/1308.1879}{{\tt arXiv:1308.1879}}].

\bibitem{Jenkins:2017dyc}
E.~E. Jenkins, A.~V. Manohar, and P.~Stoffer, {\it {Low-Energy Effective Field
  Theory below the Electroweak Scale: Anomalous Dimensions}},  {\em JHEP} {\bf
  01} (2018) 084, [\href{http://arxiv.org/abs/1711.05270}{{\tt
  arXiv:1711.05270}}].

\bibitem{Cirigliano:2012ab}
V.~Cirigliano, M.~Gonz\'alez-Alonso, and M.~L. Graesser, {\it {Non-standard
  Charged Current Interactions: beta decays versus the LHC}},  {\em JHEP} {\bf
  02} (2013) 046, [\href{http://arxiv.org/abs/1210.4553}{{\tt
  arXiv:1210.4553}}].

\bibitem{Dekens:2013zca}
W.~Dekens and J.~de~Vries, {\it {Renormalization Group Running of Dimension-Six
  Sources of Parity and Time-Reversal Violation}},  {\em JHEP} {\bf 05} (2013)
  149, [\href{http://arxiv.org/abs/1303.3156}{{\tt arXiv:1303.3156}}].

\bibitem{Bhattacharya:2015rsa}
T.~Bhattacharya, V.~Cirigliano, R.~Gupta, E.~Mereghetti, and B.~Yoon, {\it
  {Dimension-5 CP-odd operators: QCD mixing and renormalization}},  {\em Phys.
  Rev.} {\bf D92} (2015), no.~11 114026,
  [\href{http://arxiv.org/abs/1502.07325}{{\tt arXiv:1502.07325}}].

\bibitem{Davidson:2016edt}
S.~Davidson, {\it {$\mu \rightarrow e \gamma $ and matching at ${m_W}$}},  {\em
  Eur. Phys. J.} {\bf C76} (2016), no.~7 370,
  [\href{http://arxiv.org/abs/1601.07166}{{\tt arXiv:1601.07166}}].

\bibitem{Crivellin:2017rmk}
A.~Crivellin, S.~Davidson, G.~M. Pruna, and A.~Signer, {\it
  {Renormalisation-group improved analysis of $\mu\to e$ processes in a
  systematic effective-field-theory approach}},  {\em JHEP} {\bf 05} (2017)
  117, [\href{http://arxiv.org/abs/1702.03020}{{\tt arXiv:1702.03020}}].

\bibitem{Cirigliano:2017azj}
V.~Cirigliano, S.~Davidson, and Y.~Kuno, {\it {Spin-dependent $\mu \to e$
  conversion}},  {\em Phys. Lett.} {\bf B771} (2017) 242--246,
  [\href{http://arxiv.org/abs/1703.02057}{{\tt arXiv:1703.02057}}].

\bibitem{Celis:2017hod}
A.~Celis, J.~Fuentes-Mart\'in, A.~Vicente, and J.~Virto, {\it {DsixTools: The
  Standard Model Effective Field Theory Toolkit}},  {\em Eur. Phys. J.} {\bf
  C77} (2017), no.~6 405, [\href{http://arxiv.org/abs/1704.04504}{{\tt
  arXiv:1704.04504}}].

\bibitem{Aebischer:2017gaw}
J.~Aebischer, M.~Fael, C.~Greub, and J.~Virto, {\it {B physics Beyond the
  Standard Model at One Loop: Complete Renormalization Group Evolution below
  the Electroweak Scale}},  {\em JHEP} {\bf 09} (2017) 158,
  [\href{http://arxiv.org/abs/1704.06639}{{\tt arXiv:1704.06639}}].

\bibitem{Gonzalez-Alonso:2017iyc}
M.~Gonz{\'a}lez-Alonso, J.~Martin~Camalich, and K.~Mimouni, {\it
  {Renormalization-group evolution of new physics contributions to
  (semi)leptonic meson decays}},  {\em Phys. Lett.} {\bf B772} (2017) 777--785,
  [\href{http://arxiv.org/abs/1706.00410}{{\tt arXiv:1706.00410}}].

\bibitem{Falkowski:2017pss}
A.~Falkowski, M.~Gonz{\'a}lez-Alonso, and K.~Mimouni, {\it {Compilation of
  low-energy constraints on 4-fermion operators in the SMEFT}},  {\em JHEP}
  {\bf 08} (2017) 123, [\href{http://arxiv.org/abs/1706.03783}{{\tt
  arXiv:1706.03783}}].

\bibitem{Misiak:2004ew}
M.~Misiak and M.~Steinhauser, {\it {Three loop matching of the dipole operators
  for $b \to s \gamma$ and $b \to s g$}},  {\em Nucl. Phys.} {\bf B683} (2004)
  277--305, [\href{http://arxiv.org/abs/hep-ph/0401041}{{\tt hep-ph/0401041}}].

\bibitem{Czakon:2006ss}
M.~Czakon, U.~Haisch, and M.~Misiak, {\it {Four-Loop Anomalous Dimensions for
  Radiative Flavour-Changing Decays}},  {\em JHEP} {\bf 03} (2007) 008,
  [\href{http://arxiv.org/abs/hep-ph/0612329}{{\tt hep-ph/0612329}}].

\bibitem{Misiak:2017woa}
M.~Misiak, A.~Rehman, and M.~Steinhauser, {\it {NNLO QCD counterterm
  contributions to $\bar B \to X_{s\gamma}$ for the physical value of $m_c$}},
  {\em Phys. Lett.} {\bf B770} (2017) 431--439,
  [\href{http://arxiv.org/abs/1702.07674}{{\tt arXiv:1702.07674}}].

\bibitem{Feruglio:1992wf}
F.~Feruglio, {\it {The Chiral approach to the electroweak interactions}},  {\em
  Int. J. Mod. Phys.} {\bf A8} (1993) 4937--4972,
  [\href{http://arxiv.org/abs/hep-ph/9301281}{{\tt hep-ph/9301281}}].

\bibitem{Grinstein:2007iv}
B.~Grinstein and M.~Trott, {\it {A Higgs-Higgs bound state due to new physics
  at a TeV}},  {\em Phys. Rev.} {\bf D76} (2007) 073002,
  [\href{http://arxiv.org/abs/0704.1505}{{\tt arXiv:0704.1505}}].

\bibitem{Gavela:2016bzc}
B.~M. Gavela, E.~E. Jenkins, A.~V. Manohar, and L.~Merlo, {\it {Analysis of
  General Power Counting Rules in Effective Field Theory}},  {\em Eur. Phys.
  J.} {\bf C76} (2016), no.~9 485, [\href{http://arxiv.org/abs/1601.07551}{{\tt
  arXiv:1601.07551}}].

\bibitem{Alonso:2014csa}
R.~Alonso, B.~Grinstein, and J.~Martin~Camalich, {\it {$SU(2)\times U(1)$ gauge
  invariance and the shape of new physics in rare $B$ decays}},  {\em Phys.
  Rev. Lett.} {\bf 113} (2014) 241802,
  [\href{http://arxiv.org/abs/1407.7044}{{\tt arXiv:1407.7044}}].

\bibitem{Alonso:2014rga}
R.~Alonso, E.~E. Jenkins, and A.~V. Manohar, {\it {Holomorphy without
  Supersymmetry in the Standard Model Effective Field Theory}},  {\em Phys.
  Lett.} {\bf B739} (2014) 95--98, [\href{http://arxiv.org/abs/1409.0868}{{\tt
  arXiv:1409.0868}}].

\bibitem{Babu:1993qv}
K.~S. Babu, C.~N. Leung, and J.~T. Pantaleone, {\it {Renormalization of the
  neutrino mass operator}},  {\em Phys. Lett.} {\bf B319} (1993) 191--198,
  [\href{http://arxiv.org/abs/hep-ph/9309223}{{\tt hep-ph/9309223}}].

\bibitem{Antusch:2001ck}
S.~Antusch, M.~Drees, J.~Kersten, M.~Lindner, and M.~Ratz, {\it {Neutrino mass
  operator renormalization revisited}},  {\em Phys. Lett.} {\bf B519} (2001)
  238--242, [\href{http://arxiv.org/abs/hep-ph/0108005}{{\tt hep-ph/0108005}}].

\bibitem{Grinstein:1991cd}
B.~Grinstein and M.~B. Wise, {\it {Operator analysis for precision electroweak
  physics}},  {\em Phys.Lett.} {\bf B265} (1991) 326--334.

\bibitem{Alioli:2017ces}
S.~Alioli, V.~Cirigliano, W.~Dekens, J.~de~Vries, and E.~Mereghetti, {\it
  {Right-handed charged currents in the era of the Large Hadron Collider}},
  {\em JHEP} {\bf 05} (2017) 086, [\href{http://arxiv.org/abs/1703.04751}{{\tt
  arXiv:1703.04751}}].

\bibitem{Dedes:2017zog}
A.~Dedes, W.~Materkowska, M.~Paraskevas, J.~Rosiek, and K.~Suxho, {\it {Feynman
  rules for the Standard Model Effective Field Theory in $R_\xi$-gauges}},
  {\em JHEP} {\bf 06} (2017) 143, [\href{http://arxiv.org/abs/1704.03888}{{\tt
  arXiv:1704.03888}}].

\bibitem{Broncano:2003fq}
A.~Broncano, M.~B. Gavela, and E.~E. Jenkins, {\it {Neutrino physics in the
  seesaw model}},  {\em Nucl. Phys.} {\bf B672} (2003) 163--198,
  [\href{http://arxiv.org/abs/hep-ph/0307058}{{\tt hep-ph/0307058}}].

\bibitem{Broncano:2002rw}
A.~Broncano, M.~B. Gavela, and E.~E. Jenkins, {\it {The Effective Lagrangian
  for the seesaw model of neutrino mass and leptogenesis}},  {\em Phys. Lett.}
  {\bf B552} (2003) 177--184, [\href{http://arxiv.org/abs/hep-ph/0210271}{{\tt
  hep-ph/0210271}}]. [Erratum: Phys. Lett. {\bf B636} (2006) 332].

\bibitem{Broncano:2004tz}
A.~Broncano, M.~B. Gavela, and E.~E. Jenkins, {\it {Renormalization of lepton
  mixing for Majorana neutrinos}},  {\em Nucl. Phys.} {\bf B705} (2005)
  269--295, [\href{http://arxiv.org/abs/hep-ph/0406019}{{\tt hep-ph/0406019}}].

\bibitem{Galison:1983pa}
P.~Galison and A.~Manohar, {\it {Two Z's or not two Z's?}},  {\em Phys. Lett.}
  {\bf B136} (1984) 279--283.

\bibitem{Jenkins:2009dy}
E.~E. Jenkins and A.~V. Manohar, {\it {Algebraic Structure of Lepton and Quark
  Flavor Invariants and CP Violation}},  {\em JHEP} {\bf 10} (2009) 094,
  [\href{http://arxiv.org/abs/0907.4763}{{\tt arXiv:0907.4763}}].

\bibitem{Hanany:2010vu}
A.~Hanany, E.~E. Jenkins, A.~V. Manohar, and G.~Torri, {\it {Hilbert Series for
  Flavor Invariants of the Standard Model}},  {\em JHEP} {\bf 03} (2011) 096,
  [\href{http://arxiv.org/abs/1010.3161}{{\tt arXiv:1010.3161}}].

\bibitem{Henning:2017fpj}
B.~Henning, X.~Lu, T.~Melia, and H.~Murayama, {\it {Operator bases,
  $S$-matrices, and their partition functions}},  {\em JHEP} {\bf 10} (2017)
  199, [\href{http://arxiv.org/abs/1706.08520}{{\tt arXiv:1706.08520}}].

\bibitem{Kobach:fr}
A.~Kobach and J.~Song. {\tt Mathematica} code to count invariants, unpublished.

\bibitem{Heeck:2013rpa}
J.~Heeck and W.~Rodejohann, {\it {Neutrinoless Quadruple Beta Decay}},  {\em
  EPL} {\bf 103} (2013), no.~3 32001,
  [\href{http://arxiv.org/abs/1306.0580}{{\tt arXiv:1306.0580}}].

\bibitem{Arnold:2017bnh}
{\bf NEMO-3} Collaboration, R.~Arnold et~al., {\it {Search for neutrinoless
  quadruple-$\beta$ decay of $^{150}$Nd with the NEMO-3 detector}},  {\em Phys.
  Rev. Lett.} {\bf 119} (2017), no.~4 041801,
  [\href{http://arxiv.org/abs/1705.08847}{{\tt arXiv:1705.08847}}].

\bibitem{Bordone:2017anc}
M.~Bordone, G.~Isidori, and S.~Trifinopoulos, {\it {Semileptonic $B$-physics
  anomalies: A general EFT analysis within $U(2)^n$ flavor symmetry}},  {\em
  Phys. Rev.} {\bf D96} (2017), no.~1 015038,
  [\href{http://arxiv.org/abs/1702.07238}{{\tt arXiv:1702.07238}}].

\bibitem{Pich:2013lsa}
A.~Pich, {\it {Precision Tau Physics}},  {\em Prog. Part. Nucl. Phys.} {\bf 75}
  (2014) 41--85, [\href{http://arxiv.org/abs/1310.7922}{{\tt
  arXiv:1310.7922}}].

\bibitem{Lees:2013uzd}
{\bf BaBar} Collaboration, J.~P. Lees et~al., {\it {Measurement of an Excess of
  $\bar{B} \to D^{(*)}\tau^- \bar{\nu}_\tau$ Decays and Implications for
  Charged Higgs Bosons}},  {\em Phys. Rev.} {\bf D88} (2013), no.~7 072012,
  [\href{http://arxiv.org/abs/1303.0571}{{\tt arXiv:1303.0571}}].

\bibitem{Aaij:2015yra}
{\bf LHCb} Collaboration, R.~Aaij et~al., {\it {Measurement of the ratio of
  branching fractions $\mathcal{B}(\bar{B}^0 \to
  D^{*+}\tau^{-}\bar{\nu}_{\tau})/\mathcal{B}(\bar{B}^0 \to
  D^{*+}\mu^{-}\bar{\nu}_{\mu})$}},  {\em Phys. Rev. Lett.} {\bf 115} (2015),
  no.~11 111803, [\href{http://arxiv.org/abs/1506.08614}{{\tt
  arXiv:1506.08614}}]. [Erratum: Phys. Rev. Lett. {\bf 115} (2015), no. 15
  159901].

\bibitem{Hirose:2016wfn}
{\bf Belle} Collaboration, S.~Hirose et~al., {\it {Measurement of the $\tau$
  lepton polarization and $R(D^*)$ in the decay $\bar{B} \to D^* \tau^-
  \bar{\nu}_\tau$}},  {\em Phys. Rev. Lett.} {\bf 118} (2017), no.~21 211801,
  [\href{http://arxiv.org/abs/1612.00529}{{\tt arXiv:1612.00529}}].

\bibitem{Aaij:2014ora}
{\bf LHCb} Collaboration, R.~Aaij et~al., {\it {Test of lepton universality
  using $B^{+}\rightarrow K^{+}\ell^{+}\ell^{-}$ decays}},  {\em Phys. Rev.
  Lett.} {\bf 113} (2014) 151601, [\href{http://arxiv.org/abs/1406.6482}{{\tt
  arXiv:1406.6482}}].

\bibitem{Aaij:2017vbb}
{\bf LHCb} Collaboration, R.~Aaij et~al., {\it {Test of lepton universality
  with $B^{0} \rightarrow K^{*0}\ell^{+}\ell^{-}$ decays}},  {\em JHEP} {\bf
  08} (2017) 055, [\href{http://arxiv.org/abs/1705.05802}{{\tt
  arXiv:1705.05802}}].

\bibitem{Baron:2013eja}
{\bf ACME} Collaboration, J.~Baron et~al., {\it {Order of Magnitude Smaller
  Limit on the Electric Dipole Moment of the Electron}},  {\em Science} {\bf
  343} (2014) 269--272, [\href{http://arxiv.org/abs/1310.7534}{{\tt
  arXiv:1310.7534}}].

\bibitem{Bennett:2008dy}
{\bf Muon $(g-2)$} Collaboration, G.~W. Bennett et~al., {\it {An Improved Limit
  on the Muon Electric Dipole Moment}},  {\em Phys. Rev.} {\bf D80} (2009)
  052008, [\href{http://arxiv.org/abs/0811.1207}{{\tt arXiv:0811.1207}}].

\bibitem{Hanneke:2008tm}
D.~Hanneke, S.~Fogwell, and G.~Gabrielse, {\it {New Measurement of the Electron
  Magnetic Moment and the Fine Structure Constant}},  {\em Phys. Rev. Lett.}
  {\bf 100} (2008) 120801, [\href{http://arxiv.org/abs/0801.1134}{{\tt
  arXiv:0801.1134}}].

\bibitem{Bennett:2006fi}
{\bf Muon $(g-2)$} Collaboration, G.~W. Bennett et~al., {\it {Final Report of
  the Muon E821 Anomalous Magnetic Moment Measurement at BNL}},  {\em Phys.
  Rev.} {\bf D73} (2006) 072003,
  [\href{http://arxiv.org/abs/hep-ex/0602035}{{\tt hep-ex/0602035}}].

\bibitem{Mohr:2015ccw}
P.~J. Mohr, D.~B. Newell, and B.~N. Taylor, {\it {CODATA Recommended Values of
  the Fundamental Physical Constants: 2014}},  {\em Rev. Mod. Phys.} {\bf 88}
  (2016), no.~3 035009, [\href{http://arxiv.org/abs/1507.07956}{{\tt
  arXiv:1507.07956}}].

\bibitem{Blum:2013xva}
T.~Blum, A.~Denig, I.~Logashenko, E.~de~Rafael, B.~Lee~Roberts, T.~Teubner, and
  G.~Venanzoni, {\it {The Muon (g-2) Theory Value: Present and Future}},
  \href{http://arxiv.org/abs/1311.2198}{{\tt arXiv:1311.2198}}.

\bibitem{Ligeti:2016npd}
Z.~Ligeti, M.~Papucci, and D.~J. Robinson, {\it {New Physics in the Visible
  Final States of $B\to D^{(*)}\tau\nu$}},  {\em JHEP} {\bf 01} (2017) 083,
  [\href{http://arxiv.org/abs/1610.02045}{{\tt arXiv:1610.02045}}].

\bibitem{Fuentes-Martin:2020zaz}
J.~Fuentes-Mart\'in, P.~Ruiz-Femen\'ia, A.~Vicente, and J.~Virto, {\it
  {DsixTools 2.0: The Effective Field Theory Toolkit}},  {\em Eur. Phys. J. C}
  {\bf 81} (2021), no.~2 167, [\href{http://arxiv.org/abs/2010.16341}{{\tt
  arXiv:2010.16341}}].

\end{thebibliography}\endgroup


\end{document}